\documentclass[10pt,twocolumn,letterpaper]{article}
\usepackage[]{cvpr}

\definecolor{cvprblue}{rgb}{0.21,0.49,0.74}
\usepackage[pagebackref,breaklinks,colorlinks,allcolors=cvprblue]{hyperref}

\usepackage{xcolor}
\usepackage{colortbl}

\usepackage{multirow}

\usepackage{amsmath}
\usepackage{amssymb}
\usepackage{booktabs}
\usepackage{times}
\usepackage{epsfig}
\usepackage{multirow}
\usepackage{multicol}
\usepackage{arydshln}
\usepackage{caption}
\usepackage{subcaption}
\usepackage{algpseudocode}
\usepackage{booktabs}
\usepackage{bm}
\usepackage{graphicx}

\usepackage{algorithm}
\usepackage{algpseudocode}
\usepackage{wrapfig}

\newcommand{\mjr}[1]{{\color{red}{#1}}}

\definecolor{cvprblue}{rgb}{0.21,0.49,0.74}
\newcommand{\cvprb}[1]{{\color{cvprblue}{#1}}}
\definecolor{row1}{HTML}{D8EFE7}
\definecolor{row1}{HTML}{F2F2F2}  
\definecolor{row2}{HTML}{EDF7F3}

\definecolor{row1}{HTML}{97D7C2}
\definecolor{row2}{HTML}{ADDfCE}
\definecolor{row1}{HTML}{C2E7DB}
\definecolor{row4}{HTML}{D8EFE7}
\definecolor{row5}{HTML}{EDF7F3}

\definecolor{row1}{HTML}{E0F2F7} 
\definecolor{row1}{HTML}{CCEEF5} 
\definecolor{row3}{HTML}{A6DDEB} 
\definecolor{row4}{HTML}{7FCBE2} 
\definecolor{row5}{HTML}{59B9D8} 

\definecolor{row1}{HTML}{87CEEB}
\definecolor{row2}{HTML}{A4D9F0}
\definecolor{row3}{HTML}{C1E4F5}
\definecolor{row1}{HTML}{DEEFF9}
\definecolor{row5}{HTML}{EEF7FC}

\newcommand{\nocolor}{\textcolor{red}{No}}
\newcommand{\yescolor}{\textcolor{blue}{Yes}}

\usepackage{amsmath}
\usepackage{xspace,mathtools}
\newcommand{\vct}[1]{\boldsymbol{#1}\xspace}
\newcommand{\mat}[1]{\boldsymbol{#1}\xspace}

\definecolor{l_gray}{gray}{0.95}
\definecolor{s_gray}{gray}{1.0}

\newcommand{\cellbest}{\cellcolor{red!23}}
\newcommand{\cellsecond}{\cellcolor{orange!26}}
\newcommand{\cellthird}{\cellcolor{yellow!29}}

\def\paperID{8155} 

\def\confName{CVPR}
\def\confYear{2026}

\newcommand{\dataset}{{\textit{PhysGaia}}}
\newcommand{\typeone}{{rheological}}

\title{\textit{PhysGaia}: A Physics-Aware Benchmark with Multi-Body Interactions \\
for Dynamic Novel View Synthesis
}

\author{
Mijeong Kim\thanks{Both authors contributed equally} 
\and
Gunhee Kim\footnotemark[1]
\and
Jungyoon Choi
\and
Wonjae Roh
\and
Bohyung Han
\and
{Computer Vision Laboratory, Seoul National University, Korea} \\
{\tt\small \{mijeong.kim, \,gunhee2001, \,jungyoonchoi, \,no1jj, \,bhhan\}@snu.ac.kr} \\
\href{https://cv.snu.ac.kr/research/PhysGaia/}{\tt\small https://cv.snu.ac.kr/research/PhysGaia/}
}

\begin{document}

\twocolumn[{
\maketitle
\centering
\vspace{-4mm}
\includegraphics[width=0.98\linewidth]{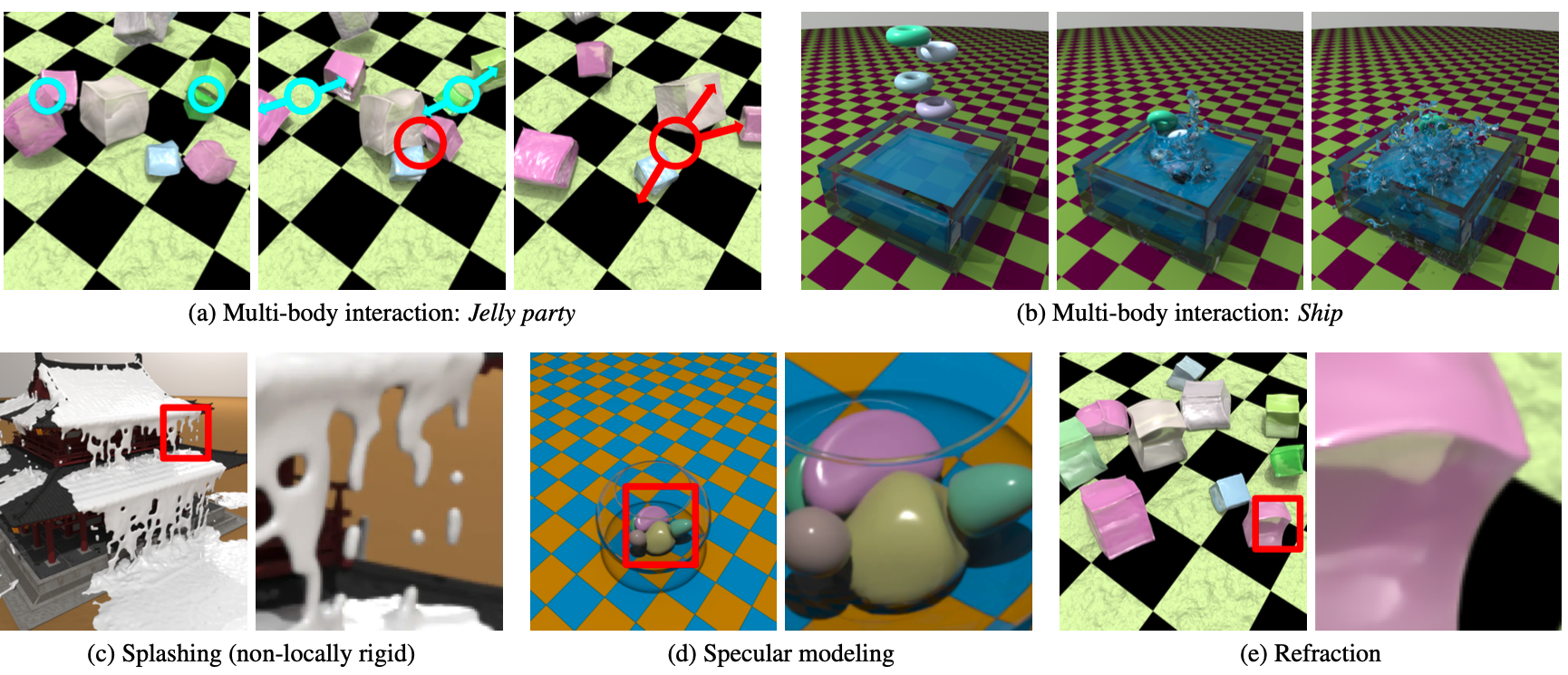}
\vspace{-1mm}
\captionof{figure}
{
Visualization of key properties of {\dataset}. 
Unlike existing benchmarks limited to single objects or materials, {\dataset} is a physics-aware benchmark featuring complex multi-body interactions across diverse substances (liquids, gases, rheological materials, and textiles). By providing ground-truth 3D trajectories and physical parameters, it uniquely enables the evaluation of physical realism alongside traditional photorealism. 
In addition to multi-body collisions, our dataset captures (c) splashing effects characterized by non-local rigid motion, as well as (d--e) complex optical phenomena such as specular reflection and refraction.
}              	
\label{fig:data_demo}
\vspace{6mm}
}]

\begingroup
\renewcommand\thefootnote{} 
\footnotetext{\protect\hspace{-0.5em}${}^*$indicates equal contribution.}
\endgroup

\setcounter{footnote}{0}
\begin{abstract}
We introduce \dataset, a novel physics-aware benchmark for Dynamic Novel View Synthesis (DyNVS) that encompasses both structured objects and unstructured physical phenomena. 
While existing datasets primarily focus on photorealistic appearance, {\dataset} is specifically designed to support physics-consistent dynamic reconstruction. 
Our benchmark features complex scenarios with rich multi-body interactions, where objects realistically collide and exchange forces. 
Furthermore, it incorporates a diverse range of materials, including liquid, gas, textile, and {\typeone} substance, moving beyond the rigid-body assumptions prevalent in prior work.
To ensure physical fidelity, all scenes in {\dataset} are generated using material-specific physics solvers that strictly adhere to fundamental physical laws. 
We provide comprehensive ground-truth information, including 3D particle trajectories and physical parameters (e.g., viscosity), enabling the quantitative evaluation of physical modeling.
To facilitate research adoption, we also provide integration pipelines for recent 4D Gaussian Splatting models along with our dataset and their results. 
By addressing the critical shortage of physics-aware benchmarks, {\dataset} can significantly advance research in dynamic view synthesis, physics-based scene understanding, and the integration of deep learning with physical simulation, ultimately enabling more faithful reconstruction and interpretation of complex dynamic scenes.
\end{abstract} 
\section{Introduction}
\label{sec:intro}

Since the emergence of Neural Radiance Fields (NeRF), Novel View Synthesis (NVS) algorithms based on deep learning have witnessed rapid advancements. 
While initial research primarily focused on static environments, recent efforts have shifted toward dynamic scene understanding to meet the demands of immersive and interactive AR/VR applications. 
This emerging direction, termed Dynamic Novel View Synthesis (DyNVS), aims to reconstruct complex 4D spatiotemporal scenes from video inputs, enabling the synthesis of photorealistic images at novel viewpoints and time steps beyond the training distribution.

The evolution of DyNVS has been closely related to the availability of diverse benchmarks. 
Early DyNVS datasets~\citep{li2022neural, pumarola2021d, yoon2020novel} primarily provided multi-view training videos, albeit with relatively constrained object motion. 
These benchmarks established the groundwork for initial research~\citep{pumarola2021d, gao2021dynamic, du2021neural, park2021nerfies, Hexplane, kplanes_2023, shao2023tensor4d, fang2022fast}, which focused on modeling temporal dynamics by deforming a canonical geometry over time. 
More recently, benchmarks featuring videos captured with handheld mobile devices have been introduced~\citep{park2021nerfies, park2021hypernerf, gao2022monocular}. 
These datasets consist of monocular sequences, prompting research to focus on mitigating overfitting under such sparse supervision. 
Building upon these foundations, subsequent works~\citep{kim2024ua4dgs, wang2024shape, zhang2024bags, huang2023sc, kwak2025modecgs, kratimenos2024dynmf, cai2024dyna, kim2026gp4dgs} have extended Gaussian Splatting (GS)~\citep{3dgs} to dynamic scenarios, aiming for high-fidelity, photorealistic rendering.

As DyNVS continues to advance, a natural next step is to move {beyond photorealism and incorporate physical realism}, enabling models not only to render how scenes look, but also to reason about how they behave.
Recent pioneering works~\cite{xie2024physgaussian, zhang2024physdreamer, liu2024physgen, borycki2024gasp, jiang2025phystwin} have begun exploring this direction by integrating differentiable physics simulation into the 4D Gaussian Splatting (4DGS) framework. 
Despite this burgeoning interest, existing research remains largely confined to 4D generation or simplified DyNVS scenarios~\cite{jiang2025phystwin, zhong2024reconstruction}, often restricted to isolated objects with a single material type. 
Consequently, more complex environments---characterized by multi-body interactions and a diverse range of physical materials such as liquids and gases---remain significantly underexplored. 
Furthermore, while real-world videos of such complex scenarios can be readily captured, they inherently lack the ground-truth physical information (e.g., 3D trajectories and material parameters) necessary for a rigorous and quantitative evaluation of physical realism.
In response, we introduce {\dataset}, a new benchmark with novel evaluation metrics designed to support and accelerate research in these challenging directions.
Figure~\ref{fig:data_demo} illustrates examples with complex physical interactions in our benchmark.
Our contributions are summarized as follows:
\begin{itemize}
\item We introduce {\dataset}, a physics-aware dataset featuring rich interactions among multiple objects and encompassing a wide range of physical materials, including liquids, gases, textiles, and {\typeone} substances.

\item {\dataset} provides essential ground-truth information, such as 3D particle trajectories and
physical parameters, enabling quantitative evaluation of physical modeling.

\item We test existing DyNVS methods on {\dataset}, revealing their fundamental limitations in achieving physical realism and demonstrating the potential for improvement in this field.
\end{itemize}

\section{Related Work}
\label{sec:related}
\subsection{Dynamic Novel View Synthesis}
In recent years, significant advances have been made in novel view synthesis~\cite{nerf, Kim_2022_CVPR, chen2022tensorf, garbin2021fastnerf, realtime_plenoctree, muller2022instant, 3dgs}. 
Although this field initially focused on static scene reconstruction, it has since extended to dynamic scenes, known as Dynamic Novel View Synthesis (DyNVS).
Early DyNVS methods were built on Neural Radiance Fields (NeRF)~\cite{gao2021dynamic, du2021neural, park2021nerfies, Hexplane, kplanes_2023, shao2023tensor4d, fang2022fast}, which usually modeled scene dynamics either by implicitly modeling with temporal inputs~\cite{gao2021dynamic, du2021neural} or directly estimating the time-wise deformation of canonical geometry through auxiliary neural networks~\cite{park2021nerfies, Hexplane, kplanes_2023, shao2023tensor4d, fang2022fast}.
Following the emergence of 3D Gaussian Splatting (3DGS)\cite{3dgs}, recent DyNVS research has shifted toward Gaussian-based representations, leading to the development of 4D Gaussian Splatting (4DGS)~\cite{huang2023sc, yang2024deformable, wu20234d, li2024spacetime, lin2024gaussian, lu20243d, guo2024motion, liang2025gaufre, lei2024mosca, duan20244d, waczynska2024d, liu2024modgs, stearns2024dynamic,kim2026gp4dgs}.
In 4DGS, an additional deformation network is employed to animate canonical Gaussian primitives over time, enabling efficient and high-quality modeling of dynamic scenes.

Modeling dynamic scenes with 4DGS is increasingly moving toward incorporating physical laws to govern motion.
PhysGaussian~\cite{xie2024physgaussian} pioneers this direction by combining an MPM~\citep{Sulsky1994} simulator with Gaussian Splatting, treating each Gaussian primitive as a particle within the particle-grid simulation framework of MPM.
This work has inspired many subsequent studies~\cite{zhong2024reconstruction, jiang2025phystwin, jiang2024vr, lin2025omniphysgs, huang2025dreamphysics, qiu2024feature} integrating physics-aware priors into 4DGS; however, these efforts remain largely confined to generation tasks~\cite{jiang2024vr, lin2025omniphysgs, huang2025dreamphysics, qiu2024feature}, with only a few addressing Dynamic Novel View Synthesis (DyNVS)~\cite{zhong2024reconstruction, jiang2025phystwin}.
Even among DyNVS methods, existing approaches are typically limited to single-object scenes with homogeneous materials, predominantly focusing on {\typeone} substances.
Consequently, physics-aware DyNVS involving rich object interactions and diverse physical materials remains largely underexplored, a gap our dataset is designed to address.

\subsection{4D Datasets for Dynamic Novel View Synthesis}

\begin{table}[t]
\centering
\caption{
{Comparison with Dynamic NVS datasets.} 
Our {\dataset} benchmark offers diverse scenes with complex multi-body interactions and complete ground-truth physical information (parameters and trajectories).
}
\label{tab:comparison_dnvs_dataset}
\vspace{-2mm}
\scalebox{0.8}{ 
\setlength\tabcolsep{4pt} 
\begin{tabular}{l|cc|ccc|c}
\toprule
\multirow{2.5}{*}{\textbf{Datasets}} & \multicolumn{2}{c|}{\textbf{Scenes}} & \multicolumn{3}{c|}{\textbf{Physics Info.}} & \multirow{2.5}{*}{\textbf{View}} \\
& Total & Phys.\textsuperscript{\dag} & Inter. & Param. & Traj. & \\
\midrule
Plenoptic~\cite{li2022neural}     & \ \ \ \ 6   & \ \ \ 1{\tiny (G)} & -- & -- & -- & Multi \\
NVIDIA Dynamic~\cite{yoon2020novel}  & \ \ \ \ 8   & \ 0            & -- & -- & -- & Mono \\
FluidNexus~\cite{gao2025fluidnexus} & 240 & 240{\tiny (G)} & \checkmark & -- & -- & Multi \\
DyCheck~\cite{gao2022monocular}   & \ \ \ \ 7   & \ \ \ 1{\tiny (T)} & -- & -- & -- & Mono \\
Nerfies~\cite{park2021nerfies}    & \ \ \ \ 4   & \ 0            & -- & -- & -- & Mono \\
HyperNeRF~\cite{park2021hypernerf} & \ \ \ \ 4   & \ 0            & -- & -- & -- & Mono \\
NeRF-DS~\cite{yan2023nerf}        & \ \ \ \ 8   & \ 0            & -- & -- & -- & Mono \\
EvDNeRF~\cite{bhattacharya2024evdnerf} & \ \ \ \ 6 & \ 0          & -- & -- & -- & Multi \\
Synthetic Soccer~\cite{lewin2023dynamic} & \ \ \ \ 3 & \ 0           & -- & -- & -- & Multi \\
HDR-HexPlane~\cite{wu2024fast}    & \ \ \ \ 8   & \ 0            & -- & -- & -- & Multi \\
D-NeRF~\cite{pumarola2021d}       & \ \ \ \ 8   & \ \ \ 1{\tiny (R)} & -- & -- & -- & Mono \\
Phystwin~\cite{jiang2025phystwin} & \ \ 22  & \ \ 22{\tiny (R)} & -- & -- & -- & Multi \\
\midrule
\textbf{\dataset~(Ours)} & \ \ \textbf{17} & \textbf{17} & \checkmark & \checkmark & \checkmark & Both \\
\bottomrule
\multicolumn{7}{l}{\textsuperscript{\dag}\small G: Gas, T: Textile, R: Rheological} \\
\end{tabular}
}
\vspace{-2mm}
\end{table}

Table~\ref{tab:comparison_dnvs_dataset} provides a comparative overview of our dataset alongside existing DyNVS datasets.
The early DyNVS datasets~\cite{li2022neural, pumarola2021d, yoon2020novel} primarily employed multiview configurations or captured scenes with very limited motion, typically involving mostly rigid objects.
These datasets paved the way for DyNVS research, with subsequent works~\cite{pumarola2021d, gao2021dynamic, du2021neural, park2021nerfies, Hexplane, kplanes_2023, shao2023tensor4d, fang2022fast} exploring deforming canonical geometries over time to model scene dynamics.
To advance DyNVS toward practical AR/VR applications, more user-friendly datasets captured using handheld mobile phones were later introduced.
Nerfies~\cite{park2021nerfies} pioneered handheld iPhone captures, though its scenes remained largely static.
HyperNeRF~\cite{park2021hypernerf} then introduced more rapid and varied motions, and DyCheck~\cite{gao2022monocular} further addressed camera teleportation artifacts observed in HyperNeRF.
These monocular datasets have motivated research aimed at mitigating overfitting to training views, with some approaches~\cite{kim2024ua4dgs, zhang2024bags, wang2024shape} leveraging additional priors such as diffusion models~\cite{rombach2022high}, depth estimators~\cite{yang2024depth}, and point trackers~\cite{yang2023cotracker}, while others~\cite{huang2023sc, wang2024shape, kratimenos2024dynmf, kwak2025modecgs, cai2024dyna} constrain deformation via motion factorization~\cite{tomasi1992shape} or As-Rigid-As-Possible (ARAP) regularization~\cite{sorkine2007as}.
Nonetheless, the primary objective across these datasets remains photorealistic reconstruction, with limited emphasis on physics-aware dynamic modeling.

Unlike existing DyNVS datasets, {\dataset} provides scenes featuring multi-body interactions across diverse physical materials, as shown in Table~\ref{tab:comparison_dnvs_dataset}.
Although some datasets~\cite{li2022neural, pumarola2021d, gao2022monocular, gao2025fluidnexus, jiang2025phystwin} include scenes with physical phenomena, they are limited in scale, lack accurate physics simulation, or do not capture complex multi-object interactions.
In this regard, {\dataset} occupies a unique position with strong potential to advance physics-aware dynamic scene modeling.


\subsection{4D Datasets from Physics Simulator}

\begin{table}[t]
\centering
\caption{
{Comparison with physics-simulated datasets for the DyNVS task.} 
Our {\dataset} benchmark offers diverse materials, multi-body interactions, and complete physical ground truth (parameters and trajectories).
}
\label{tab:comparison_physics_dataset}
\vspace{-2mm}
\scalebox{0.8}{
\setlength\tabcolsep{4pt} 
\begin{tabular}{l|c|cccc|cc}
\toprule
\multirow{2.5}{*}{\textbf{Datasets}} & \textbf{Multi-body} & \multicolumn{4}{c|}{\textbf{Materials\textsuperscript{\dag}}} & \multicolumn{2}{c}{\textbf{Physics Info.}} \\
& \textbf{Interactions} & L & G & T & R & Params. & 3D Traj. \\
\midrule
DNG~\cite{zhang2021dynamic}                & \yescolor & & & $\checkmark$ & & \nocolor & \yescolor \\
CLOTH4D~\cite{zou2023cloth4d}              & \yescolor & & & $\checkmark$ & & \nocolor & \yescolor \\
4D-DRESS~\cite{wang20244d}                 & \yescolor & & & $\checkmark$ & & \nocolor & \yescolor \\
Rasheed \textit{et al.}~\cite{rasheed2020learning} & \nocolor  & & & $\checkmark$ & & \yescolor & \nocolor \\
Deng \textit{et al.}~\cite{deng2023learning}    & \nocolor  & $\checkmark$ & & & & \nocolor & \nocolor \\
ScalarFlow~\cite{ScalarFlow2019}           & \nocolor  & & $\checkmark$ & & & \yescolor & \yescolor \\
PAC-NeRF~\cite{li2023pac}                  & \nocolor  & $\checkmark$ & & & $\checkmark$ & \yescolor & \yescolor \\
Spring-Gaus~\cite{zhong2024reconstruction} & \nocolor  & & & & $\checkmark$ & \yescolor & \yescolor \\
\midrule
\textbf{\dataset~(Ours)} & \yescolor & $\checkmark$ & $\checkmark$ & $\checkmark$ & $\checkmark$ & \yescolor & \yescolor \\
\bottomrule
\multicolumn{8}{l}{\textsuperscript{\dag}\small L: Liquid, G: Gas, T: Textile, R: Rheological} \\
\end{tabular}
}
\vspace{-2mm}
\end{table}

To provide a comprehensive comparison, we also compare {\dataset} with existing 4D datasets generated via physics simulation, as summarized in Table~\ref{tab:comparison_physics_dataset}.
Since DyNVS tasks require multiview RGB imagery, we focus on datasets that provide such data.

Some existing datasets support multi-object interactions, but either lack rich interaction dynamics or are limited to a single material type.
For instance, clothed human datasets~\cite{zhang2021dynamic, zou2023cloth4d, wang20244d} naturally exhibit textile-body interactions, yet are restricted to textile materials and human-centric motions.
In terms of ground-truth physical information, a few datasets~\cite{li2023pac, zhong2024reconstruction, ScalarFlow2019} provide both physical parameters and 3D trajectories; however, they remain confined to single-material settings and do not capture rich multi-object interactions.
PAC-NeRF~\cite{li2023pac} focuses on liquids and viscoelastic materials, but its liquid scenarios are constrained to highly viscous flows, making them behaviorally similar to viscoelastic materials.
Spring-Gaus~\cite{zhong2024reconstruction} is restricted to viscoelasticity, while ScalarFlow~\cite{ScalarFlow2019} concentrates on gas.
In contrast, {\dataset} directly addresses these limitations by offering rich multi-object interactions, a diverse range of physical materials, and accurate simulation parameters and trajectories, making it a uniquely valuable resource for advancing physics-aware understanding of dynamic scenes.

\begin{figure*}[t]
    \begin{center}
        \centering
                        \begin{subfigure}[b]{\linewidth}
        \centering
        \includegraphics[width=0.161\linewidth]{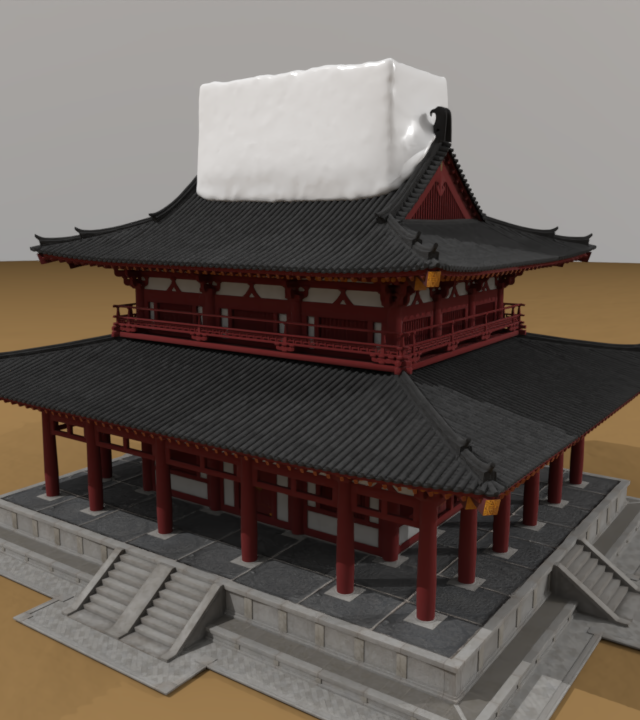}
        \includegraphics[width=0.161\linewidth]{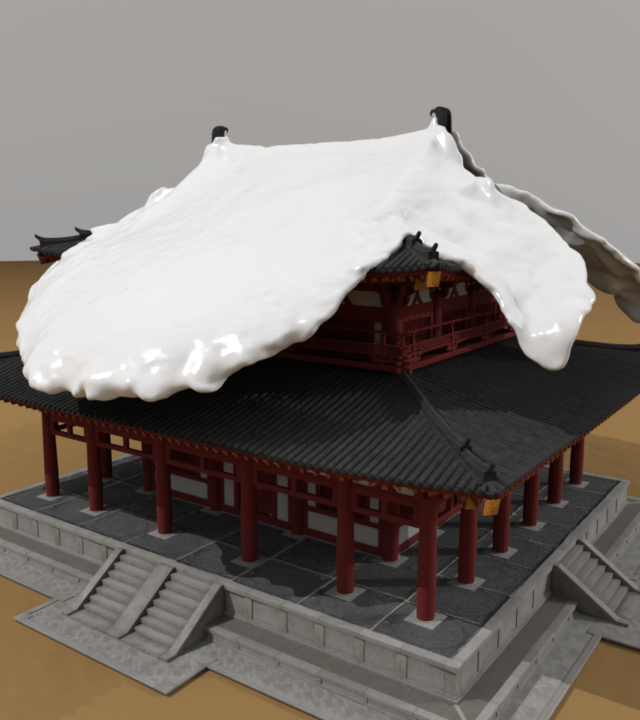}
        \includegraphics[width=0.161\linewidth]{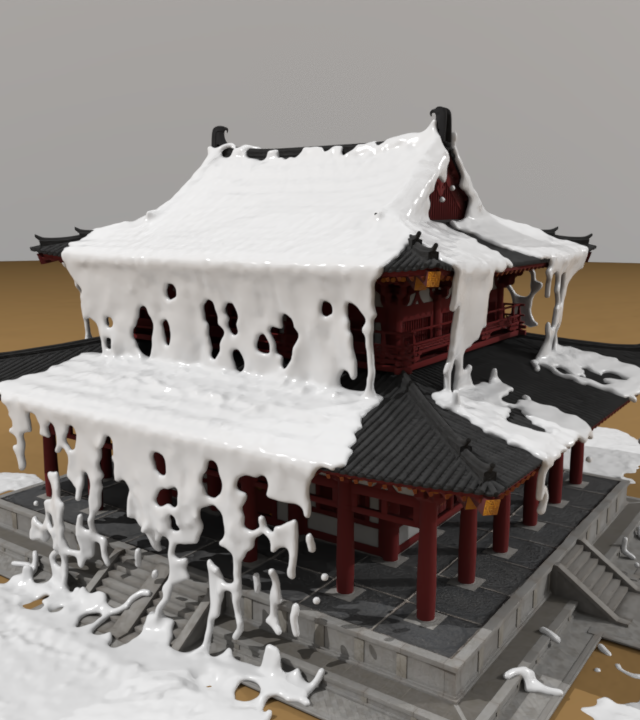}
        \includegraphics[width=0.161\linewidth]{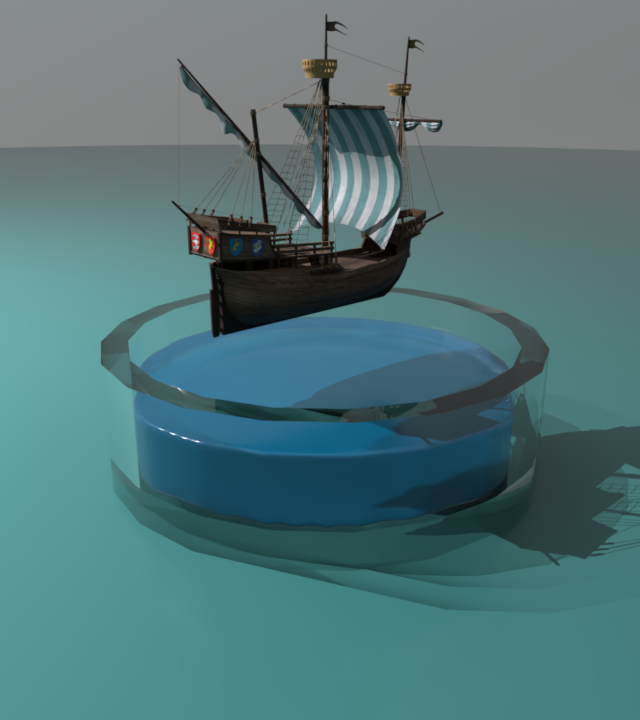}
        \includegraphics[width=0.161\linewidth]{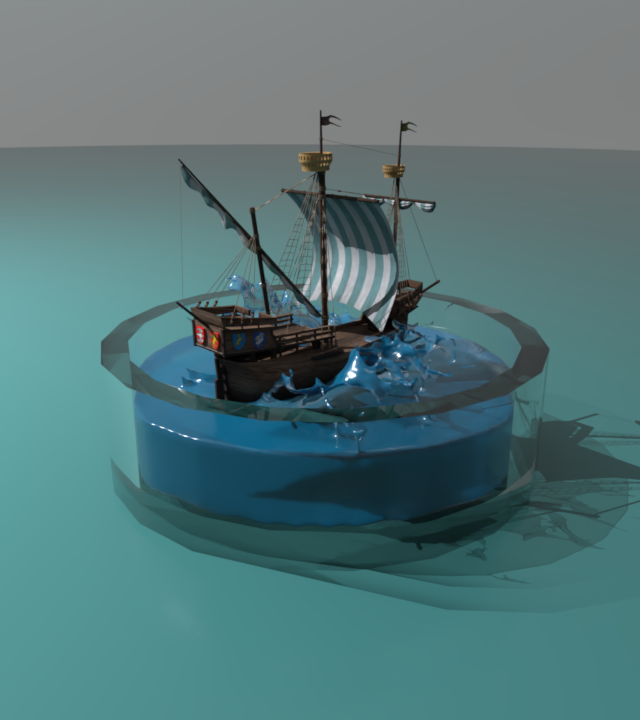}
        \includegraphics[width=0.161\linewidth]{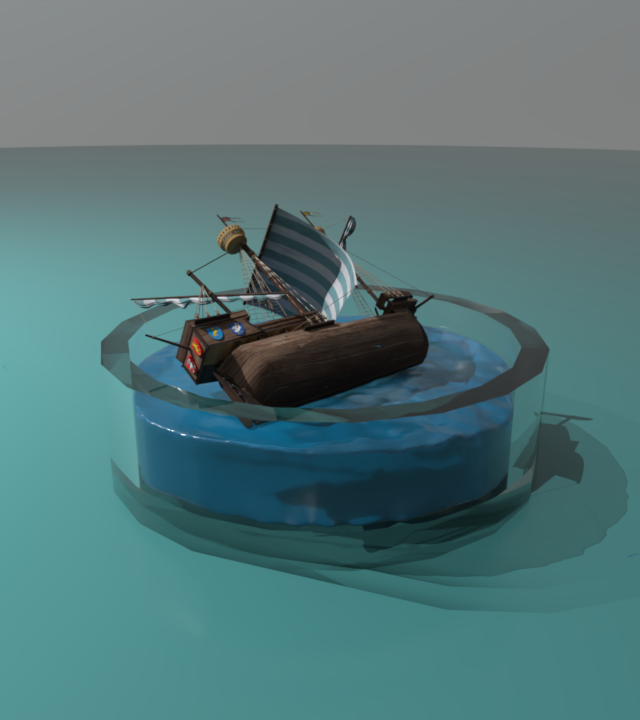}
        \caption{Liquid with FLIP solver: \textit{hanok} (left) and \textit{ship} (right)}
               \hfill \vspace{-1mm}
        \end{subfigure}
         \begin{subfigure}[b]{\linewidth}
        \centering
         \includegraphics[width=0.161\linewidth]{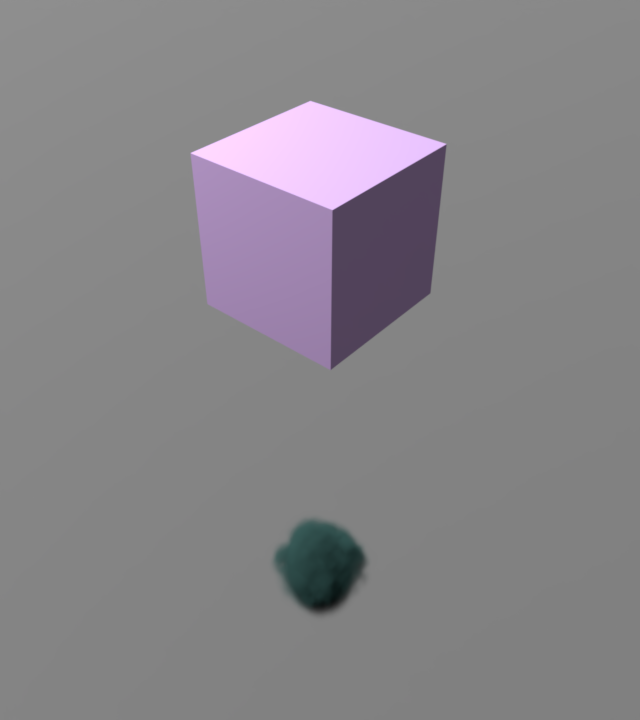}
         \includegraphics[width=0.161\linewidth]{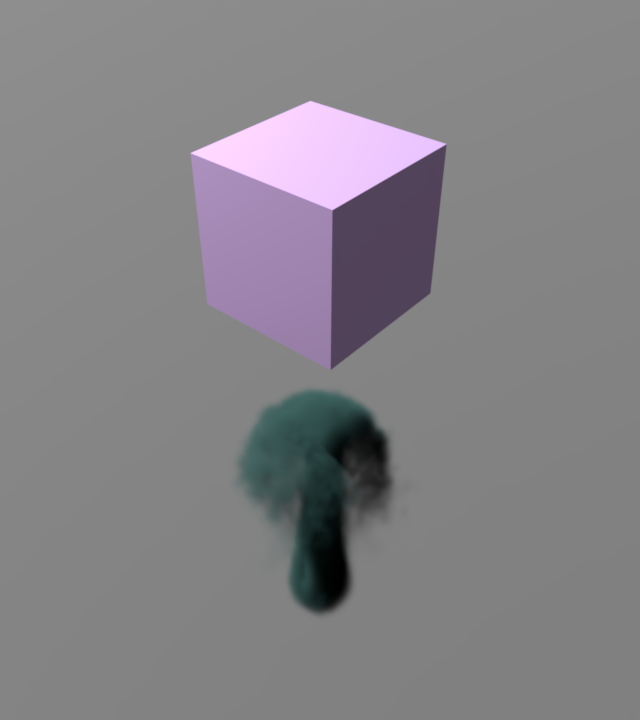}
         \includegraphics[width=0.161\linewidth]{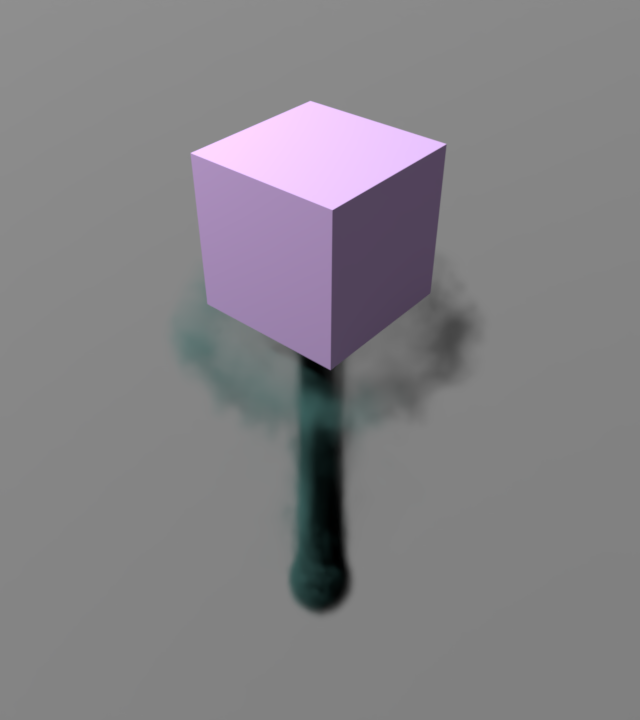}
        \includegraphics[width=0.161\linewidth]{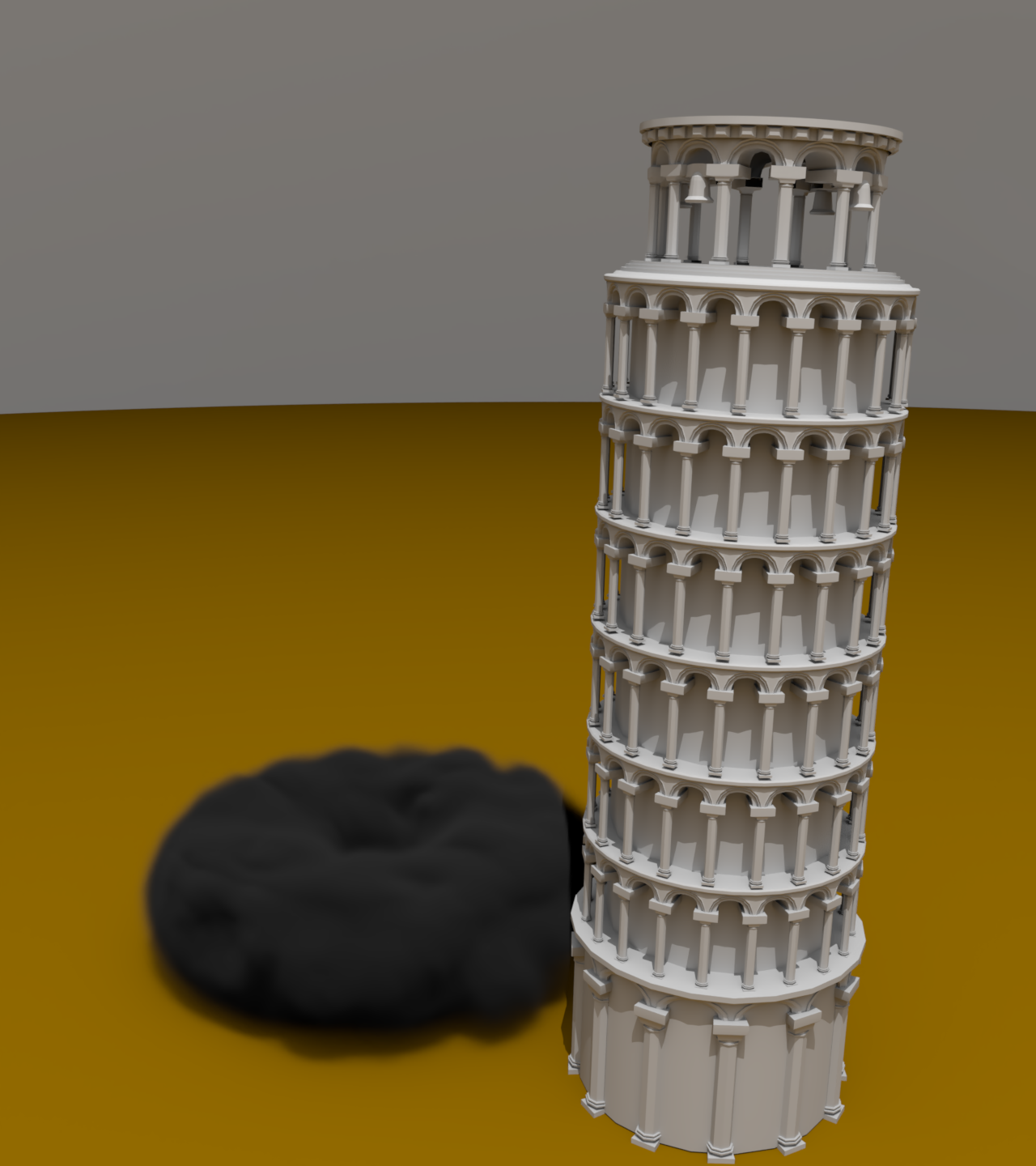}
        \includegraphics[width=0.161\linewidth]{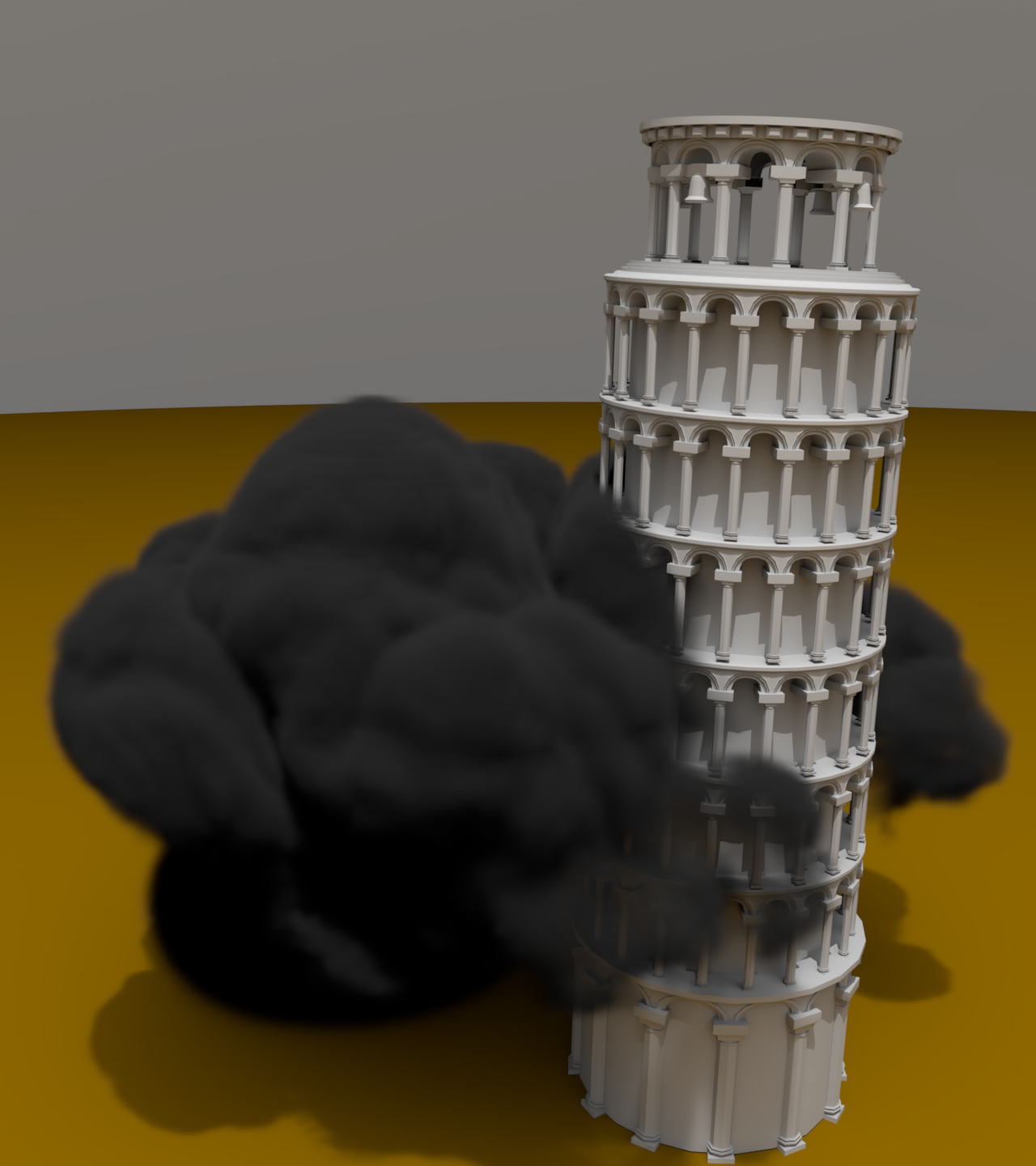}
        \includegraphics[width=0.161\linewidth]{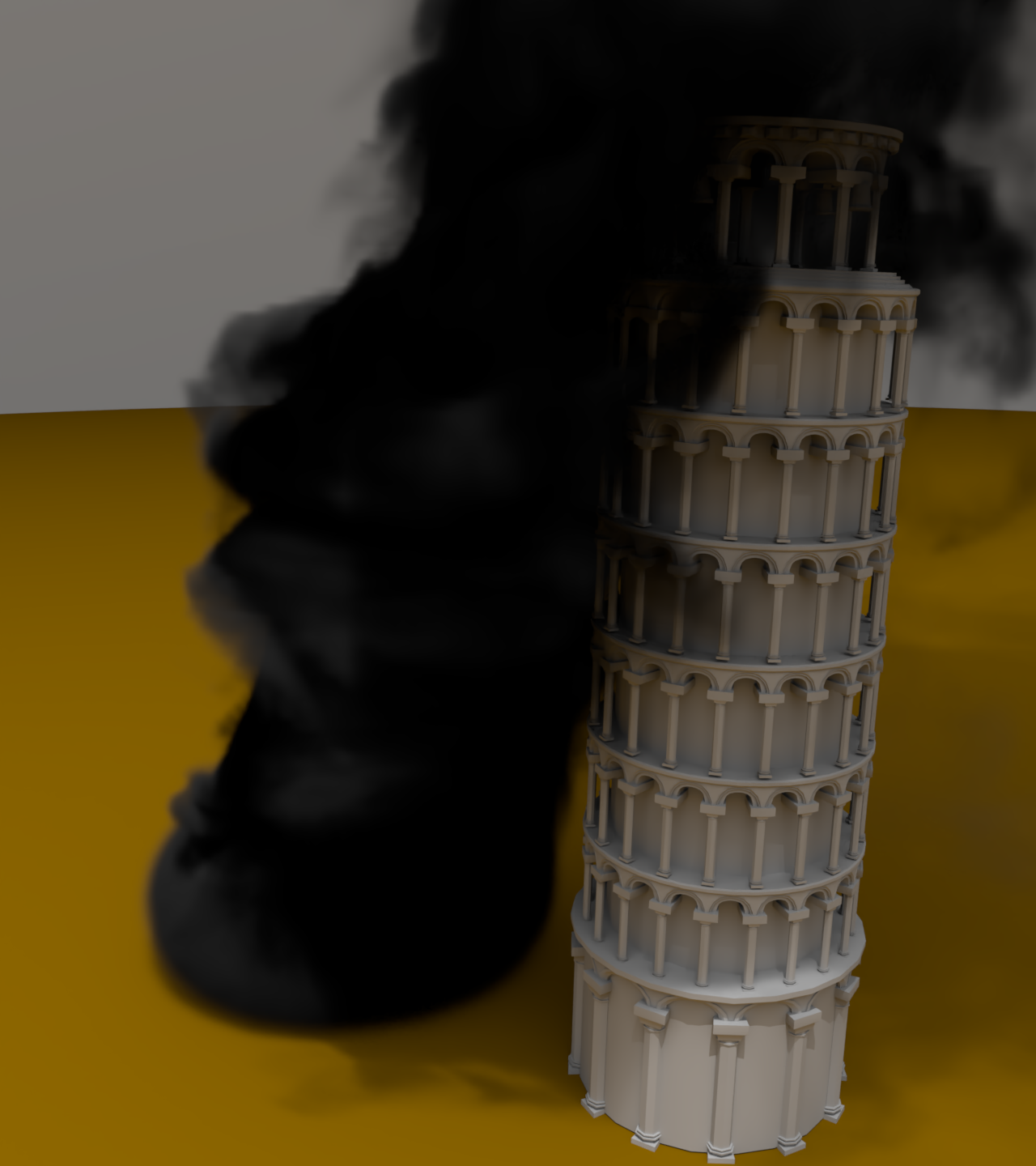}
        \centering
        \caption{Gas with Pyro solver: \textit{box-smoke} (left) and \textit{pisa} (right)}
        \hfill \vspace{-1mm}
        \end{subfigure}     
                \begin{subfigure}[b]{\linewidth}
        \centering
        \includegraphics[width=0.161\linewidth]{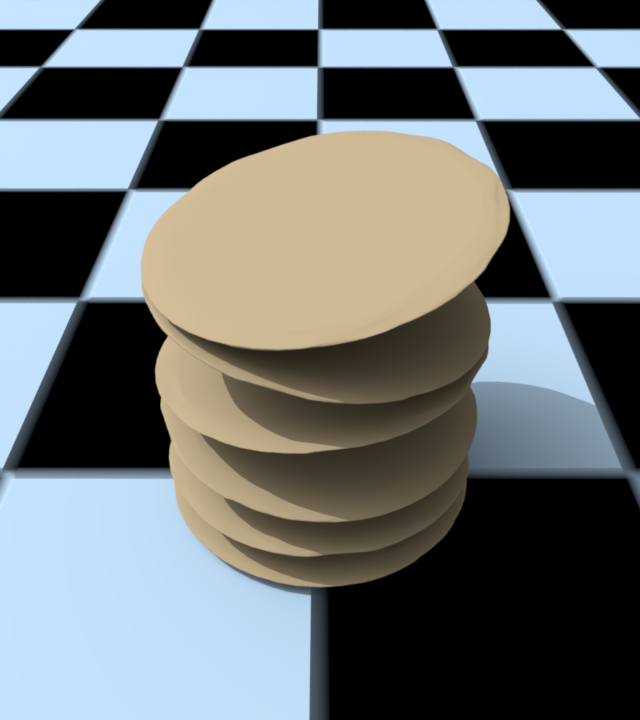}
        \includegraphics[width=0.161\linewidth]{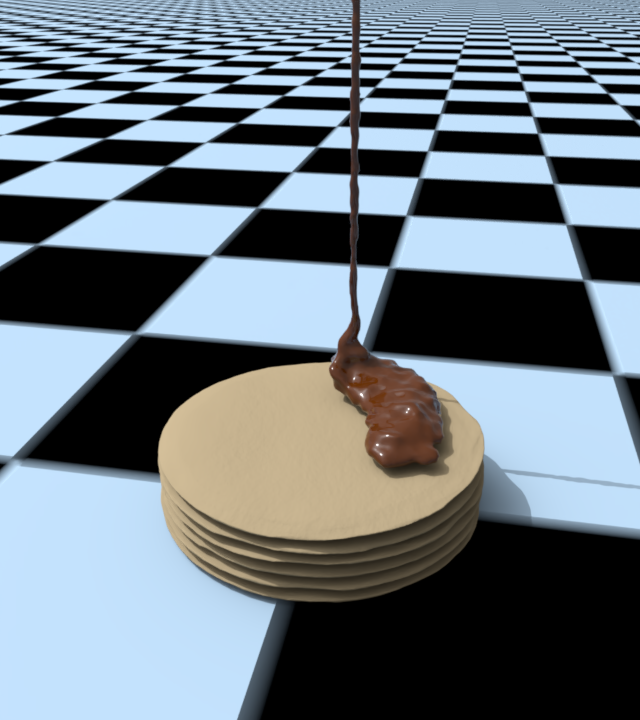}
        \includegraphics[width=0.161\linewidth]{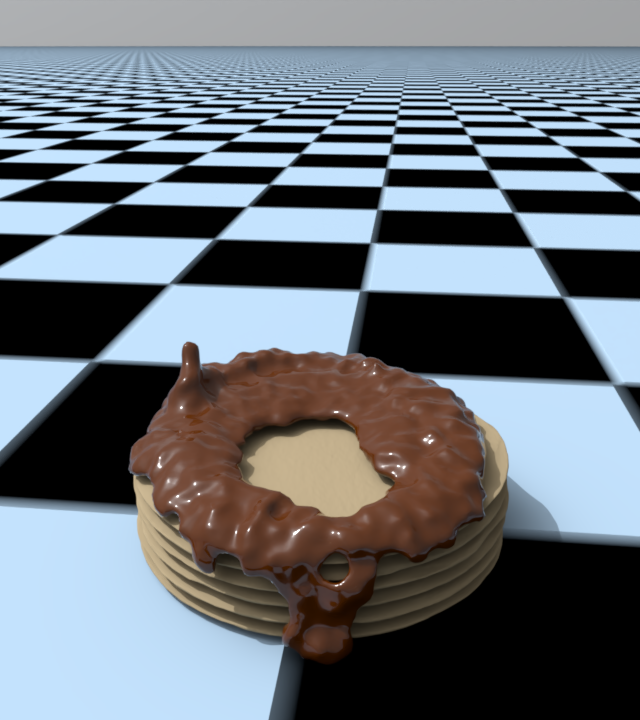}
        \includegraphics[width=0.161\linewidth]{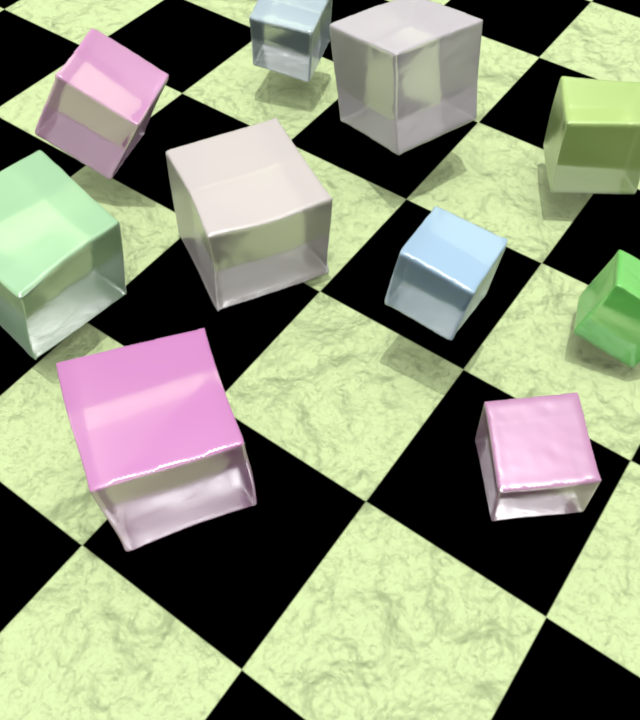}
        \includegraphics[width=0.161\linewidth]{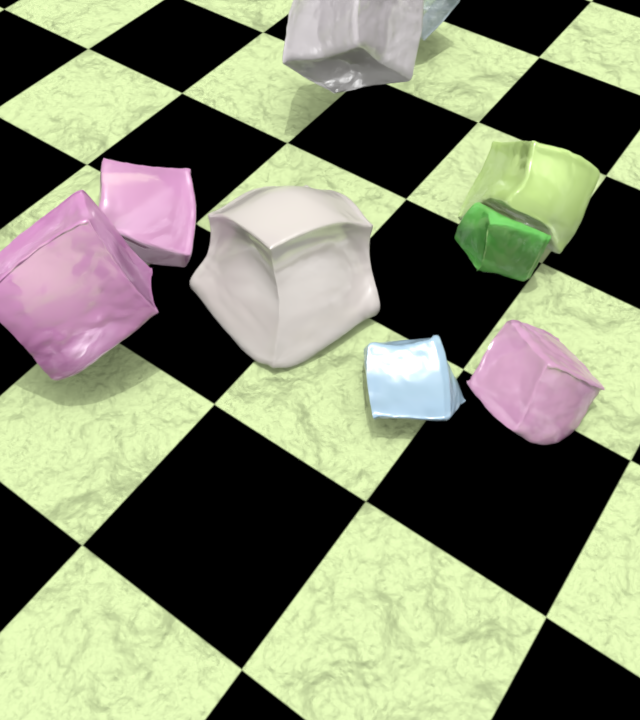}
        \includegraphics[width=0.161\linewidth]{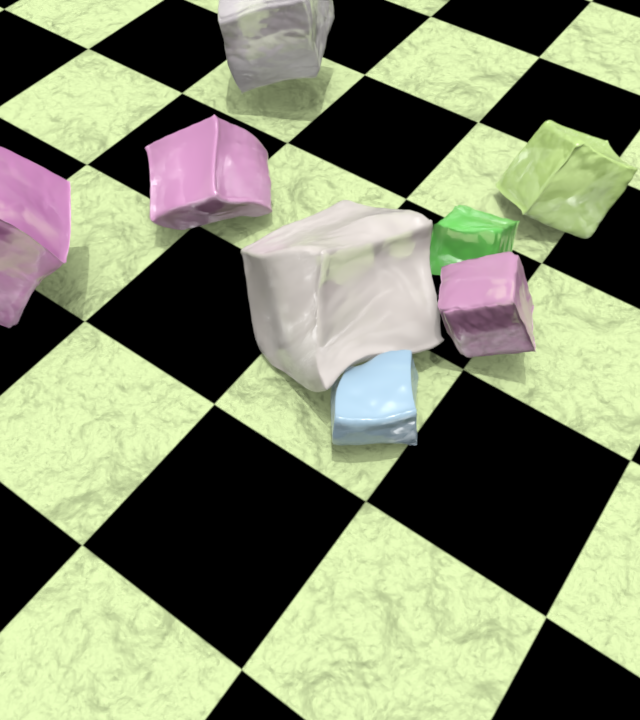}
        \caption{Rheological substances with MPM solver: \textit{pancake} (left) and \textit{jelly party} (right)}
         \hfill \vspace{-1mm}
        \end{subfigure}
                        \begin{subfigure}[b]{\linewidth}
        \centering
        \includegraphics[width=0.161\linewidth]{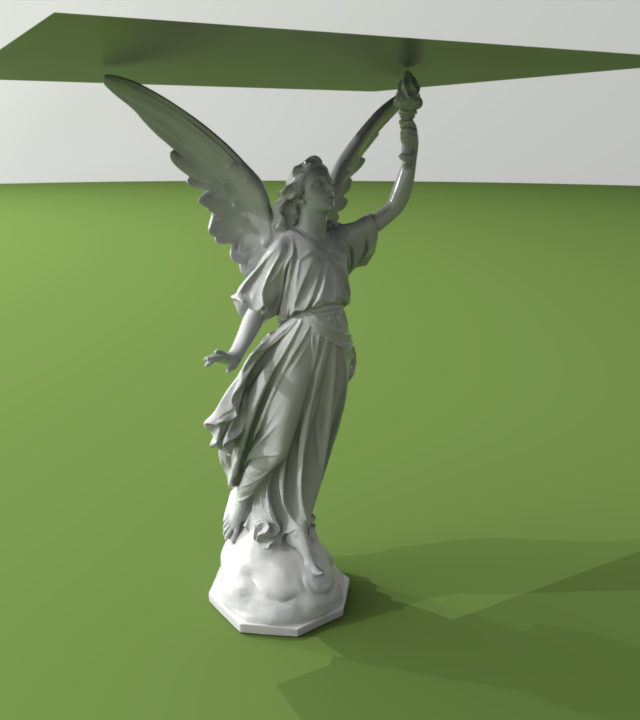}
        \includegraphics[width=0.161\linewidth]{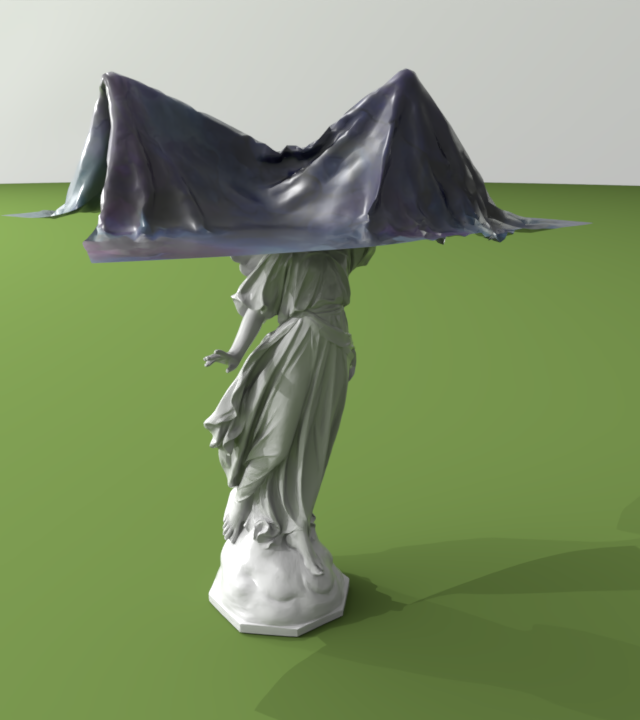}
        \includegraphics[width=0.161\linewidth]{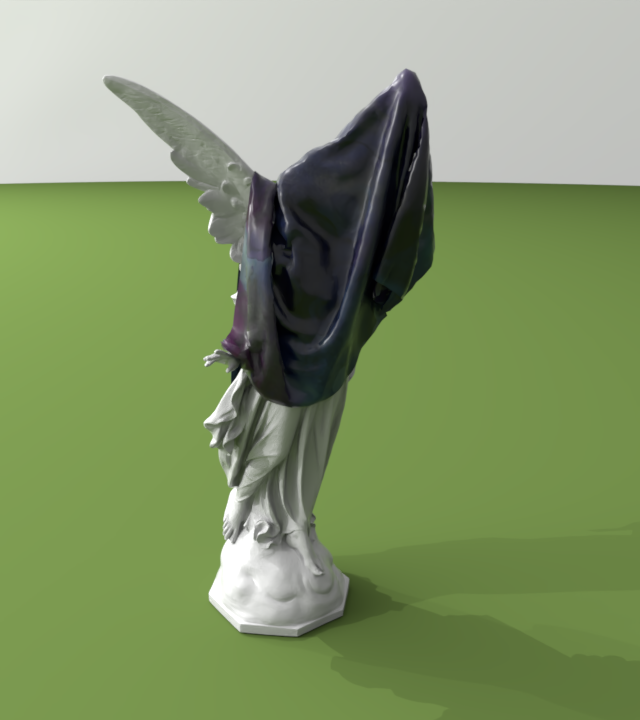}
        \includegraphics[width=0.161\linewidth]{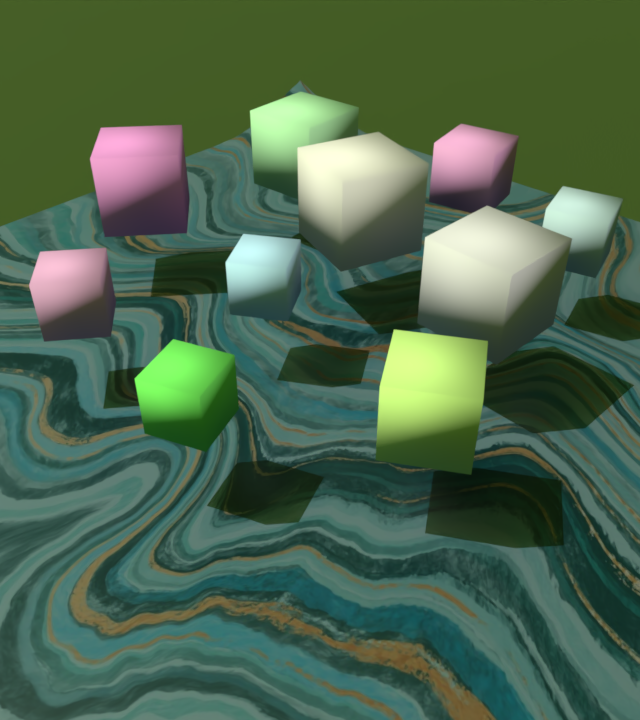}
        \includegraphics[width=0.161\linewidth]{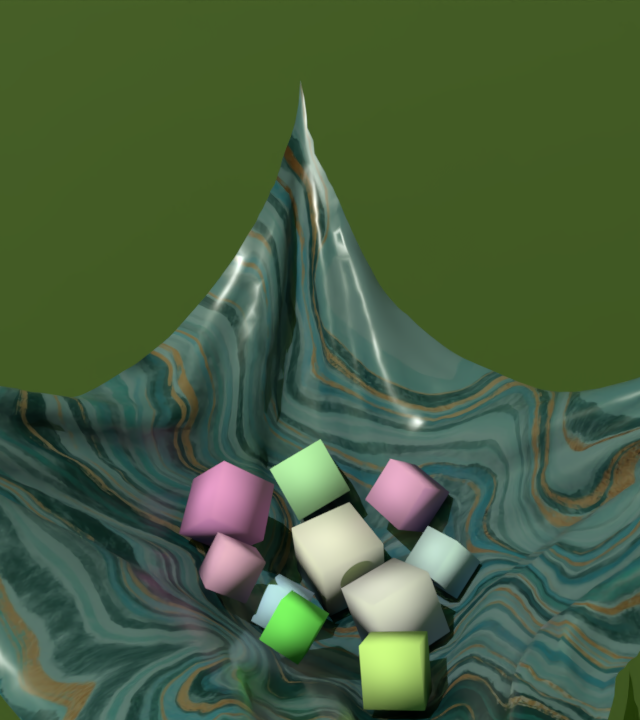}
        \includegraphics[width=0.161\linewidth]{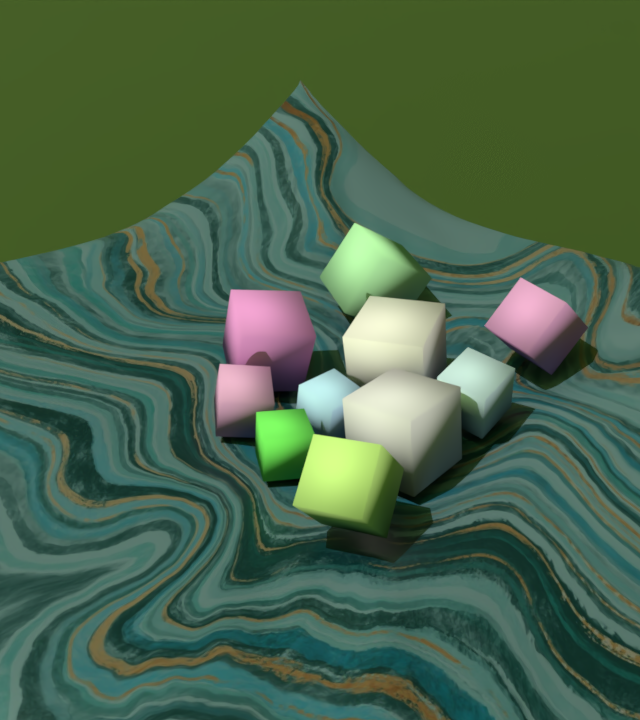}
        \centering
        \caption{Textile with Vellum solver:  \textit{lucy} (left) and \textit{basin} (right)}
           \hfill \vspace{-1mm}
        \end{subfigure}     
       \hfill\vspace{-7mm}
  \caption{
  Examples from the proposed physics-aware benchmark, {\dataset}. 
  They exhibit complex physical interactions between multiple objects composed of diverse materials such as liquid, gas, rheological substance, and textile.
  More importantly, our benchmark enables evaluation of physics realism to foster physics reasoning in dynamic scenes.}
  	\label{fig:scene_visualization}
    \end{center}%
    \vspace{-4mm}
\end{figure*}

\vspace{-1mm}
\section{PhysGaia}
\label{sec:contribution}

We propose \textit{{\dataset}} to advance physically realistic reconstruction in DyNVS, moving beyond mere photorealism.
As shown in Tables~\ref{tab:comparison_dnvs_dataset} and~\ref{tab:comparison_physics_dataset}, {\dataset} uniquely features complex multi-body interactions across a diverse range of physical materials, distinguishing it from existing DyNVS datasets.
We believe {\dataset} holds significant potential to deepen the understanding of physics in dynamic scenes.

\subsection{Benchmark Properties}

\paragraph{Complex scenarios with physics-aware dynamics}
Our dataset consists of 17 scenes featuring multi-object interactions, as visualized in Figure~\ref{fig:scene_visualization}.
To ensure that each scene adheres to physical laws with accurately calculated force exchange among objects, we carefully select material-specific solvers: Fluid-Implicit Particle (FLIP)~\cite{Brackbill1986} for liquids, Pyro~\cite{houdini_pyrosolver} for gases, Vellum~\cite{houdini_vellumsolver} for textiles, and Material Point Method (MPM)~\cite{Sulsky1994} for {\typeone} materials. 
Beyond multi-body interactions, our dataset also captures various physical phenomena such as non-locally rigid motion commonly observed in liquid and gas scenes, specular reflection, and refraction, as shown in Figure~\ref{fig:data_demo}.
These properties enhance the realism of our dataset and support a wide range of downstream tasks and research applications.

\vspace{-2mm}
\paragraph{Providing physics parameters}
In contrast to real-world video datasets, where the ease of capture is offset by inaccessible underlying physics, our simulated dataset offers complete access to all physical information.
This includes 3D particle trajectories and physics parameters such as viscosity, Young's modulus, Poisson's ratio, and temperature for gas scenes\footnote{Temperature is a crucial component when adopting smoke-related simulators for physical reasoning, particularly for buoyancy calculation.}.
This comprehensive provision enables precise evaluation of physical reasoning in dynamic scenes, directly facilitating the future research directions outlined in Section~\ref{subsec:potential_reasearch}.

\vspace{-2mm}
\paragraph{Supporting diverse DyNVS tasks}
Our benchmark uniquely supports both multi-view and monocular dynamic novel view synthesis (DyNVS) tasks, as shown in Table~\ref{tab:comparison_dnvs_dataset}.
Unlike most existing multi-view datasets~\citep{li2022neural, bhattacharya2024evdnerf, lewin2023dynamic, wu2024fast} with videos captured from fixed camera positions, \dataset~provides several moving monocular video sequences with independent trajectories.
This allows for diverse training configurations; using full sequences corresponds to a multiview DyNVS task, while using a single sequence corresponds to a monocular DyNVS task. 
For the evaluation, we use two static cameras with a large baseline.

\vspace{-2mm}
\paragraph{Customizability}

\begin{figure}[t]
    \begin{center}
        \centering
        \centering
        \begin{subfigure}[b]{0.24\linewidth}
        \centering
        \includegraphics[width=\linewidth]{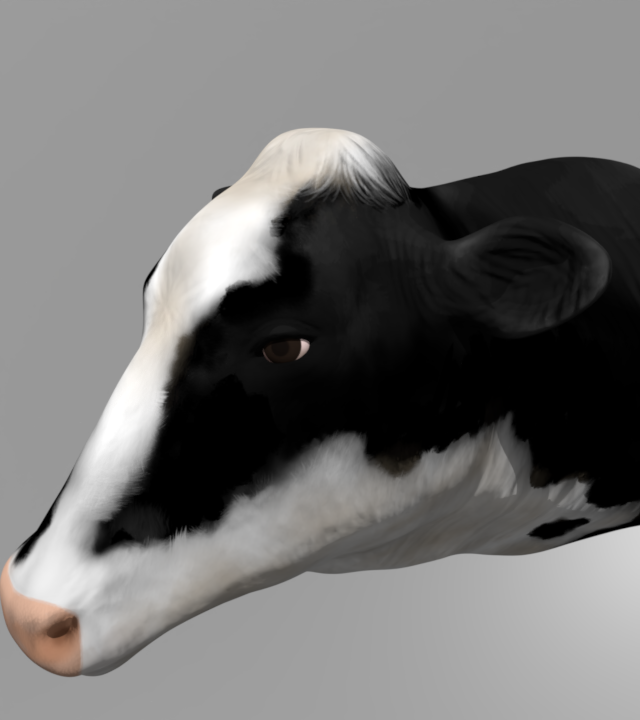}
        \caption{RGB}
        \end{subfigure}
        \begin{subfigure}[b]{0.24\linewidth}
        \centering
        \includegraphics[width=\linewidth]{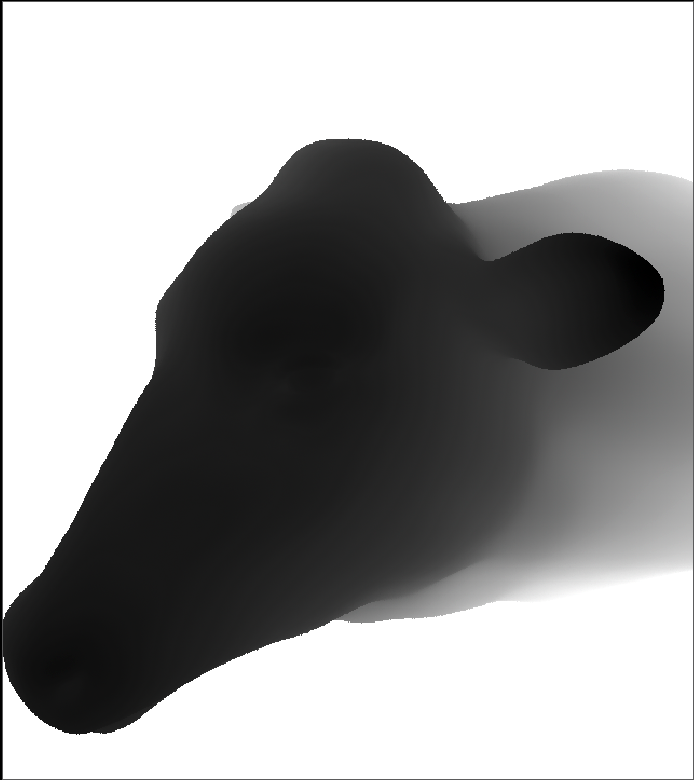}
        \caption{Depth}
        \end{subfigure}
         \begin{subfigure}[b]{0.24\linewidth}
        \centering
        \includegraphics[width=\linewidth]{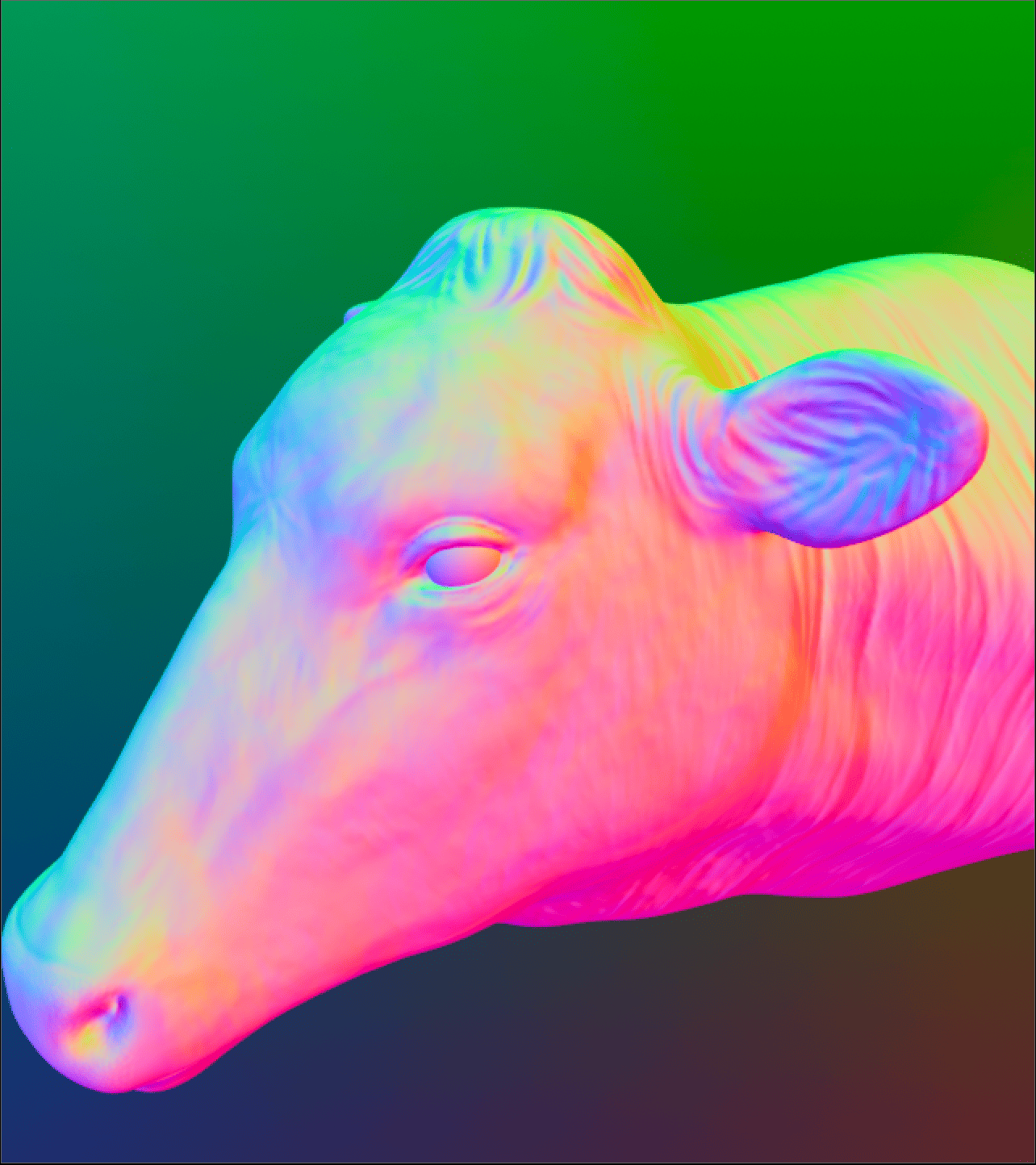}
        \caption{Normal}
        \end{subfigure}
          \begin{subfigure}[b]{0.24\linewidth}
        \centering
        \includegraphics[width=\linewidth]{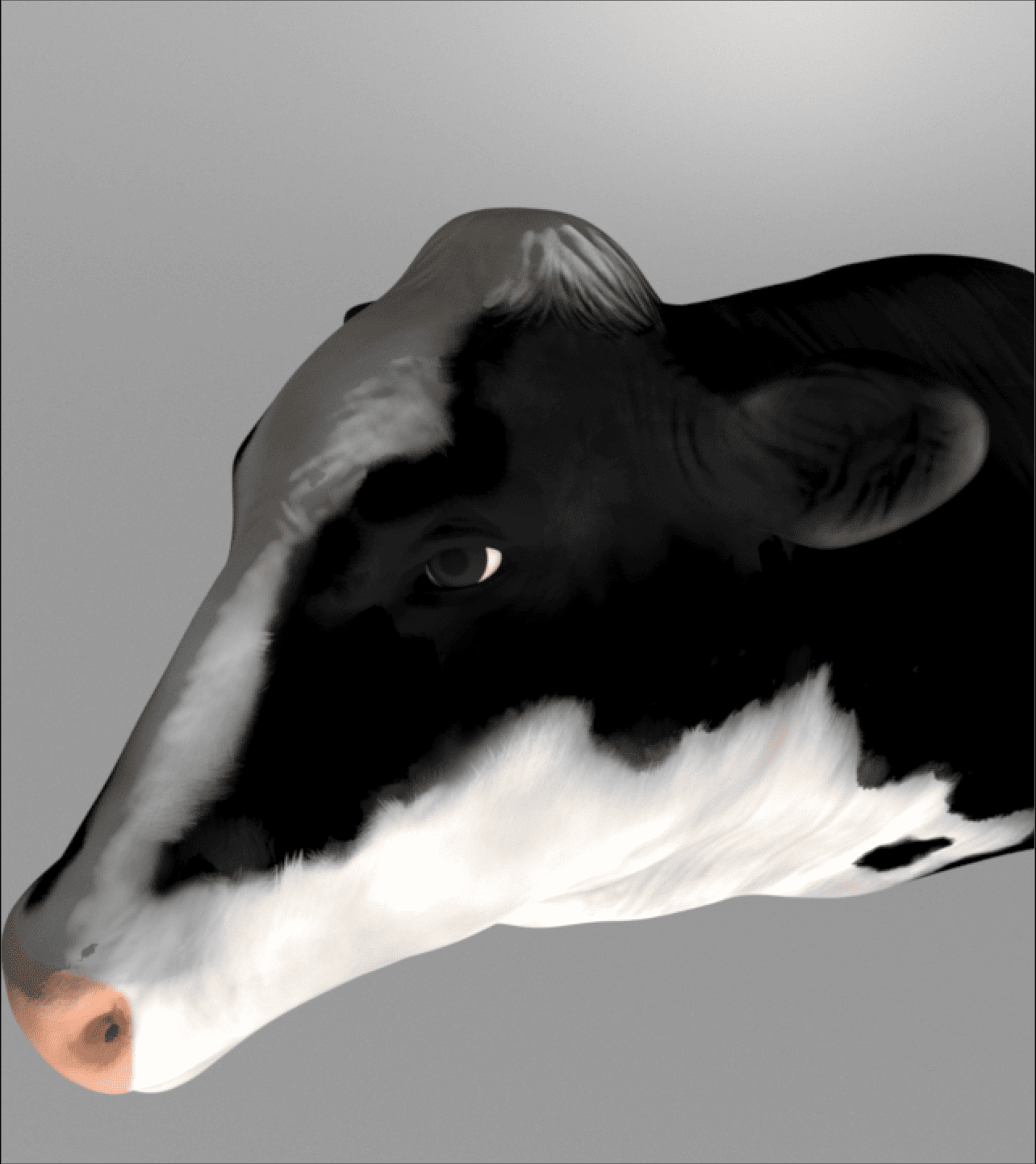}
        \caption{Re-lighted}
        \label{fig:relighting}
        \end{subfigure}
          \caption{
	Examples of diverse modalities that users can generate from the provided simulation node graphs.
	This can facilitate adaptation to specific downstream tasks.
          }
  	\label{fig:downstream}
	    \vspace{-7mm}
    \end{center}%
\end{figure}
We provide complete source files encompassing physics solvers, source geometries, materials, and texture controls. 
Such an open and comprehensive framework enables researchers to generate customized datasets with higher resolution imagery and diverse modalities including depth maps, surface normals, and relighted scenes, facilitating adaptation to specific downstream tasks, as shown in Figure~\ref{fig:downstream}.

\vspace{-2mm}
\paragraph{Accessibility}
Our \textit{{\dataset}} benchmark is designed for research-friendly and easy access, providing integration pipelines that enable the use of state-of-the-art 4D Gaussian Splatting models~\cite{wang2024shape, yang2024deformable, wu20234d, li2024spacetime, lei2024mosca} with our data.
Furthermore, we include COLMAP-reconstructed point clouds for each scene. These provisions aim to facilitate the adoption of our dataset by researchers working with state-of-the-art DyNVS models.

\subsection{Potential Research}
\label{subsec:potential_reasearch}
This subsection highlights the potential impact of our {{\dataset}} dataset by outlining several promising future research directions it uniquely enables.

\vspace{-2mm}
\paragraph{Physical reasoning of dynamic scenes}

Since our {\dataset} dataset provides ground-truth physics information, it facilitates precise evaluation of physical reasoning in dynamic scenes.
For instance, ground-truth physics parameters like viscosity can be used to evaluate inverse physics estimation methods, where differentiable simulators are employed to optimize these parameters.
Furthermore, unlike existing 4DGS research that primarily focuses on photorealism and relies on ground-truth RGB images, our dataset offers ground-truth 3D trajectories, enabling evaluation of the actual motion of individual Gaussian primitives in 4DGS.
We believe this unique feature establishes our dataset as a valuable benchmark for developing and evaluating physics-aware DyNVS models.

\vspace{-2mm}
\paragraph{Multi-body interaction}

While recent research integrates physics into DyNVS~\cite{jiang2025phystwin, zhong2024reconstruction, cai2024gic,chen2025vid2sim}, it largely remains limited to single materials and often single objects within scenes. 
As a result, the crucial aspect of physics reasoning for interactions between multiple objects in DyNVS – particularly the estimation of force exchange and deformation during contact – remains largely unexplored.
We believe our novel dataset, specifically designed with complex multi-object interactions, will be instrumental in enabling significant future research in multi-physics modeling and adaptive representations for handling hybrid scenes.

\vspace{-2mm}
\paragraph{Integration of material-specific physics solver} 
The dominant approach for integrating physics into DyNVS algorithms currently involves adopting differentiable simulators~\cite{Sulsky1994, hu2019taichi, warp2022, Genesis, hu2019difftaichi, hu2018mlsmpmcpic}, treating Gaussian primitives as particles in simulators.
However, we emphasize that different physical phenomena are best captured by different physics solvers, a principle reflected in our dataset construction process detailed in our supplementary documents.
This can guide researchers seeking to integrate more appropriate solvers tailored to specific material behaviors, such as fluids, cloth, or smoke.
For example, the FLIP~\cite{Brackbill1986} solver excels at simulating incompressible fluids due to its hybrid particle-grid representation, offering greater stability and realism compared to purely particle-based SPH-based solver. 
Similarly, thermodynamic effects like temperature and buoyancy are crucial for smoke simulation, typically represented using voxel-based grids. Integrating such volumetric solvers into particle-based frameworks like Gaussian splatting, however, remains largely unexplored.

\section{Metrics for Physical Realism}
\label{sec:evaluation}

When reconstructing 4D scenes, faithfully capturing actual 3D particle flows can be viewed as a proxy for physically realistic reconstruction.
To this end, we introduce metrics that quantify how well the motions of Gaussian primitives in 4DGS align with ground-truth physical trajectories.

\vspace{-2mm}
\paragraph{Trajectory Distance (TD)} 
Trajectory distance measures the spatial deviation along entire trajectories.
Let $M$ denote the number of reconstructed primitives, $T$ the total number of frames, $X_i^{t,\mathrm{recon}}$ the 3D position of reconstructed primitive $i$ at time $t$, and $X_j^{t,\mathrm{gt}}$ the 3D position of ground-truth trajectory $j$. 
TD is defined as the average Euclidean distance between matched trajectories as follows:
\begin{align}
\mathrm{TD} = \frac{1}{M T} \sum_{i=0}^{M-1} \sum_{t=0}^{T-1} \| X_i^{t,\mathrm{recon}} - X_{j(i)}^{t,\mathrm{gt}} \|_2,
\end{align}
where $j(i)$ denotes the closest ground-truth trajectory matched to primitive $i$ in the initial frame. 
Lower TD values indicate closer alignment with ground-truth motion, reflecting higher physical realism.

\vspace{-2mm}
\paragraph{Area Under the Outlier Percentages (AUOP)} 
To complement TD, we introduce a percentage-based metric that captures outlier behavior holistically across the entire sequence.
For each primitive $i$ at timestep $t$, an outlier indicator $O_i^t$ is defined as follows:
\begin{align}   
O_i^t =
\begin{cases}
1 & \text{if } O_i^{t-1} = 1 \text{ or } \|X_i^{t,\mathrm{recon}} - X_{j(i)}^{t,\mathrm{gt}}\|_2 > \delta, \\
0 & \text{otherwise},
\end{cases}
\end{align}
where $1$ indicates the primitive is marked as an outlier, and $\delta$ is a deviation threshold.
Once a trajectory exceeds this threshold, it remains marked as an outlier for all subsequent frames.
The percentage of outlier trajectories is computed at each timestep, yielding a curve over time as shown in Figure~\ref{fig:outlier}. 
To summarize this temporal behavior, we propose AUOP, which quantifies the area under this curve and provides a single comprehensive value for the entire sequence.
Lower AUOP values indicate better compliance with ground-truth physics, reflecting fewer and shorter deviations throughout the sequence.

\vspace{-2mm}
\paragraph{Discussion}
TD and AUOP provide complementary perspectives: TD measures absolute trajectory accuracy, while AUOP captures the persistence of large deviations, together enabling a more robust evaluation of physical realism in reconstructed flows.

\section{Empirical Analysis with {\dataset}}

\label{sec:analysis}
\vspace{-1mm}
In this section, we demonstrate the characteristics of our benchmark and highlight critical yet largely overlooked limitations of existing DyNVS approaches.

\subsection{Comparison with Physics-based Benchmarks}

\begin{table}[t]
        \centering
        \caption{
Comparison with existing physics-based benchmarks on motion complexity and photorealism.
{\dataset} achieves the highest dynamic score~\cite{liao2024evaluation} and lowest FID~\cite{heusel2017gans} and KID~\cite{binkowski2018demystifying}, demonstrating richer dynamics and greater visual realism than existing DyNVS datasets.
Note that PAC-NeRF sequences are too short to compute a dynamic score.
}
        \scalebox{0.8}{
        \centering
        \setlength\tabcolsep{6.2pt} 
        \begin{tabular}{l|cc|cc}
        \toprule
        \multirow{2}{*}{\textbf{Benchmark}}			&  \multicolumn{2}{c}{\textbf{Motion complexity}} &  \multicolumn{2}{c}{\textbf{Photorealism}}\\
        & Multi-body & Dynamic score $\uparrow$  &  FID $\downarrow$  & KID $\downarrow$ \\
        \hline
        ScalarFlow~\cite{ScalarFlow2019} &\textcolor{red}{No}& 0.391 &293.5 				&0.255\\ 
        PAC-NeRF~\cite{li2023pac} & \textcolor{red}{No} & N/A& 242.6                        & 0.164   \\
        Spring-Gaus~\cite{zhong2024reconstruction} & \textcolor{red}{No}& 0.372          & 261.8                        & 0.171\\
        \hdashline
        {\dataset} & \textcolor{blue}{Yes} &\textbf{0.444} &      \textbf{207.8}                        & \textbf{0.118}  \\
        \bottomrule
        \end{tabular}
    }
        \label{tab:reality_comparison}
\end{table}

\begin{figure}[t]
            \centering
                \begin{subfigure}[b]{0.357\linewidth}
            \centering
            \includegraphics[width=0.485\linewidth]{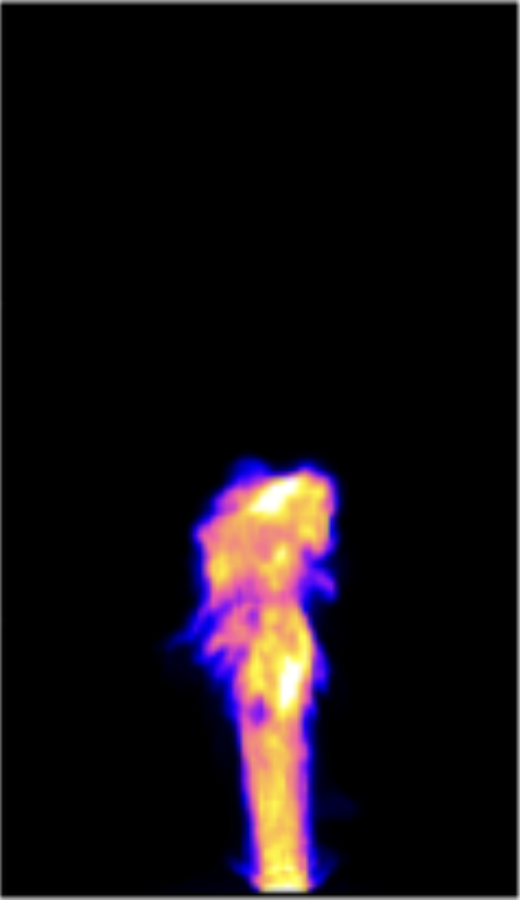}
            \includegraphics[width=0.485\linewidth]{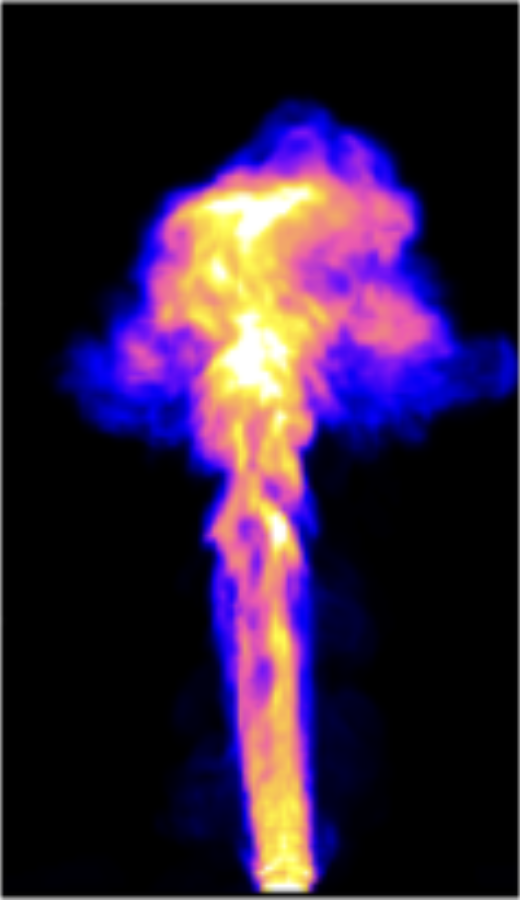}
            \caption{ScalarFlow~\cite{ScalarFlow2019}}
    	       \label{subfig:scalarflow}
            \end{subfigure}
            \begin{subfigure}[b]{0.30\linewidth}
            \centering
            \includegraphics[width=\linewidth]{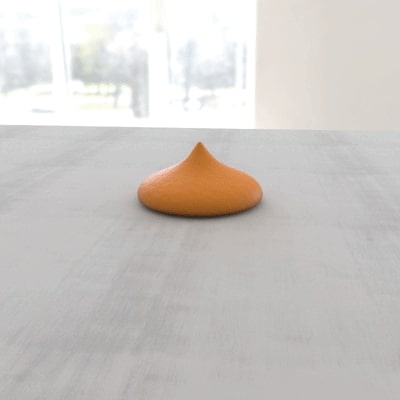}
            \caption{PAC-NeRF~\cite{li2023pac} }
            \label{subfig:pac}
            \end{subfigure}
            \begin{subfigure}[b]{0.30\linewidth}
            \centering
            \includegraphics[width=\linewidth]{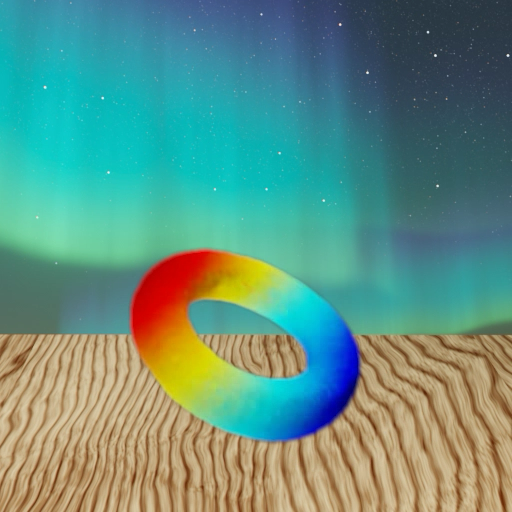}
            \caption{Spring-Gaus~\cite{zhong2024reconstruction}}
            \label{subfig:spring_gaus}
            \end{subfigure}
            \vspace{-1mm}
            \caption{ 
            {Limitations of existing datasets. While ScalarFlow, PAC-NeRF, and Spring-Gaus address physical phenomena, they are limited in narrow coverage of physical materials, overly simplified dynamics, and an absence of rich multi-object interactions.}
            }
      	 \label{fig:dataset_comparison}
	 \vspace{-2mm}
\end{figure}

Existing physics-based benchmarks \cite{ScalarFlow2019, li2023pac, zhong2024reconstruction}\footnote{These datasets provide reliable ground-truth physics, including trajectories and physical parameters, as detailed in Table \ref{tab:comparison_physics_dataset}.} often exhibit limited motion diversity and simplistic dynamics, as visualized in Figure \ref{fig:dataset_comparison}. 
Table \ref{tab:reality_comparison} demonstrates that {\dataset} significantly bridges the gap in both motion complexity and realism.
To evaluate motion complexity, we employ the dynamic score \cite{liao2024evaluation}, which quantifies motion through feature-level distances and inter-frame variance. {\dataset} achieves the highest score, reflecting its capacity to capture intricate multi-body interactions.
Regarding photorealism, we compute FID \cite{heusel2017gans} and KID \cite{binkowski2018demystifying} by flattening video sequences into image sets to measure distributional alignment with TinyImageNet.
Our dataset yields the lowest scores across both metrics, confirming superior visual fidelity. These results validate that {\dataset} effectively bridges the gap in existing benchmarks by integrating physical accuracy, high photorealism, and complex motion dynamics.

\subsection{Analysis on Dynamic Novel View Synthesis}

We analyze recent DyNVS approaches (D-3DGS~\citep{yang2024deformable}, 4DGS~\citep{wu20234d}, STG~\citep{li2024spacetime}, MoSca~\citep{lei2024mosca}, and SoM~\citep{wang2024shape}) on {\dataset} in terms of both photorealism and physical realism.

\subsubsection{Evaluation of Photorealism}


\begin{table*}[t]
\centering
\caption{ 
Average quantitative results for each material category across all algorithms.
While performance is generally high for textile, it deteriorates significantly for rheological substances, which typically exhibit complex dynamics involving multiple interacting components.}
\label{tab:main_results}
\scalebox{0.8}{
\centering
\setlength\tabcolsep{4pt} 
\begin{tabular}{cl|ccc|ccc|ccc|ccc}
\toprule
\multirow{2}{*}{{View}} & \multirow{2}{*}{{Method}} & \multicolumn{3}{c|}{{Liquid}} & \multicolumn{3}{c|}{{Gas}} & \multicolumn{3}{c|}{{Rheological materials}} & \multicolumn{3}{c}{{Textile}} \\
& & PSNR $\uparrow$ & SSIM $\uparrow$ & LPIPS $\downarrow$ & PSNR $\uparrow$ & SSIM $\uparrow$ & LPIPS $\downarrow$  & PSNR $\uparrow$ & SSIM $\uparrow$ & LPIPS $\downarrow$  & PSNR $\uparrow$ & SSIM $\uparrow$ & LPIPS $\downarrow$ \\
\midrule
\multirow{5}{*}{Monocular}  
& D-3DGS~\citep{yang2024deformable} 	& \cellsecond22.7 & \cellbest0.87 & \cellbest0.22  & \cellbest21.9 & \cellbest0.89 & \cellbest0.16  & \cellbest20.1 & \cellbest0.84 & \cellbest0.15 & \cellsecond22.1 & \cellsecond0.83 & \cellbest0.18 \\
& 4DGS~\citep{wu20234d} 	& \cellbest24.2 & \cellbest0.87 & \cellsecond0.23 & \cellsecond21.7 & \cellsecond0.88 & \cellsecond0.17 & \cellsecond19.5 & \cellsecond0.82 & \cellsecond0.18 & \cellbest24.9 & \cellbest0.84 & \cellbest0.18 \\
& STG~\citep{li2024spacetime} 	& 19.2 & 0.72 & 0.39 & \cellbest21.9 & \cellthird0.85 & \cellthird0.24 & 13.6 & 0.63 & 0.40 & \cellthird21.9 & \cellbest0.84 & \cellsecond0.21 \\
& MoSca~\citep{lei2024mosca}             & \cellthird20.5 & \cellthird0.78 & 0.39 & \cellthird21.2 & 0.79 & 0.35 & \cellthird17.8 & 0.74 & \cellthird0.23 & 18.6 & 0.74 & 0.36 \\
& SoM~\citep{wang2024shape}              & 19.6 & \cellsecond0.80 & \cellthird0.32 & 20.0 & 0.84 & 0.27 & 16.7 & \cellthird0.75 & \cellthird0.23 & 20.7 & \cellthird0.79 & \cellthird0.22 \\
\midrule
\multirow{3}{*}{Multiview}  
& D-3DGS~\citep{yang2024deformable} 	& \cellsecond22.2 & \cellsecond0.87 & \cellsecond0.24 & 23.7 & \cellbest0.91 & \cellbest0.13 & \cellbest22.2 & \cellbest0.89 & \cellbest0.10 & \cellbest27.7 & \cellbest0.90 & \cellbest0.12 \\
& 4DGS~\citep{wu20234d} 	& \cellbest25.1 & \cellbest0.88 & \cellbest0.22 & \cellsecond24.2 & \cellsecond0.89 & \cellsecond0.17 & \cellsecond21.0 & \cellsecond0.85 & \cellsecond0.15 & \cellsecond26.6 & \cellsecond0.87 & \cellsecond0.15 \\
& STG~\citep{li2024spacetime} 	& 20.8 & 0.75 & 0.40 & \cellbest25.0 & \cellbest0.91 & 0.19 & 17.2 & 0.70 & 0.36 & 21.1 & 0.81 & 0.25 \\
\bottomrule
\end{tabular}
}
\label{tab:quantitative_average_material}
\end{table*}

\begin{figure*}[t]
    \begin{center}
        \centering
        \begin{subfigure}[b]{0.161\linewidth}
        \centering
	\includegraphics[width=\linewidth]{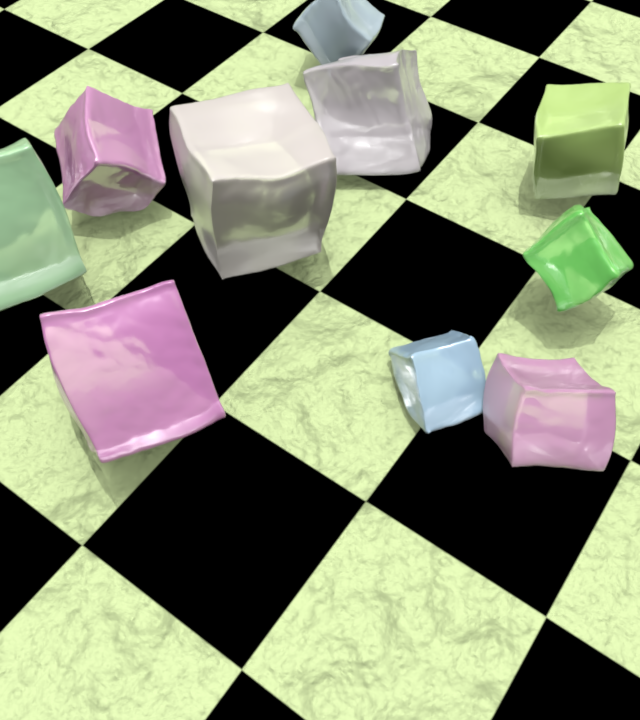}
        \caption{GT image}
        \end{subfigure}
                        \begin{subfigure}[b]{0.161\linewidth}
        \centering
	\includegraphics[width=\linewidth]{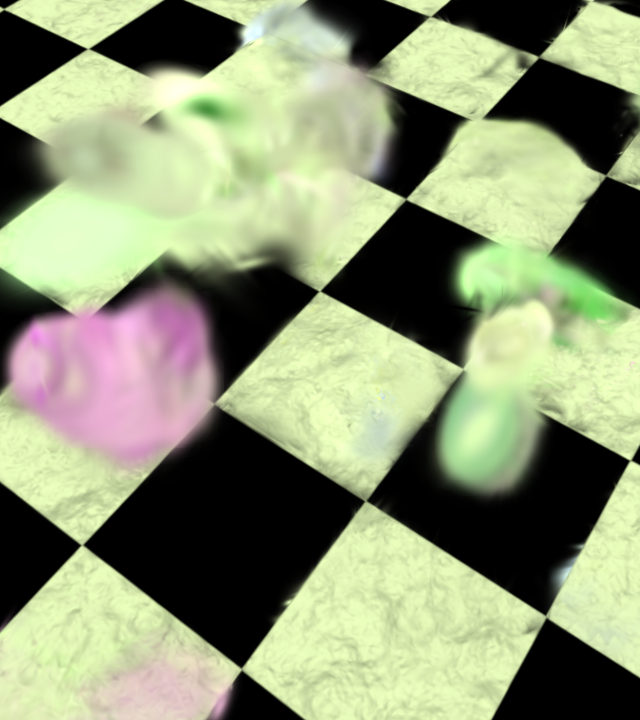}
         \caption{D-3DGS~\cite{yang2024deformable}}
        \end{subfigure}
                        \begin{subfigure}[b]{0.161\linewidth}
        \centering
	\includegraphics[width=\linewidth]{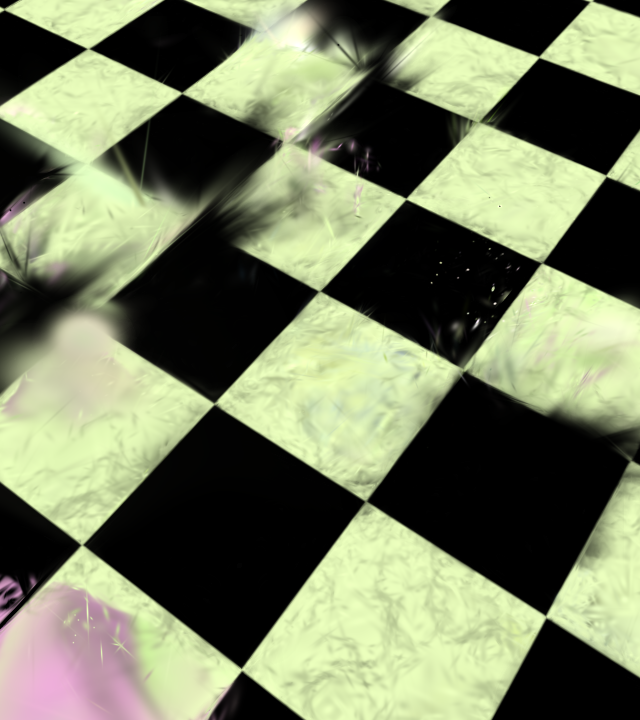}
          \caption{4DGS~\cite{wu20234d}}
        \end{subfigure}
                        \begin{subfigure}[b]{0.161\linewidth}
        \centering
	\includegraphics[width=\linewidth]{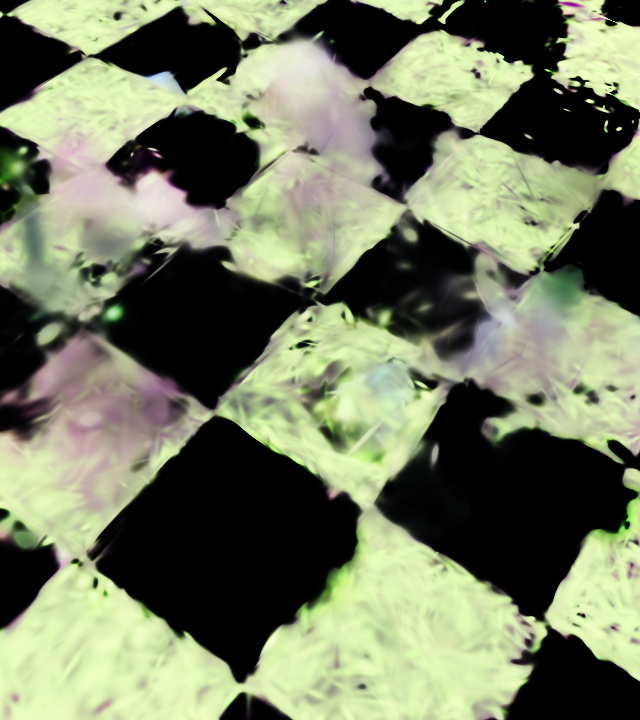}

        \caption{STG~\cite{li2024spacetime}}
        \end{subfigure}
            \begin{subfigure}[b]{0.161\linewidth}
        \centering
	\includegraphics[width=\linewidth]{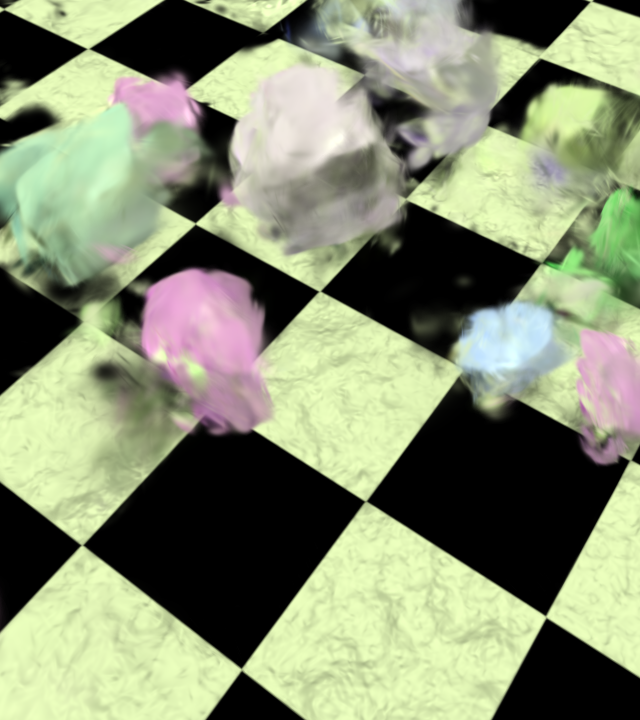}
         \caption{MoSca~\cite{lei2024mosca}}
        \end{subfigure}
                        \begin{subfigure}[b]{0.161\linewidth}
        \centering
	\includegraphics[width=\linewidth]{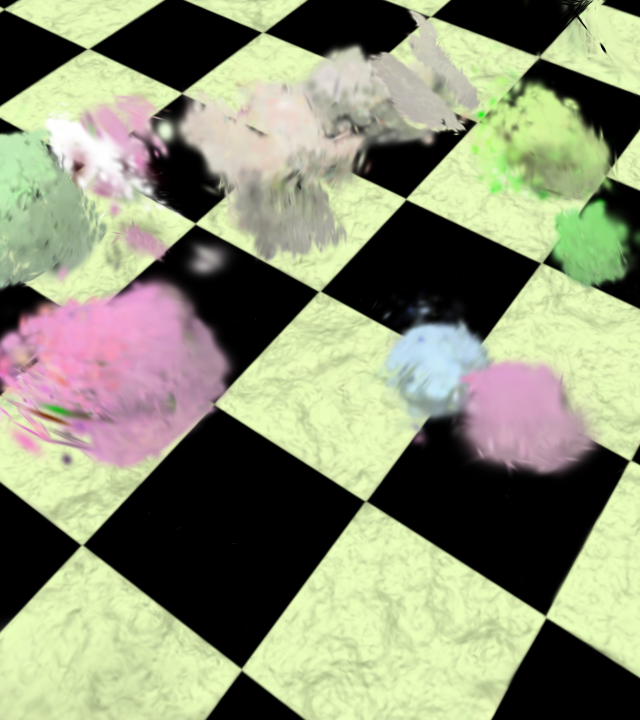}
         \caption{SoM~\cite{wang2024shape}}
        \end{subfigure}
        \vspace{-1mm}
  \caption{
Qualitative results of recent DyNVS methods on the \textit{jelly party} scene with monocular setup.
All methods struggle to accurately capture multi-body interactions by frequently exhibiting needle-like artifacts and failing to reconstruct dynamic elements accurately.
Please refer to our supplementary documents for more qualitative results.}
  	\label{fig:qualitative_result_main}
	\vspace{-5mm}
    \end{center}%
\end{figure*}

\begin{figure}[t]
    \begin{center}
        \centering
	\includegraphics[width=\linewidth]{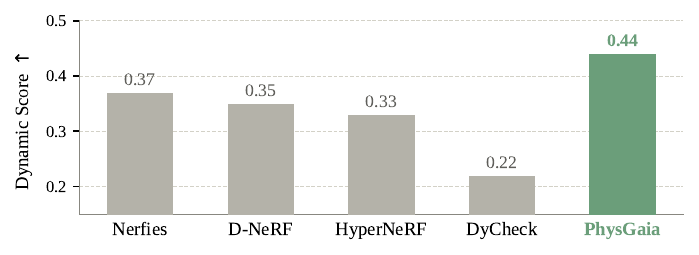}
	  \caption{
	Comparison of motion diversity across DyNVS datasets. Dynamic Score (DS)~\cite{liao2024evaluation} measures feature-level distances across frames, serving as a proxy for 	motion complexity. {\dataset} achieves the highest score, reflecting the rich multi-object interactions present in our benchmark.}
   	\label{fig:comparison_entropy}
	\vspace{-7mm}
    \end{center}%
\end{figure}

We adopt standard image quality metrics for photorealism evaluation: PSNR, SSIM~\citep{wang2004image}, and LPIPS~\citep{Zhang_2018_CVPR}.
Tables~\ref{tab:main_results} and~\ref{tab:quantitative_average} show that existing methods struggle on {\dataset}, unlike their performance on existing DyNVS benchmarks~\citep{yoon2020novel, gao2022monocular}.
We attribute this to the motion complexity from multi-object interactions, evidenced by the significantly higher motion entropy of {\dataset} (Figure~\ref{fig:comparison_entropy}).
Each method faces distinct challenges: MLP- and grid-based representations (D-NeRF~\citep{pumarola2021d}, 4DGS~\citep{wu20234d}) fail on fine-grained dynamics such as fluid splashing, polynomial motion models (STG~\citep{li2024spacetime}) are too restrictive for complex trajectories, and motion factorization or ARAP-based constraints (MoSca~\citep{lei2024mosca}, SoM~\citep{wang2024shape}) break down under rich multi-object interactions.
Figure~\ref{fig:qualitative_result_main} illustrates these failures, where all methods produce needle-like artifacts in the \textit{jelly party} scene.
Additional qualitative results are provided in the supplementary material.

\begin{table}[t]
\centering
\caption{Average quantitative results for both monocular and multiview settings, averaged across all 17 scenes. 
While multiview setups generally offer better reconstruction performance than monocular ones, even multiview results achieve PSNR scores below 30.
This highlights the substantial difficulty in reconstructing the complex multi-body interactions in our benchmark.
}
\scalebox{0.8}{
\setlength\tabcolsep{3pt} 
\begin{tabular}{l|ccc|ccc}
\toprule
\multirow{2}{*}{\textbf{Method}}			&  \multicolumn{3}{c}{\textbf{Monocular}} &  \multicolumn{3}{c}{\textbf{Multiview}}\\
&				PSNR $\uparrow$  & SSIM $\uparrow$  & LPIPS $\downarrow$ &PSNR $\uparrow$  & SSIM $\uparrow$  & LPIPS $\downarrow$\\
\midrule
D-3DGS~\citep{yang2024deformable} & \cellsecond21.7& \cellbest0.86 & \cellbest0.18 & \cellsecond24.2 & \cellbest0.89& \cellbest0.14 \\
4DGS~\citep{wu20234d} 	&\cellbest22.7& \cellsecond0.85 & \cellsecond0.19 & \cellbest24.4 &  \cellsecond0.87 &  \cellsecond0.17 \\
STG~\citep{li2024spacetime} 	&19.3&0.76&0.30 &21.0&0.79&0.30\\
MoSca ~\citep{lei2024mosca}  &  \cellthird19.5 &0.76 &0.34&N/A&N/A&N/A\\
SoM~\citep{wang2024shape} 	&19.3& \cellthird0.80 & \cellthird0.26 &N/A&N/A&N/A\\
\bottomrule
\end{tabular}
}
\label{tab:quantitative_average}
\vspace{-2mm}
\end{table}

\subsubsection{Evaluation of Physical Realism}

\begin{figure*}[t]
    \begin{center}
        \centering
                        \begin{subfigure}[b]{0.161\linewidth}
        \centering
        \includegraphics[width=\linewidth]{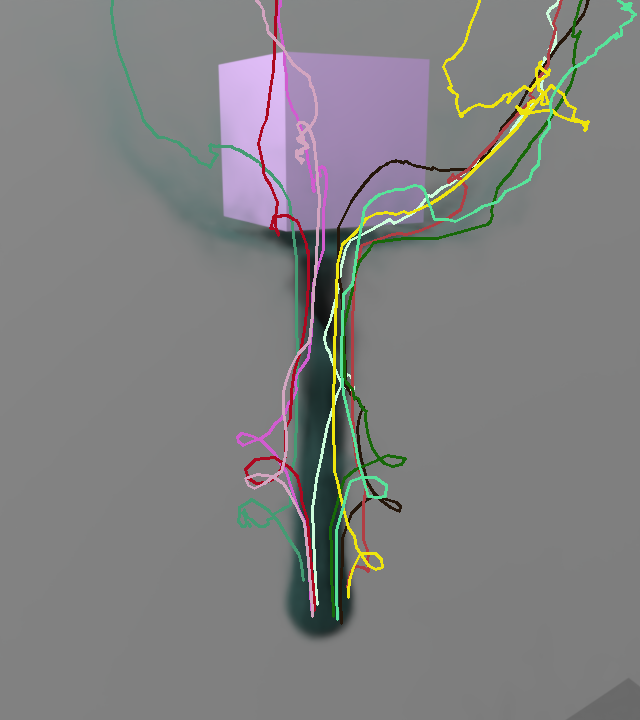}
        \includegraphics[width=\linewidth]{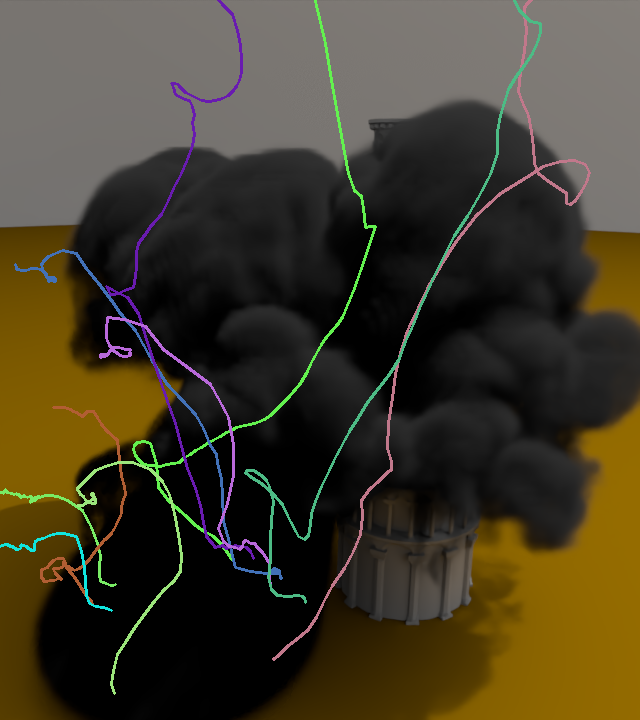}
        \caption{GT motion}
        \end{subfigure}
                        \begin{subfigure}[b]{0.161\linewidth}
        \centering
        \includegraphics[width=\linewidth]{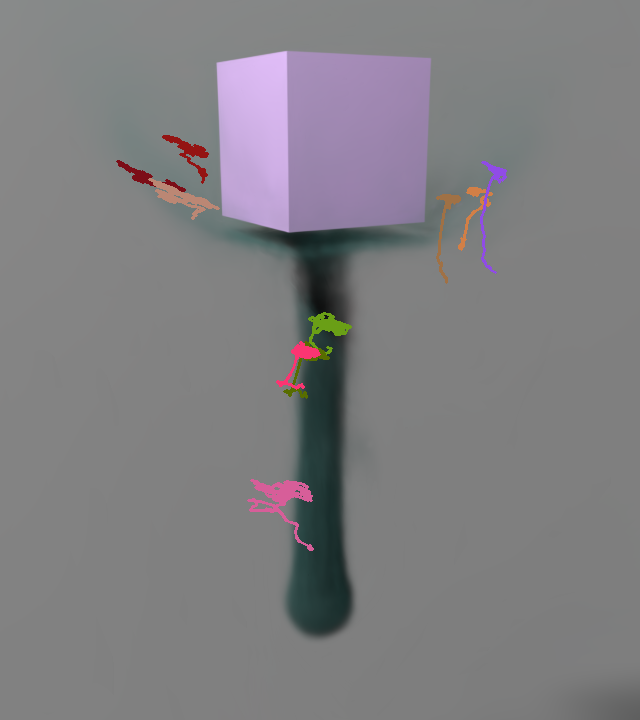}
         \includegraphics[width=\linewidth]{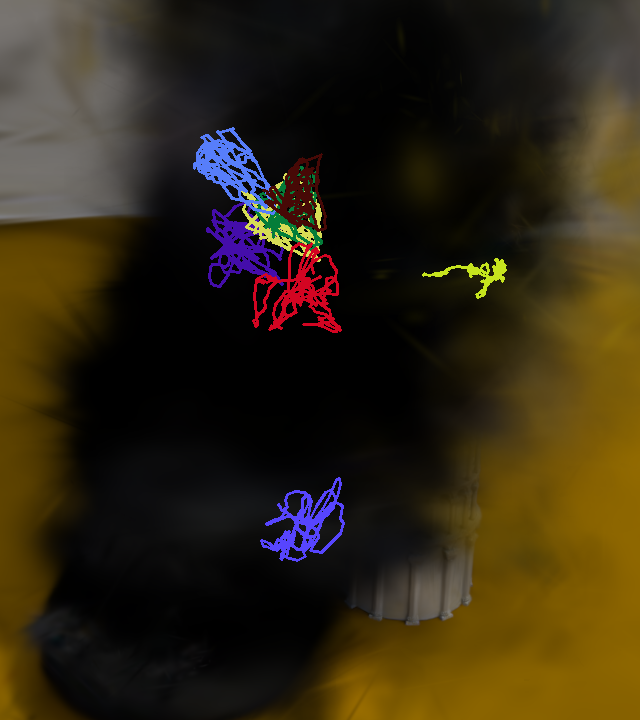}
         \caption{{D-3DGS~\citep{yang2024deformable}}}
        \end{subfigure}
                        \begin{subfigure}[b]{0.161\linewidth}
        \centering
                \includegraphics[width=\linewidth]{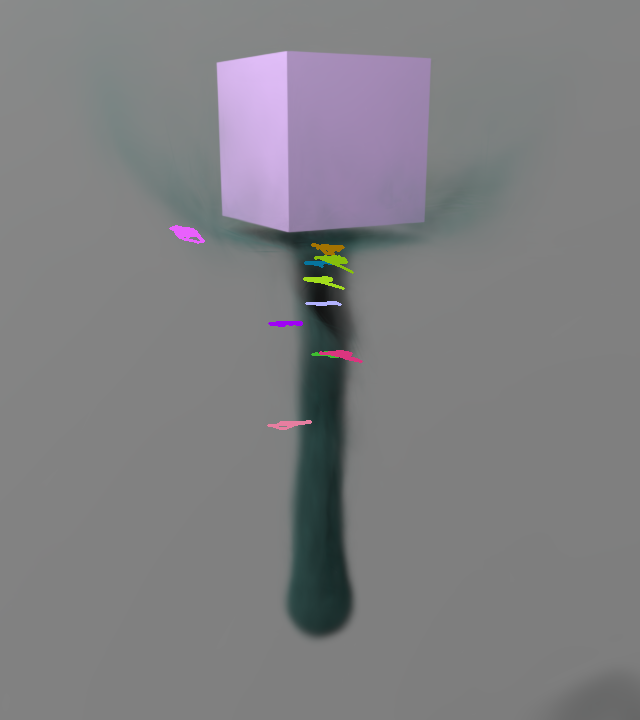}
         \includegraphics[width=\linewidth]{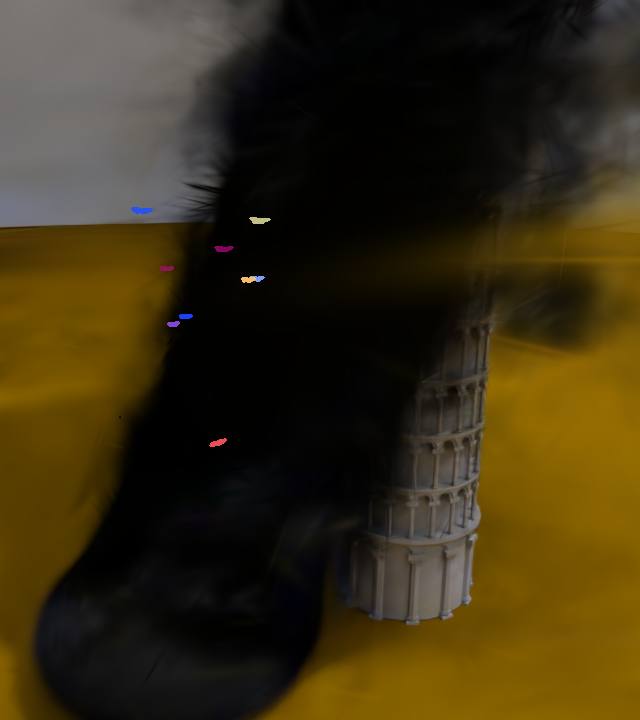}
          \caption{4DGS~\citep{wu20234d} }
        \end{subfigure}
                        \begin{subfigure}[b]{0.161\linewidth}
        \centering        
        \includegraphics[width=\linewidth]{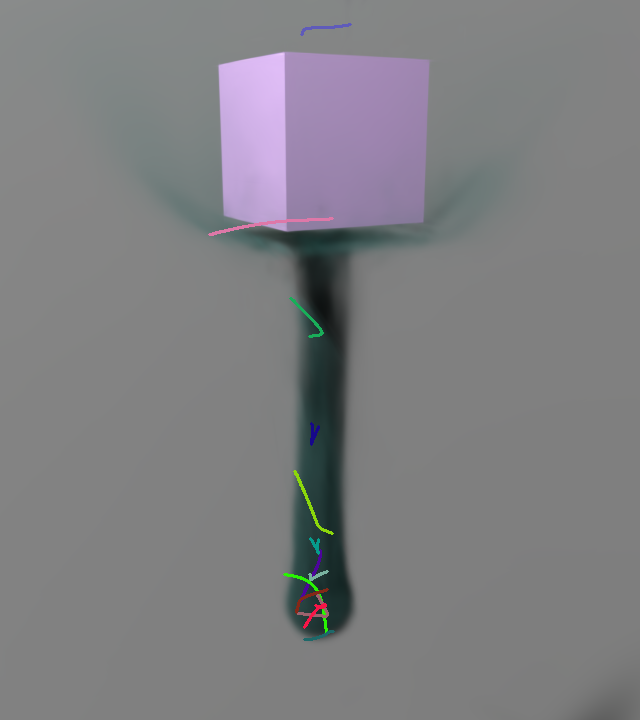}
         \includegraphics[width=\linewidth]{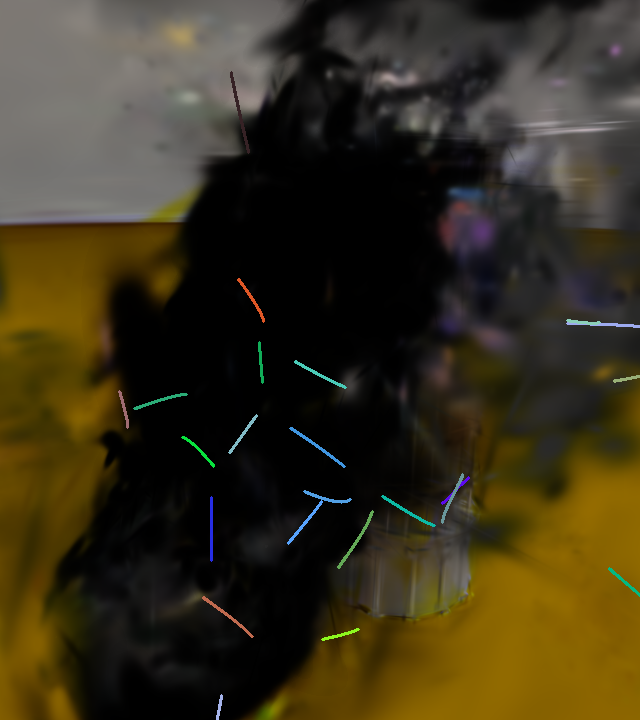}
        \caption{STG~\citep{li2024spacetime}}
        \end{subfigure}
                        \begin{subfigure}[b]{0.161\linewidth}
        \centering
         \includegraphics[width=\linewidth]{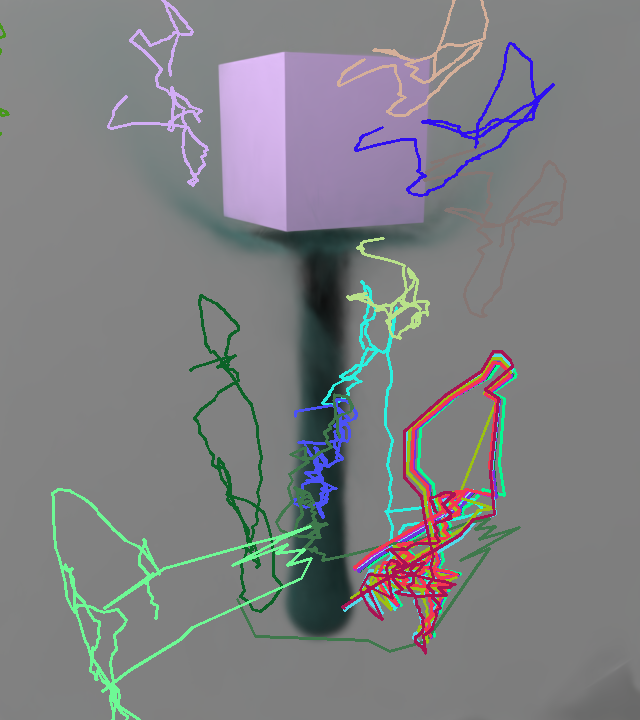}
         \includegraphics[width=\linewidth]{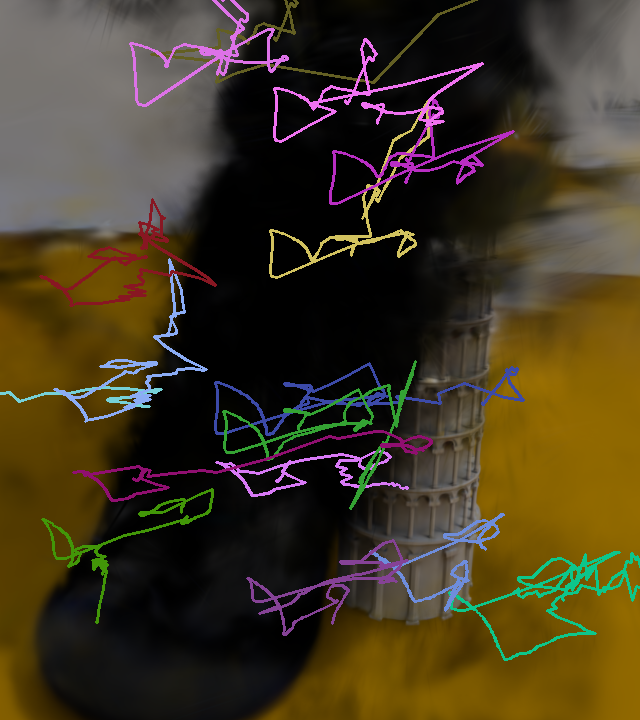}
         \caption{MoSca~\citep{lei2024mosca}}
        \end{subfigure}
                        \begin{subfigure}[b]{0.161\linewidth}
        \centering
         \includegraphics[width=\linewidth]{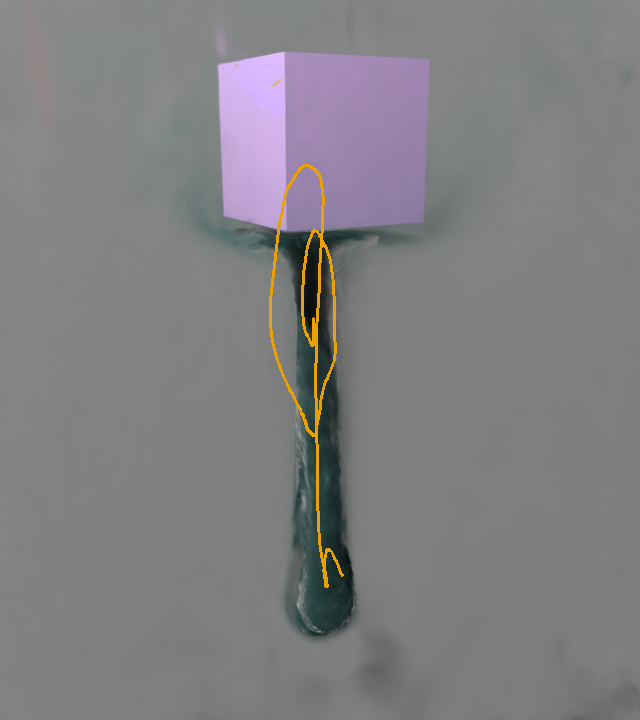}
         \includegraphics[width=\linewidth]{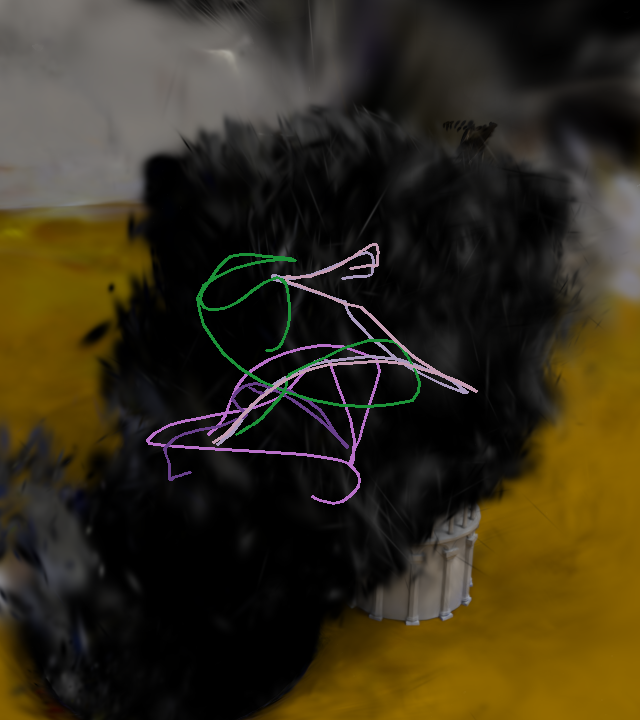}
         \caption{SoM~\citep{wang2024shape}}
        \end{subfigure}                
        \vspace{-1mm}
  \caption{Comparison of reconstructed flows and ground truth on the \textit{Box-smoke} and \textit{Pisa} scenes.
Reconstructed flows deviate significantly from ground truth, where photorealism is achieved through local surface fluctuations rather than following actual physical motion.}
  	\label{fig:physical_realism}
    \end{center}%
    \vspace{-4mm}
\end{figure*}

\begin{figure}[t]
    \begin{center}
        \centering
        \includegraphics[width=0.89\linewidth]{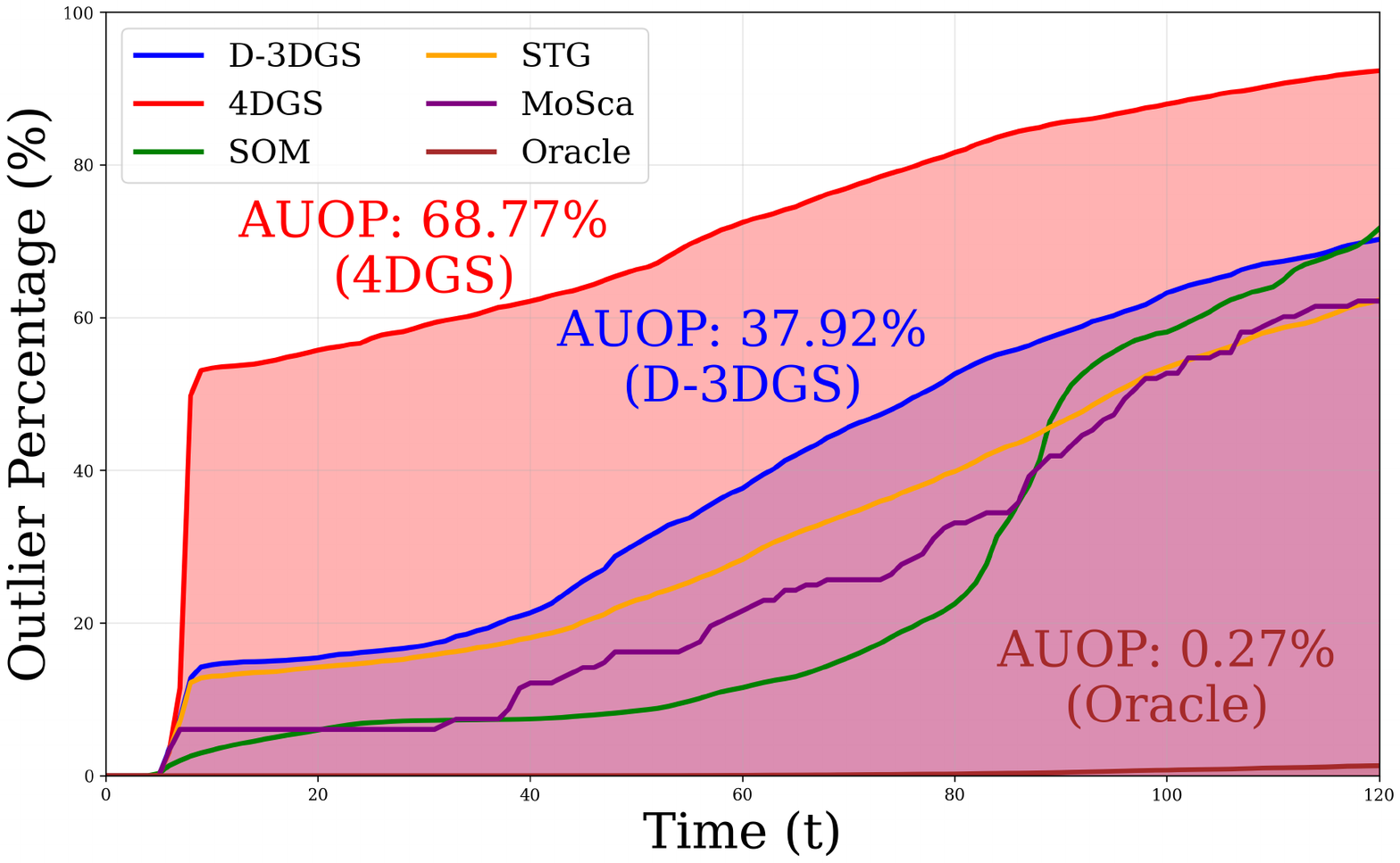}       
            \vspace{-5mm}                         
  \caption{Outlier percentage over time on the \textit{box-smoke} scene with AUOP values indicated.
The oracle (GT-GT) achieves near-zero outliers, while existing methods exhibit rapidly increasing outlier percentages, resulting in significantly higher AUOP values.
This highlights their inability to track physically plausible flows.}
  	\label{fig:outlier}
    \end{center}%
    \vspace{-3mm}
\end{figure}

\begin{table}[t]
\centering
    \centering
    \caption{Performance for physics realism. The box-smoke scene shows higher TD and AUOP; existing methods are hard to capture physics with simplified visuals.
    }
    \label{tab:auop}
    \scalebox{0.8}{
    \setlength\tabcolsep{12pt} 
    \begin{tabular}{l|cc|cc}
    \toprule
    \multirow{2}{*}{\textbf{Method}} & \multicolumn{2}{c|}{\textbf{Cow}} & \multicolumn{2}{c}{\textbf{Box-smoke}} \\ 
    & TD $\downarrow$ & AUOP $\downarrow$ & TD $\downarrow$ & AUOP $\downarrow$ \\ 
    \midrule
    D-3DGS~\cite{yang2024deformable} & 0.07 & ~~0.0 & 4.01 & 37.9 \\
    4DGS~\cite{wu20234d} & 0.90 & ~~0.0 & 6.00 & 68.8 \\
    STG~\cite{li2024spacetime} & 0.30 & ~~0.7 & 3.62 & 30.2 \\
    MoSca~\cite{lei2024mosca} & 0.33 & ~~0.7 & 3.67 & 28.3 \\
    SoM~\cite{wang2024shape} & 0.61 & 17.7 & 4.45 & 25.4 \\
    \hdashline
    Oracle & \textbf{0.03} & ~~\textbf{0.0} & \textbf{0.84} & ~~\textbf{0.3} \\
    \bottomrule
    \end{tabular}
    }
\vspace{-2mm} 
\end{table}

To evaluate whether Gaussian primitives follow ground-truth physical flows, we report TD and AUOP for two scenes in Table~\ref{tab:auop}\footnote{Oracle denotes results from an additional set of ground-truth trajectories (GT-GT), serving as an upper bound on performance.}: the less dynamic \textit{cow} scene and the highly dynamic \textit{box-smoke} scene.
While all methods perform reasonably on the \textit{cow} scene, they exhibit substantially higher TD and AUOP on the \textit{box-smoke} scene---optimizing solely for RGB reconstruction achieves photorealism but fails to ensure physical plausibility.
Figures~\ref{fig:physical_realism} and~\ref{fig:outlier} further confirm this: reconstructed trajectories deviate significantly from ground truth, with primitives merely fluctuating near the surface rather than following actual physical motion, and outlier percentages increasing rapidly over time.
These results underscore the necessity of physics-aware evaluation and position {\dataset} as a critical benchmark for driving progress in this direction.

\subsection{Analysis on Inverse-physics Problems}

\begin{table}[t]
\centering
    \centering
\caption{Comparison of estimated physics parameters (Young's modulus $E$ and Poisson's ratio $\nu$) against ground truth. 
Both PAC-NeRF~\cite{li2023pac}  and GIC~\cite{cai2024gic} methods substantially underestimate material stiffness in multi-object scenarios.}
\centering
\scalebox{0.8}{
\centering
\setlength\tabcolsep{25pt} 
\begin{tabular}{lcc}
\toprule
\textbf{Method} & {$E$} & \textbf{$\nu$} \\
\midrule
Ground Truth & $100.0\times 10^3$ & 0.50\\
\hdashline
PAC-NeRF~\cite{li2023pac} & $~~~~1.7\times 10^3$ & 0.42 \\
GIC~\cite{cai2024gic} & $~~23.4\times 10^3$ & 0.37 \\
\bottomrule
\end{tabular}
}
\label{tab:pacnerf}
\vspace{-7mm}
\end{table}

Beyond DyNVS tasks, {\dataset} also supports physics parameter estimation~\citep{cai2024gic,zhong2024reconstruction,chen2025vid2sim} since it is faithfully generated using physics simulators with known ground-truth parameters.
We evaluate both PAC-NeRF~\citep{li2023pac} and GIC~\citep{cai2024gic} on the \textit{bouncing balls} scene.
As shown in Table~\ref{tab:pacnerf}, both methods exhibit similar trends and significantly underestimate Young's modulus 
$E$.
This behavior arises because both approaches were designed for single-object scenarios: when applied to multi-object scenes with complex motions and interactions, the algorithms favor lower $E$ values, allowing greater deformation to fit the observations.
These results demonstrate that our benchmark exposes underexplored challenges in multi-object physics parameter estimation, providing a valuable testbed for advancing this research direction.

\section{Conclusion}
\label{sec:conclusion}

We propose a novel physics-aware dataset, specifically designed to understand physics in dynamic scenes, particularly for DyNVS.
Our benchmark  captures complex multi-body interactions with a wide variety of materials.
Each scene is faithfully generated using material-specific physics solvers, ensuring adherence to physical laws and providing rich ground-truth physics data including 3D flows and configurations. 
This ground-truth data uniquely enables the evaluation of physical reasoning.
We also test several tasks on {\dataset}, revealing their fundamental limitations in achieving physical realism and handling multi-objects, which highlights significant potential for improvement.

\paragraph{Acknowledgements}
This work was partly supported by 
Samsung Research, Samsung Electronics Co., Ltd., 
the National Research Foundation of Korea (NRF) grant [RS-2022-NR070855, Trustworthy Artificial Intelligence],
and the Institute of Information \& communications Technology Planning \& Evaluation (IITP) grants 
[RS-2025-25442338, AI star Fellowship Support Program (Seoul National University);  
RS-2022-II220959 (No.2022-0-00959), (Part 2) Few-Shot Learning of Causal Inference in Vision and Language for Decision Making; 
No.RS-2021-II211343, Artificial Intelligence Graduate School Program (Seoul National University)] 
funded by the Korea government (MSIT).

{
    \small
    \bibliographystyle{ieeenat_fullname}
    \bibliography{references}

@String(CVPR  = {IEEE Conf. Comput. Vis. Pattern Recog.})

@String(ICCV  = {Int. Conf. Comput. Vis.})

@String(ECCV  = {Eur. Conf. Comput. Vis.})

@String(NeurIPS = {Adv. Neural Inform. Process. Syst.})

@String(ICML  = {Int. Conf. Mach. Learn.})

@String(ICLR  = {Int. Conf. Learn. Represent.})

@String(AAAI  = {AAAI})

@String(TOG   = {ACM Trans. Graph.})

@String(TIP   = {IEEE Trans. Image Process.})

@String(TCSVT = {IEEE Trans. Circuit Syst. Video Technol.})

@String(CVPR  = {CVPR})

@String(ICCV  = {ICCV})

@String(ECCV  = {ECCV})

@String(NeurIPS = {NeurIPS})

@String(ICML  = {ICML})

@String(ICLR  = {ICLR})

@String(TOG   = {ACM TOG})

@String(TIP   = {IEEE TIP})

@String(TCSVT = {IEEE TCSVT})

@inproceedings{li2023pac,
  title={PAC-NeRF: Physics Augmented Continuum Neural Radiance Fields for Geometry-Agnostic System Identification},
  author={Li, Xuan and Qiao, Yi-Ling and Chen, Peter Yichen and Jatavallabhula, Krishna Murthy and Lin, Ming and Jiang, Chenfanfu and Gan, Chuang},
  booktitle={ICLR},
  year={2023},
}

@inproceedings{zhong2024reconstruction,
  title={Reconstruction and simulation of elastic objects with spring-mass 3d gaussians},
  author={Zhong, Licheng and Yu, Hong-Xing and Wu, Jiajun and Li, Yunzhu},
  booktitle={ECCV},
  year={2024},
}

@inproceedings{ScalarFlow2019, 
  author = {Eckert, Marie-Lena and Um, Kiwon and Thuerey, Nils},
  title = {ScalarFlow: A Large-Scale Volumetric Data Set of Real-world Scalar Transport Flows for Computer Animation and Machine Learning},
  booktitle={TOG}, 
  year={2019}
}

@misc{Stanford3DRepo,
  author       = {Stanford Computer Graphics Laboratory},
  title        = {The Stanford 3D Scanning Repository},
  year         = {1994},
  howpublished = {\url{http://graphics.stanford.edu/data/3Dscanrep/}},
}

@article{park2021hypernerf,
  author = {Park, Keunhong and Sinha, Utkarsh and Hedman, Peter and Barron, Jonathan T. and Bouaziz, Sofien and Goldman, Dan B and Martin-Brualla, Ricardo and Seitz, Steven M.},
  title = {HyperNeRF: A Higher-Dimensional Representation for Topologically Varying Neural Radiance Fields},
  journal = {ACM Trans. Graph.},
  publisher = {ACM},
  year = {2021},
}

@inproceedings{pumarola2021d,
  title={D-nerf: Neural radiance fields for dynamic scenes},
  author={Pumarola, Albert and Corona, Enric and Pons-Moll, Gerard and Moreno-Noguer, Francesc},
  booktitle={CVPR},
  year={2021}
}

@inproceedings{gao2022monocular,
  title={Monocular dynamic view synthesis: A reality check},
  author={Gao, Hang and Li, Ruilong and Tulsiani, Shubham and Russell, Bryan and Kanazawa, Angjoo},
  booktitle={NeurIPS},
  year={2022}
}

@inproceedings{park2021nerfies,
  title={Nerfies: Deformable neural radiance fields},
  author={Park, Keunhong and Sinha, Utkarsh and Barron, Jonathan T and Bouaziz, Sofien and Goldman, Dan B and Seitz, Steven M and Martin-Brualla, Ricardo},
  booktitle={ICCV},
  year={2021}
}

@inproceedings{xie2024physgaussian,
  title={Physgaussian: Physics-integrated 3d gaussians for generative dynamics},
  author={Xie, Tianyi and Zong, Zeshun and Qiu, Yuxing and Li, Xuan and Feng, Yutao and Yang, Yin and Jiang, Chenfanfu},
  booktitle={CVPR},
  year={2024}
}

@inproceedings{zhang2024physdreamer,
  title={Physdreamer: Physics-based interaction with 3d objects via video generation},
  author={Zhang, Tianyuan and Yu, Hong-Xing and Wu, Rundi and Feng, Brandon Y and Zheng, Changxi and Snavely, Noah and Wu, Jiajun and Freeman, William T},
  booktitle={ECCV},
  year={2024},
}

@inproceedings{liu2024physgen,
  title={Physgen: Rigid-body physics-grounded image-to-video generation},
  author={Liu, Shaowei and Ren, Zhongzheng and Gupta, Saurabh and Wang, Shenlong},
  booktitle={ECCV},
  year={2024},
}

@misc{borycki2024gasp,
  title={Gasp: Gaussian splatting for physic-based simulations},
  author={Borycki, Piotr and Smolak, Weronika and Waczy{\'n}ska, Joanna and Mazur, Marcin and Tadeja, S{\l}awomir and Spurek, Przemys{\l}aw},
  note={arXiv preprint arXiv:2409.05819},
  year={2024}
}

@misc{jiang2025phystwin,
  title={PhysTwin: Physics-Informed Reconstruction and Simulation of Deformable Objects from Videos},
  author={Jiang, Hanxiao and Hsu, Hao-Yu and Zhang, Kaifeng and Yu, Hsin-Ni and Wang, Shenlong and Li, Yunzhu},
  note={arXiv preprint arXiv:2503.17973},
  year={2025}
}

@inproceedings{lin2025omniphysgs,
  title={OmniPhysGS: 3D Constitutive Gaussians for General Physics-Based Dynamics Generation},
  author={Lin, Yuchen and Lin, Chenguo and Xu, Jianjin and Mu, Yadong},
  booktitle={ICLR},
  year={2025}
}

@inproceedings{qiu2024feature,
  title={Feature splatting: Language-driven physics-based scene synthesis and editing},
  author={Qiu, Ri-Zhao and Yang, Ge and Zeng, Weijia and Wang, Xiaolong},
  booktitle={ECCV},
  year={2024}
}

@inproceedings{gao2025fluidnexus,
  title={FluidNexus: 3D fluid reconstruction and prediction from a single video},
  author={Gao, Yue and Yu, Hong-Xing and Zhu, Bo and Wu, Jiajun},
  booktitle={CVPR},
  year={2025}
}

@inproceedings{jiang2024vr,
  title={Vr-gs: A physical dynamics-aware interactive gaussian splatting system in virtual reality},
  author={Jiang, Ying and Yu, Chang and Xie, Tianyi and Li, Xuan and Feng, Yutao and Wang, Huamin and Li, Minchen and Lau, Henry and Gao, Feng and Yang, Yin and others},
  booktitle={SIGGRAPH},
  year={2024}
}

@inproceedings{huang2025dreamphysics,
  title={DreamPhysics: Learning Physics-Based 3D Dynamics with Video Diffusion Priors},
  author={Huang, Tianyu and Zhang, Haoze and Zeng, Yihan and Zhang, Zhilu and Li, Hui and Zuo, Wangmeng and Lau, Rynson WH},
  booktitle={AAAI},
  year={2025}
}

@inproceedings{cai2024gic,
  title={Gic: Gaussian-informed continuum for physical property identification and simulation},
  author={Cai, Junhao and Yang, Yuji and Yuan, Weihao and He, Yisheng and Dong, Zilong and Bo, Liefeng and Cheng, Hui and Chen, Qifeng},
  booktitle={NeurIPS},
  year={2024}
}

@inproceedings{chen2025vid2sim,
  title={Vid2Sim: Generalizable, Video-based Reconstruction of Appearance, Geometry and Physics for Mesh-free Simulation},
  author={Chen, Chuhao and Dou, Zhiyang and Wang, Chen and Huang, Yiming and Chen, Anjun and Feng, Qiao and Gu, Jiatao and Liu, Lingjie},
  booktitle={CVPR},
  year={2025}
}

@inproceedings{rasheed2020learning,
  title={Learning to measure the static friction coefficient in cloth contact},
  author={Rasheed, Abdullah Haroon and Romero, Victor and Bertails-Descoubes, Florence and Wuhrer, Stefanie and Franco, Jean-S{\'e}bastien and Lazarus, Arnaud},
  booktitle={CVPR},
  year={2020}
}

@inproceedings{deng2023learning,
  title={Learning vortex dynamics for fluid inference and prediction},
  author={Deng, Yitong and Yu, Hong-Xing and Wu, Jiajun and Zhu, Bo},
  booktitle={ICML},
  year={2023},
}

@inproceedings{li2022neural,
  title={Neural 3d video synthesis from multi-view video},
  author={Li, Tianye and Slavcheva, Mira and Zollhoefer, Michael and Green, Simon and Lassner, Christoph and Kim, Changil and Schmidt, Tanner and Lovegrove, Steven and Goesele, Michael and Newcombe, Richard and others},
  booktitle={CVPR},
  year={2022}
}

@inproceedings{yan2023nerf,
  title={Nerf-ds: Neural radiance fields for dynamic specular objects},
  author={Yan, Zhiwen and Li, Chen and Lee, Gim Hee},
  booktitle={CVPR},
  year={2023}
}

@inproceedings{lewin2023dynamic,
  title={Dynamic NeRFs for soccer scenes},
  author={Lewin, Sacha and Vandegar, Maxime and Hoyoux, Thomas and Barnich, Olivier and Louppe, Gilles},
  booktitle={Multimedia Content Analysis in Sports},
  year={2023}
}

@inproceedings{wu2024fast,
  title={Fast high dynamic range radiance fields for dynamic scenes},
  author={Wu, Guanjun and Yi, Taoran and Fang, Jiemin and Liu, Wenyu and Wang, Xinggang},
  booktitle={3DV},
  year={2024},
}

@inproceedings{yoon2020novel,
  title={Novel view synthesis of dynamic scenes with globally coherent depths from a monocular camera},
  author={Yoon, Jae Shin and Kim, Kihwan and Gallo, Orazio and Park, Hyun Soo and Kautz, Jan},
  booktitle={CVPR},
  year={2020}
}

@inproceedings{wang20244d,
  title     = {{4D-DRESS}: A 4D Dataset of Real-World Human Clothing With Semantic Annotations},
  author    = {Wenbo Wang and Hsuan-I Ho and Chen Guo and Boxiang Rong and Artur Grigorev and Jie Song and Juan Jose Zarate and Otmar Hilliges},
  booktitle = {CVPR},
  year      = {2024},
}

@inproceedings{zou2023cloth4d,
  title={CLOTH4D: A Dataset for Clothed Human Reconstruction},
  author={Zou, Xingxing and Han, Xintong and Wong, Waikeung},
  booktitle={CVPR},
  year={2023},
}

@article{zhang2021dynamic,
  title={Dynamic Neural Garments},
  author={Zhang, Meng and Wang, Tuanfeng Y. and Ceylan, Duygu and Mitra, Niloy J.},
  journal={TOG},
  year={2021},
}

@inproceedings{bhattacharya2024evdnerf,
  title     = {EvDNeRF: Reconstructing Event Data with Dynamic Neural Radiance Fields},
  author    = {Bhattacharya, Anish and Madaan, Ratnesh and Cladera, Fernando and Vemprala, Sai and Bonatti, Rogerio and Daniilidis, Kostas and Kapoor, Ashish and Kumar, Vijay and Matni, Nikolai and Gupta, Jayesh K},
  booktitle = {WACV},
  year      = {2024}
}

@inproceedings{macklin2016xpbd,
  title={XPBD: position-based simulation of compliant constrained dynamics},
  author={Macklin, Miles and M{\"u}ller, Matthias and Chentanez, Nuttapong},
  booktitle={Proceedings of the 9th International Conference on Motion in Games},
  year={2016}
}

@article{muller2007position,
  author={M{\"u}ller, Matthias and Heidelberger, Bruno and Hennix, Marcus and Ratcliff, John},
  title={Position based dynamics},
  journal={Journal of Visual Communication and Image Representation},
  year={2007},
}

@inproceedings{li2024spacetime,
  title={Spacetime gaussian feature splatting for real-time dynamic view synthesis},
  author={Li, Zhan and Chen, Zhang and Li, Zhong and Xu, Yi},
  booktitle={CVPR},
  year={2024},
}

@inproceedings{wang2024shape,
  author       = {Qianqian Wang and Vickie Ye and Hang Gao and Jake Austin and Zhengqi Li and Angjoo Kanazawa},
  title        = {Shape of Motion: 4D Reconstruction from a Single Video},
  booktitle = {ICCV}, 
  year         = {2025},
}

@misc{houdini_pyrosolver,
  title = {Pyro Solver},
  author = {{SideFX Software}},
  year = {2012},
  howpublished = {\url{https://www.sidefx.com/docs/houdini/pyro/intro.html}},
}

@misc{houdini_vellumsolver,
  title = {Vellum Solver},
  author = {{SideFX Software}},
  year = {2017},
  howpublished = {\url{https://www.sidefx.com/docs/houdini/vellum/overview.html}},
}

@article{Sulsky1994,
  author = {Sulsky, Deborah and Chen, Zhen and Schreyer, Howard L.},
  title = {A particle method for history-dependent materials},
  journal = {Computer Methods in Applied Mechanics and Engineering},
  year = {1994},
}

@article{Brackbill1986,
  author = {J. U. Brackbill},
  title = {FLIP: A low-dissipation, particle-in-cell method for fluid flow},
  journal = {Journal of Computational Physics},
  year = {1986},
}

@inproceedings{hu2019taichi,
  title={Taichi: a language for high-performance computation on spatially sparse data structures},
  author={Hu, Yuanming and Li, Tzu-Mao and Anderson, Luke and Ragan-Kelley, Jonathan and Durand, Fr{\'e}do},
  booktitle={TOG},
  year={2019},
}

@misc{warp2022,
  title        = {Warp: A High-performance Python Framework for GPU Simulation and Graphics},
  author       = {Miles Macklin},
  month        = {March},
  year         = {2022},
  note         = {NVIDIA GPU Technology Conference (GTC)},
  howpublished = {\url{https://github.com/nvidia/warp}}
}

@software{Genesis,
  author = {Genesis Authors},
  title = {Genesis: A Universal and Generative Physics Engine for Robotics and Beyond},
  year = {2024},
  url = {https://github.com/Genesis-Embodied-AI/Genesis}
}

@inproceedings{hu2019difftaichi,
  title={DiffTaichi: Differentiable Programming for Physical Simulation},
  author={Hu, Yuanming and Anderson, Luke and Li, Tzu-Mao and Sun, Qi and Carr, Nathan and Ragan-Kelley, Jonathan and Durand, Fr{\'e}do},
  booktitle={ICLR},
  year={2020}
}

@inproceedings{hu2018mlsmpmcpic,
  title={A Moving Least Squares Material Point Method with Displacement Discontinuity and Two-Way Rigid Body Coupling},
  author={Hu, Yuanming and Fang, Yu and Ge, Ziheng and Qu, Ziyin and Zhu, Yixin and Pradhana, Andre and Jiang, Chenfanfu},
  booktitle={TOG},
  year={2018},
}

@inproceedings{wang2004image,
  title={Image quality assessment: from error visibility to structural similarity},
  author={Wang, Zhou and Bovik, Alan C and Sheikh, Hamid R and Simoncelli, Eero P},
  booktitle={TIP},
  year={2004},
}

@inproceedings{Zhang_2018_CVPR,
  title = {The Unreasonable Effectiveness of Deep Features as a Perceptual Metric},
  author = {Zhang, Richard and Isola, Phillip and Efros, Alexei A. and Shechtman, Eli and Wang, Oliver},
  booktitle = {CVPR},
  year = {2018}
}

@inproceedings{nerf,
 title={NeRF: Representing Scenes as Neural Radiance Fields for View Synthesis},
 author={Ben Mildenhall and Pratul P. Srinivasan and Matthew Tancik and Jonathan T. Barron and Ravi Ramamoorthi and Ren Ng},
 year={2020},
 booktitle={ECCV},
}

@inproceedings{chen2022tensorf,
  title={Tensorf: Tensorial radiance fields},
  author={Chen, Anpei and Xu, Zexiang and Geiger, Andreas and Yu, Jingyi and Su, Hao},
  booktitle={ECCV},
  year={2022},
}

@inproceedings{garbin2021fastnerf,
  title={Fastnerf: High-fidelity neural rendering at 200fps},
  author={Garbin, Stephan J and Kowalski, Marek and Johnson, Matthew and Shotton, Jamie and Valentin, Julien},
  booktitle={ICCV},
  year={2021}
}

@inproceedings{realtime_plenoctree,
    author    = {Wang, Liao and Zhang, Jiakai and Liu, Xinhang and Zhao, Fuqiang and 
        Zhang, Yanshun and Zhang, Yingliang and Wu, Minye and Yu, Jingyi and Xu, Lan},
    title     = {Fourier PlenOctrees for Dynamic Radiance Field Rendering in Real-Time},
    booktitle = {CVPR},
    year      = {2022},
}

@inproceedings{muller2022instant,
  title={Instant neural graphics primitives with a multiresolution hash encoding},
  author={M{\"u}ller, Thomas and Evans, Alex and Schied, Christoph and Keller, Alexander},
  booktitle={ACM TOG},
  year={2022},
}

@inproceedings{3dgs,
  title={3d gaussian splatting for real-time radiance field rendering},
  author={Kerbl, Bernhard and Kopanas, Georgios and Leimk{\"u}hler, Thomas and Drettakis, George},
  booktitle={ACM ToG},
  year={2023}
}

@inproceedings{gao2021dynamic,
  title={Dynamic view synthesis from dynamic monocular video},
  author={Gao, Chen and Saraf, Ayush and Kopf, Johannes and Huang, Jia-Bin},
  booktitle={ICCV},
  year={2021}
}

@inproceedings{du2021neural,
  title={Neural radiance flow for 4d view synthesis and video processing},
  author={Du, Yilun and Zhang, Yinan and Yu, Hong-Xing and Tenenbaum, Joshua B and Wu, Jiajun},
  booktitle={ICCV},
  year={2021},
}

@inproceedings{Hexplane,
    author    = {Cao, Ang and Johnson, Justin},
    title     = {HexPlane: A Fast Representation for Dynamic Scenes},
    booktitle = {CVPR},
    year      = {2023},
    }

@inproceedings{kplanes_2023,
    title={K-Planes: Explicit Radiance Fields in Space, Time, and Appearance},
    author={Sara Fridovich-Keil and Giacomo Meanti and Frederik Rahbæk Warburg and Benjamin Recht and Angjoo Kanazawa},
    year={2023},
    booktitle={CVPR},
}

@inproceedings{shao2023tensor4d,
  title={Tensor4d: Efficient neural 4d decomposition for high-fidelity dynamic reconstruction and rendering},
  author={Shao, Ruizhi and Zheng, Zerong and Tu, Hanzhang and Liu, Boning and Zhang, Hongwen and Liu, Yebin},
  booktitle={CVPR},
  year={2023}
}

@inproceedings{fang2022fast,
  title={Fast dynamic radiance fields with time-aware neural voxels},
  author={Fang, Jiemin and Yi, Taoran and Wang, Xinggang and Xie, Lingxi and Zhang, Xiaopeng and Liu, Wenyu and Nie{\ss}ner, Matthias and Tian, Qi},
  booktitle={SIGGRAPH Asia},
  year={2022}
}

@inproceedings{yang2024deformable,
  title={Deformable 3d gaussians for high-fidelity monocular dynamic scene reconstruction},
  author={Yang, Ziyi and Gao, Xinyu and Zhou, Wen and Jiao, Shaohui and Zhang, Yuqing and Jin, Xiaogang},
  booktitle={CVPR},
  year={2024}
}

@inproceedings{huang2023sc,
  title={SC-GS: Sparse-Controlled Gaussian Splatting for Editable Dynamic Scenes},
  author={Huang, Yi-Hua and Sun, Yang-Tian and Yang, Ziyi and Lyu, Xiaoyang and Cao, Yan-Pei and Qi, Xiaojuan},
  booktitle={CVPR},
  year={2024}
}

@inproceedings{wu20234d,
  title={4d gaussian splatting for real-time dynamic scene rendering},
  author={Wu, Guanjun and Yi, Taoran and Fang, Jiemin and Xie, Lingxi and Zhang, Xiaopeng and Wei, Wei and Liu, Wenyu and Tian, Qi and Wang, Xinggang},
  booktitle={CVPR},
  year={2024}
}

@inproceedings{kim2024ua4dgs,
  title     = {UA-4DGS: 4D Gaussian Splatting in the Wild with Uncertainty-Aware Regularization},
  author    = {Mijeong Kim and Jongwoo Lim and Bohyung Han},
  booktitle = {NeurIPS},
  year      = {2024},
}

@misc{cai2024dyna,
  title       = {DynaSurfGS: Dynamic Surface Reconstruction with Planar-based Gaussian Splatting},
  author      = {Weiwei Cai and Weicai Ye and Peng Ye and Tong He and Tao Chen},
  year        = {2024},
  note        = {arXiv preprint arXiv:2408.13972}
}

@misc{zhang2024bags,
  title        = {BAGS: Building Animatable Gaussian Splatting from a Monocular Video with Diffusion Priors},
  author       = {Tingyang Zhang and Qingzhe Gao and Weiyu Li and Libin Liu and Baoquan Chen},
  year         = {2024},
  note.        = {arXiv preprint arXiv:2403.11427},
}

@inproceedings{kratimenos2024dynmf,
  title     = {DynMF: Neural Motion Factorization for Real-time Dynamic View Synthesis with 3D Gaussian Splatting},
  author    = {Agelos Kratimenos and Jiahui Lei and Kostas Daniilidis},
  booktitle = {ECCV},
  year      = {2024},
}

@misc{kwak2025modecgs,
  title   = {MoDec-GS: Global-to-Local Motion Decomposition and Temporal Interval Adjustment for Compact Dynamic 3D Gaussian Splatting},
  author  = {Sangwoon Kwak and Joonsoo Kim and Jun Young Jeong and Won-Sik Cheong and Jihyong Oh and Munchurl Kim},
  note = {arXiv preprint arXiv:2501.03714},
  year    = {2025},
}

@inproceedings{rombach2022high,
  title     = {High-Resolution Image Synthesis with Latent Diffusion Models},
  author    = {Robin Rombach and Andreas Blattmann and Dominik Lorenz and Patrick Esser and Björn Ommer},
  booktitle = {CVPR},
  year      = {2022},
}

@inproceedings{lin2024gaussian,
  title={Gaussian-flow: 4d reconstruction with dynamic 3d gaussian particle},
  author={Lin, Youtian and Dai, Zuozhuo and Zhu, Siyu and Yao, Yao},
  booktitle={CVPR},
  year={2024}
}

@inproceedings{lu20243d,
  title={3d geometry-aware deformable gaussian splatting for dynamic view synthesis},
  author={Lu, Zhicheng and Guo, Xiang and Hui, Le and Chen, Tianrui and Yang, Min and Tang, Xiao and Zhu, Feng and Dai, Yuchao},
  booktitle={CVPR},
  year={2024}
}

@inproceedings{guo2024motion,
  title={Motion-aware 3d gaussian splatting for efficient dynamic scene reconstruction},
  author={Guo, Zhiyang and Zhou, Wengang and Li, Li and Wang, Min and Li, Houqiang},
  booktitle={TCSVT},
  year={2024},
}

@inproceedings{liang2025gaufre,
  title={Gaufre: Gaussian deformation fields for real-time dynamic novel view synthesis},
  author={Liang, Yiqing and Khan, Numair and Li, Zhengqin and Nguyen-Phuoc, Thu and Lanman, Douglas and Tompkin, James and Xiao, Lei},
  booktitle={WACV},
  year={2025},
}

@misc{lei2024mosca,
  title={Mosca: Dynamic gaussian fusion from casual videos via 4d motion scaffolds},
  author={Lei, Jiahui and Weng, Yijia and Harley, Adam and Guibas, Leonidas and Daniilidis, Kostas},
  note={arXiv preprint arXiv:2405.17421},
  year={2024}
}

@inproceedings{duan20244d,
  title={4d-rotor gaussian splatting: towards efficient novel view synthesis for dynamic scenes},
  author={Duan, Yuanxing and Wei, Fangyin and Dai, Qiyu and He, Yuhang and Chen, Wenzheng and Chen, Baoquan},
  booktitle={SIGGRAPH},
  year={2024}
}

@inproceedings{waczynska2024d,
  title={D-miso: Editing dynamic 3d scenes using multi-gaussians soup},
  author={Waczynska, Joanna and Borycki, Piotr and Kaleta, Joanna and Tadeja, Slawomir and Spurek, Przemys{\l}aw},
  booktitle={NeurIPS},
  year={2024}
}

@inproceedings{liu2024modgs,
  title={MoDGS: Dynamic Gaussian Splatting from Casually-captured Monocular Videos},
  author={Liu, Qingming and Liu, Yuan and Wang, Jiepeng and Lyv, Xianqiang and Wang, Peng and Wang, Wenping and Hou, Junhui},
  booktitle={ICLR},
  year={2025}
}

@inproceedings{stearns2024dynamic,
  title={Dynamic gaussian marbles for novel view synthesis of casual monocular videos},
  author={Stearns, Colton and Harley, Adam and Uy, Mikaela and Dubost, Florian and Tombari, Federico and Wetzstein, Gordon and Guibas, Leonidas},
  booktitle={SIGGRAPH},
  year={2024}
}

@inproceedings{yang2024depth,
  title     = {Depth Anything: Unleashing the Power of Large-Scale Unlabeled Data},
  author    = {Lihe Yang and Bingyi Kang and Zilong Huang and Xiaogang Xu and Jiashi Feng and Hengshuang Zhao},
  booktitle = {CVPR},
  year      = {2024},
}

@article{yang2023cotracker,
  title   = {CoTracker: Transformers for Tracking Any Point},
  author  = {Zhengqi Yang and Yilun Du and Deqing Sun and Varun Jampani and Ce Liu and William T. Freeman and Joshua B. Tenenbaum and Jiajun Wu},
  note = {arXiv preprint arXiv:2303.06583},
  year    = {2023},
}

@article{sorkine2007as,
  title   = {As-Rigid-As-Possible Surface Modeling},
  author  = {Olga Sorkine and Marc Alexa},
  journal = {TOG},
  year    = {2007},
}

@article{tomasi1992shape,
  title   = {Shape and Motion from Image Streams under Orthography: A Factorization Method},
  author  = {Carlo Tomasi and Takeo Kanade},
  journal = {International Journal of Computer Vision},
  year    = {1992},
}

@article{gingold1977smoothed,
  title={Smoothed particle hydrodynamics: theory and application to non-spherical stars},
  author={Gingold, Robert A. and Monaghan, Joseph J.},
  journal={Monthly Notices of the Royal Astronomical Society},
  year={1977},
}

@inproceedings{liao2024evaluation,
  title={Evaluation of text-to-video generation models: A dynamics perspective},
  author={Liao, Mingxiang and Ye, Qixiang and Zuo, Wangmeng and Wan, Fang and Wang, Tianyu and Zhao, Yuzhong and Wang, Jingdong and Zhang, Xinyu and others},
  booktitle={NeurIPS},
  year={2024}
}

@inproceedings{heusel2017gans,
  title={GANs Trained by a Two Time-Scale Update Rule Converge to a Local Nash Equilibrium},
  author={Heusel, Martin and Ramsauer, Hubert and Unterthiner, Thomas and Nessler, Bernhard and Hochreiter, Sepp},
  booktitle={NeurIPS},
  year={2017}
}

@inproceedings{binkowski2018demystifying,
  title={Demystifying MMD GANs},
  author={Binkowski, Miko{\l}aj and Sutherland, Danica J. and Arbel, Michael and Gretton, Arthur},
  booktitle={ICLR},
  year={2018}
}

@inproceedings{kim2026gp4dgs,
  title     = {GP-4DGS: Probabilistic 4D Gaussian Splatting from Monocular Video
via Variational Gaussian Processes},
  author    = {Mijeong Kim and Jungtaek Kim and Bohyung Han},
  booktitle = {CVPR},
  year      = {2026},
}

@inproceedings{Kim_2022_CVPR,
    author    = {Kim, Mijeong and Seo, Seonguk and Han, Bohyung},
    title     = {InfoNeRF: Ray Entropy Minimization for Few-Shot Neural Volume Rendering},
    booktitle = {CVPR},
    year      = {2022},
}
}

\onecolumn

\setcounter{section}{0}
\setcounter{table}{0}
\setcounter{figure}{0}
\setcounter{equation}{0}
\setcounter{algorithm}{0}
\renewcommand\thesection{\Alph{section}}
\renewcommand\thetable{\Alph{table}}
\renewcommand\thefigure{\Alph{figure}}
\renewcommand\theequation{\Alph{equation}}
\renewcommand\thealgorithm{\Alph{algorithm}}

\section{Preliminary: 4D Gaussian Splatting}
\subsection{Gaussian primitive}

3D Gaussian Splatting has recently achieved real-time rendering with state-of-the-art quality on static scenes.
It adopts an explicit representation of the 3D scene through a collection of Gaussian ellipsoids $\Gamma=\left\{  \gamma_1, ..., \gamma_K \right\}$. 
Each primitive $\gamma_k$ is defined by an unnormalized 3D Gaussian kernel $\mathcal G_k(\vct x)$, parameterized by its mean $\vct \mu_k$ and covariance $\mat \Sigma_k$:
\begin{align}
\label{eq:gaussian}
\mathcal G_k(\vct x; \vct \mu_k, \mat \Sigma_k)\coloneqq \exp\left(-\frac{1}{2}(\vct x-\vct\mu_k)^\top\mat\Sigma_k^{-1}(\vct x-\vct\mu_k)\right), 
\end{align}
where $\vct \mu_k \in \mathbb{R}^3$ denotes the primitive’s center position, $\mat \Sigma_k \in \mathbb{R}^{3\times 3}$ is an anisotropic covariance matrix, and $\vct x \in \mathbb{R}^3$ represents an arbitrary spatial coordinate.
Since $\mat \Sigma_k$ must remain positive semi-definite during optimization—a constraint that is difficult to enforce directly—it is instead factorized into a rotation matrix $\mat R_k$ and a scaling matrix $\mat S_k$:
\begin{align}
\label{formula:covariance decomposition}
    \mat\Sigma_k \coloneqq \mat R_k \mat S_k \mat S_k^\top \mat R_k^\top.
\end{align}
Beyond these Gaussian parameters $(\vct \mu_k, \mat R_k, \mat S_k)$, each primitive also requires an opacity value $\alpha_k \in [0,1]$ and a feature vector $\vct f_k \in \mathbb{R}^d$, which typically encodes RGB colors or spherical harmonic (SH) coefficients.
Thus, a Gaussian primitive is fully specified as
$\gamma_k \coloneqq (\vct \mu_k, \mat R_k, \mat S_k, \alpha_k, \vct f_k)$.

\subsection{Differentiable rasterization}

Before rendering the Gaussian primitives $\Gamma$ into the image space, each 3D Gaussian kernel, $\mathcal G_k(\vct x ; \vct \mu_k, \mat \Sigma_k)$, is first projected onto the 2D image space, resulting in a 2D Gaussian kernel, $\mathcal G^\pi_k(\vct r; \vct \mu_k^\pi, \mat \Sigma_k^\pi)$.
Here, $\pi:\mathbb R^3\to\mathbb R^2$ denotes the projection from world coordinates to image space. 
In the projected Gaussian representation, $\vct r \in \mathbb{R}^2$ indicates a pixel location in an image, and the 2D mean $ \vct\mu_k^\pi \in\mathbb R^2$ and covariance $\mat\Sigma_k^\pi \in \mathbb{R}^{2\times 2}$ are given by
\begin{align}
   \vct\mu_k^\pi \coloneqq \pi (\vct \mu_k) \quad\quad \text{and}  \quad\quad  \mat \Sigma_k^\pi \coloneqq \mat J \mat W \mat  \Sigma_k \mat W^\top \mat J^\top,
\end{align}
where $\mat J$ is the Jacobian of the affine approximation of the projective transformation and $\mat W$ is the world-to-camera transformation matrix.
When rendering the primitives in $\Gamma$ from a target camera view, they are sorted according to their depth relative to the camera center.
The color of a pixel $\vct r$ is then determined by $\alpha$-blending:
\begin{align}
\label{eq:color_render}
     \mathbf{\hat I}(\vct r) \coloneqq  \sum_{k=1}^K  \omega^\pi_k(\vct r) c( \vct f_k, \vct r),
\end{align}
where $\omega^\pi_k(\vct r)$ denotes the contribution of primitive $\gamma_k$ to pixel $\vct r$, and $c(\vct f_k, \vct r)$ is its corresponding color.
If $\vct f_k$ encodes spherical harmonics coefficients, the color is evaluated using the view direction associated with $\vct r$; otherwise, $\vct f_k$ can directly represent the primitive’s RGB values.
Following the $\alpha$-blending formulation in 3DGS~\citep{3dgs}, the visibility weight is defined as
\begin{align}
\label{eq:visible_score}
   \omega^\pi_k(\vct r)\coloneqq\alpha_k\mathcal G^\pi_k(\vct r ; \vct \mu_k^\pi, \mat \Sigma_k^\pi)\prod_{j=1}^{k-1}\left(1-\alpha_j\mathcal G^\pi_j(\vct r ; \vct \mu_j^\pi, \mat \Sigma_j^\pi)\right),
\end{align}
where $\alpha_k \mathcal G^\pi_k(\vct r ; \vct \mu_k^\pi, \mat \Sigma_k^\pi)$ represents the opacity of the $k^\text{th}$ projected primitive at pixel $\vct r$, and the product term encodes the transmittance, i.e., how much light passes through the preceding primitives along the ray. Further details can be found in the original Gaussian Splatting paper~\citep{3dgs}.

\subsection{Deformation for Dynamic Modelings}

To represent 4D scenes with Gaussian Splatting, recent methods~\citep{yang2024deformable, huang2023sc, wu20234d, li2024spacetime, wang2024shape, lei2024mosca} learn time-dependent deformations that map canonical 3D Gaussian primitives to their posed states over time. 
At time $t$, the position, rotation, and scale of the $k$-th primitive are updated as
\begin{align}
(\vct \mu_k^t, \mat R_k^t, \mat S_k^t) 
= \big(\vct \mu_k + \phi_\mu(\cdot,t),\;
       \phi_r(\cdot,t)\,\mat R_k,\;
       \mat S_k + \phi_s(\cdot,t)\big),
\end{align}
where the deformation functions $\phi_\mu(\cdot,t)$, $\phi_r(\cdot,t)$, and $\phi_s(\cdot,t)$ denote learnable fields that may take as input various combinations of canonical primitive attributes, spatial coordinates, or time embeddings, depending on the parameterization. 
The rotation update $\phi_r(\cdot,t)$ is implemented as a residual rotation, typically parameterized via an exponential map to ensure $\phi_r(\cdot,t) \in SO(3)$. 
These functions are commonly instantiated with MLPs, learnable control points~\citep{ wang2024shape, lei2024mosca, huang2023sc}, HexPlane~\citep{wu20234d}, or low-order polynomials~\cite{li2024spacetime}. 
The resulting time-varying primitive is then given by 
$\gamma_k (t) \coloneqq (\vct{\mu}^t_k, \mat{R}^t_k, \mat{S}^t_k, \alpha_k, \vct{f}_k)$.


\vspace{8mm}
\section{Scene Composition}
Our benchmark consists of 17 scenes, categorized into four material types: liquids, gases, rheological materials, and textiles. 
Each category contains 4 to 5 scenes, as listed in Table~\ref{tab:scene_list}. 
This section details the solver properties for each material and provides key simulation details regarding multi-body interactions.
Please refer to our supplemental document for other simulation configurations.

\begin{table}[h]
\vspace{2mm}
\centering
\caption{List of scenes included in our {\dataset} benchmark. Our benchmark consists of 17 scenes, categorized into four material types: liquids, gases, rheological materials, and textiles.}
\scalebox{0.8}{
\setlength\tabcolsep{3pt} 
\begin{tabular}{c|c|c|c|cc}
\toprule
& \textbf{Liquid}		& \textbf{Gas} & \textbf{Rheological substance}&\textbf{Textile}			\\
\midrule
\multirow{5}{*}{Scene name} &Cereal & Pisa & Jelly party & Lucy\\
&Ship&Box-smoke & Pancake & Basin \\
&Hanok & Single smoke & Bouncing balls & Flags\\
&Ice & Falling & Cow & Single flag\\
&-- & --  &-- & Tube \\
\bottomrule
\end{tabular}
}
\label{tab:scene_list}
\vspace{4mm}
\end{table}

\begin{figure*}[h]
    \begin{center}
        \centering
                        \begin{subfigure}[b]{\linewidth}
        \centering
        \includegraphics[width=0.161\linewidth]{./asset/phystrack/flip_hanok/1_022.png}
        \includegraphics[width=0.161\linewidth]{./asset/phystrack/flip_hanok/1_031.png}
        \includegraphics[width=0.161\linewidth]{./asset/phystrack/flip_hanok/1_070.png}
        \includegraphics[width=0.161\linewidth]{./asset/phystrack/flip_ship/1_003.png}
        \includegraphics[width=0.161\linewidth]{./asset/phystrack/flip_ship/1_035.png}
        \includegraphics[width=0.161\linewidth]{./asset/phystrack/flip_ship/1_100.png}
        \includegraphics[width=0.161\linewidth]{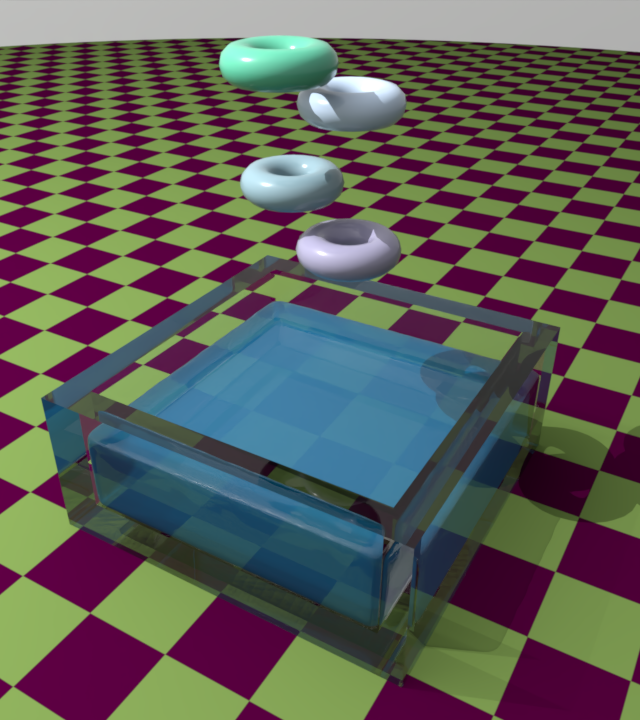}
        \includegraphics[width=0.161\linewidth]{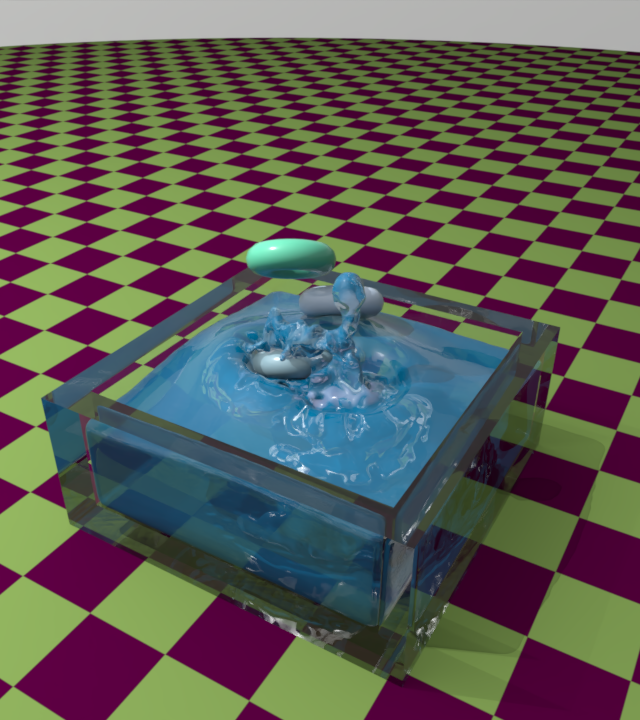}
        \includegraphics[width=0.161\linewidth]{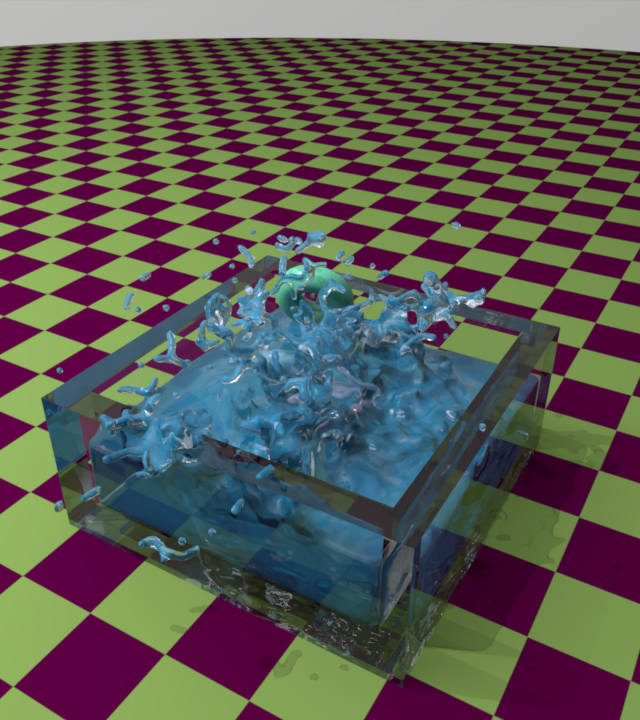}
        \includegraphics[width=0.161\linewidth]{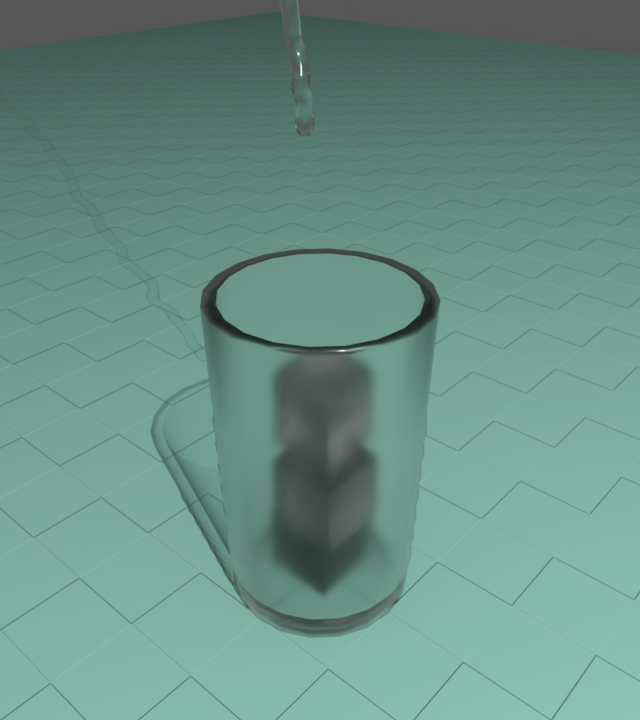}
        \includegraphics[width=0.161\linewidth]{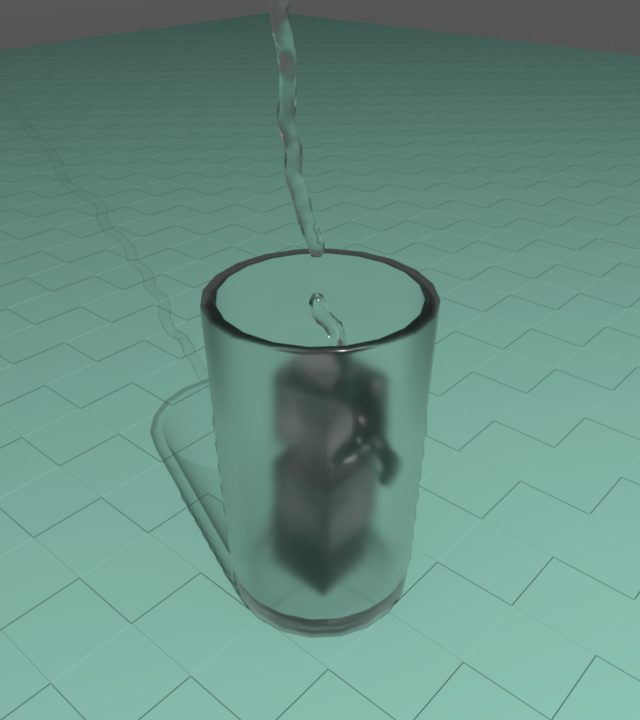}
        \includegraphics[width=0.161\linewidth]{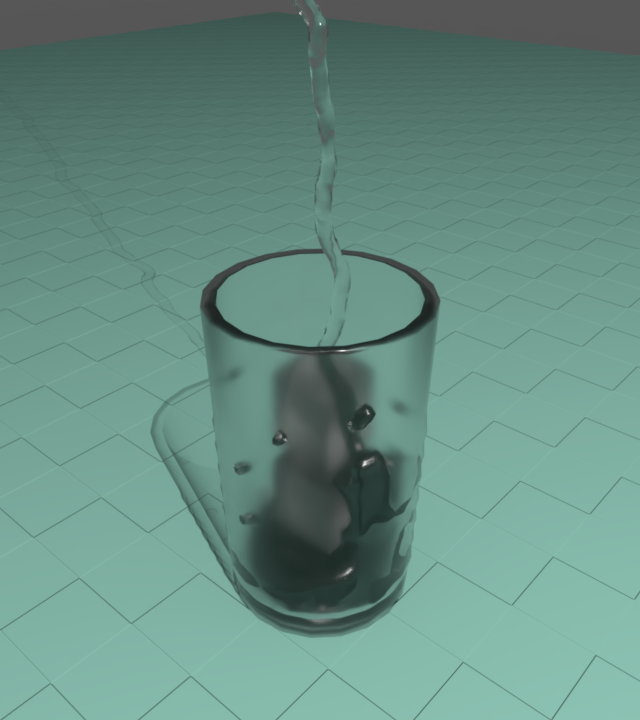}
        \caption{Liquid: \textit{hanok} (top-left), \textit{ship} (top-right), \textit{cereal} (bottom-left), \textit{ice} (bottom-right)}
               \hfill \vspace{-1mm}
        \end{subfigure}
                 \begin{subfigure}[b]{\linewidth}
        \centering
         \includegraphics[width=0.161\linewidth]{./asset/phystrack/pyro_box/2_006.png}
         \includegraphics[width=0.161\linewidth]{./asset/phystrack/pyro_box/2_016.png}
         \includegraphics[width=0.161\linewidth]{./asset/phystrack/pyro_box/2_026.png}
        \includegraphics[width=0.161\linewidth]{./asset/phystrack/pyro_pisa/2_006.png}
        \includegraphics[width=0.161\linewidth]{./asset/phystrack/pyro_pisa/2_026.png}
        \includegraphics[width=0.161\linewidth]{./asset/phystrack/pyro_pisa/2_141.png}
         \includegraphics[width=0.161\linewidth]{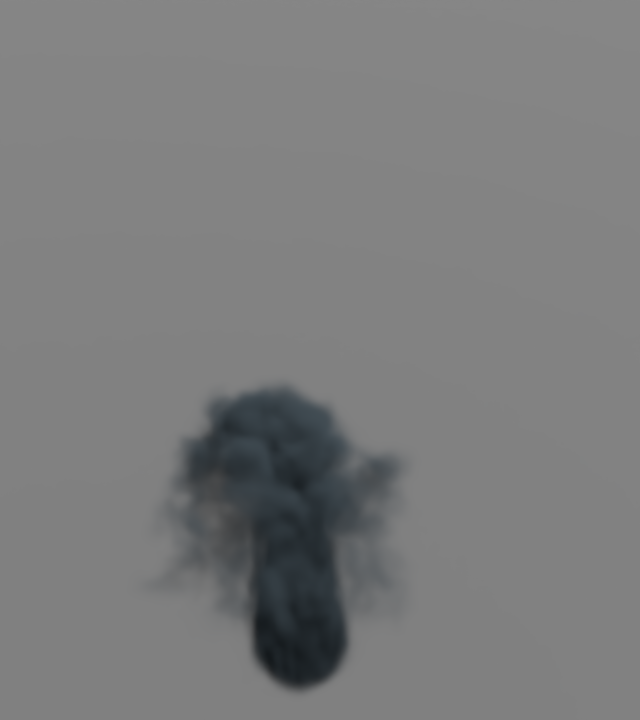}
         \includegraphics[width=0.161\linewidth]{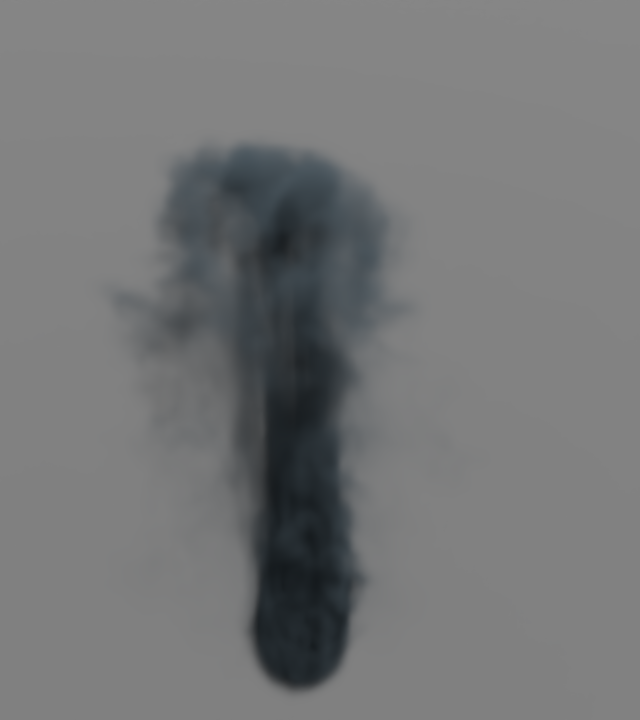}
         \includegraphics[width=0.161\linewidth]{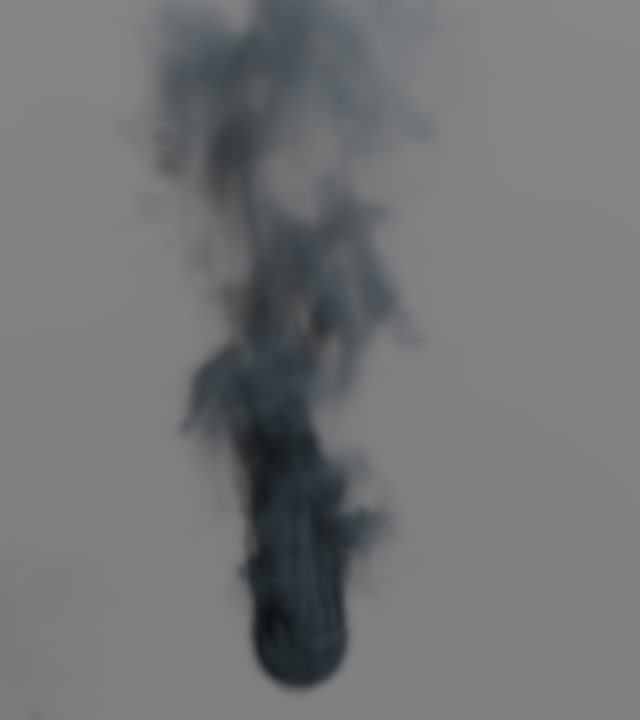}
         \includegraphics[width=0.161\linewidth]{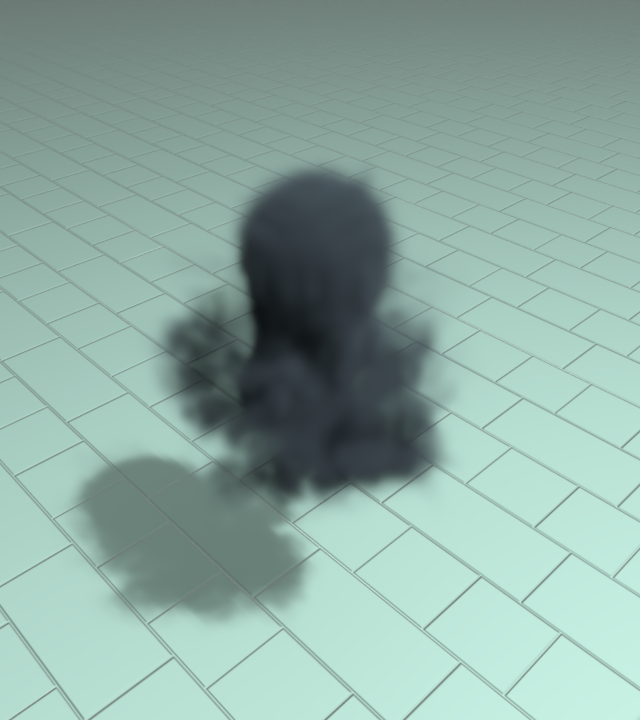}
         \includegraphics[width=0.161\linewidth]{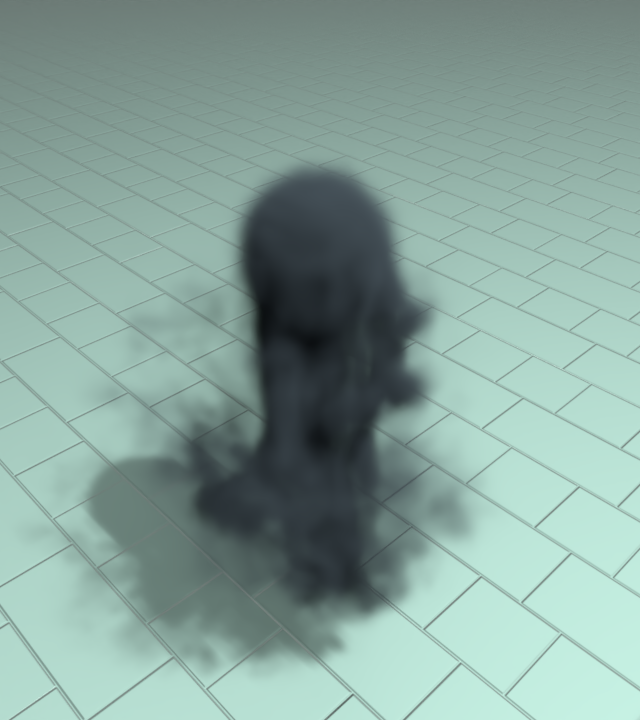}
         \includegraphics[width=0.161\linewidth]{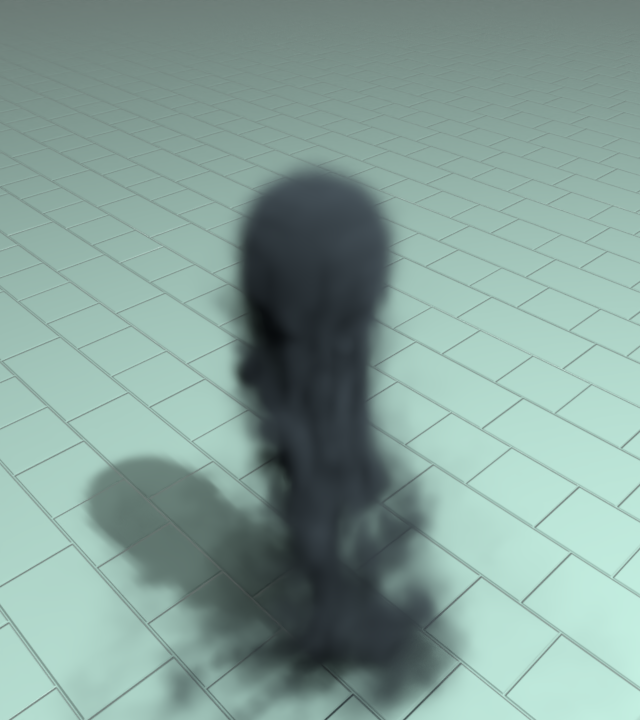}
        \centering
        \caption{Gas: \textit{box-smoke} (top-left), \textit{pisa} (top-right)), \textit{simple smoke} (bottom-left), and \textit{falling} (bottom-right)}
        \hfill \vspace{-1mm}
        \end{subfigure}
       \hfill\vspace{-6mm}
  \caption{
  Examples from the proposed physics-aware dataset, {\dataset}. 
  They exhibit complex physical interactions between multiple bodies composed of diverse materials such as liquid, gas, viscoelastic substance, and textile.
This dataset will foster physics reasoning in dynamic scenes.
}
  	\label{fig:data_viz_1}
    \end{center}%
\end{figure*}

\begin{figure*}[h]
    \begin{center}
        \centering     
                        \begin{subfigure}[b]{\linewidth}
        \centering
        \includegraphics[width=0.161\linewidth]{./asset/phystrack/mpm_pancake/0_003.png}
        \includegraphics[width=0.161\linewidth]{./asset/phystrack/mpm_pancake/0_044.png}
        \includegraphics[width=0.161\linewidth]{./asset/phystrack/mpm_pancake/0_105.png}
        \includegraphics[width=0.161\linewidth]{./asset/phystrack/mpm_jellyparty/1_010.png}
        \includegraphics[width=0.161\linewidth]{./asset/phystrack/mpm_jellyparty/1_031.png}
        \includegraphics[width=0.161\linewidth]{./asset/phystrack/mpm_jellyparty/1_049.png}
        \includegraphics[width=0.161\linewidth]{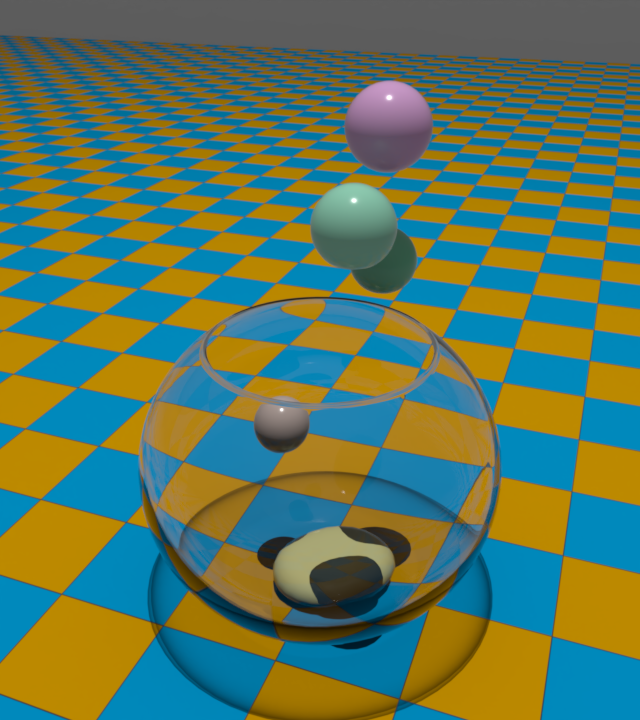}
        \includegraphics[width=0.161\linewidth]{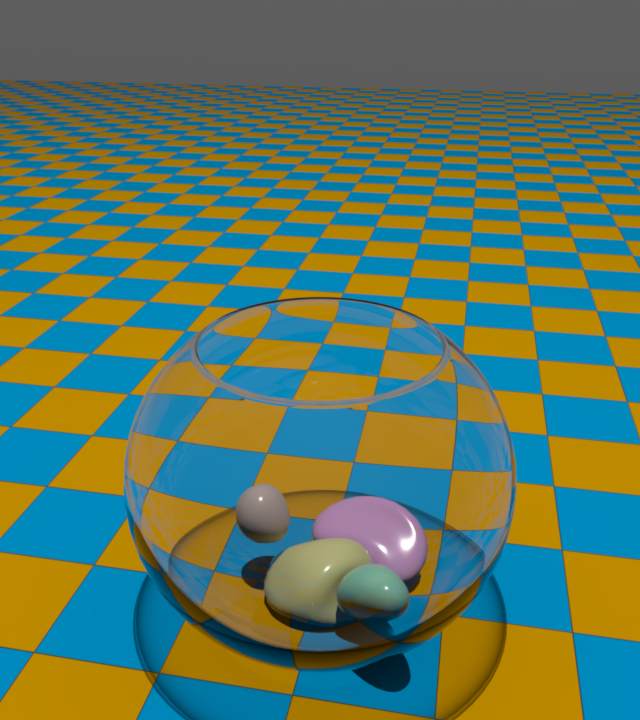}
        \includegraphics[width=0.161\linewidth]{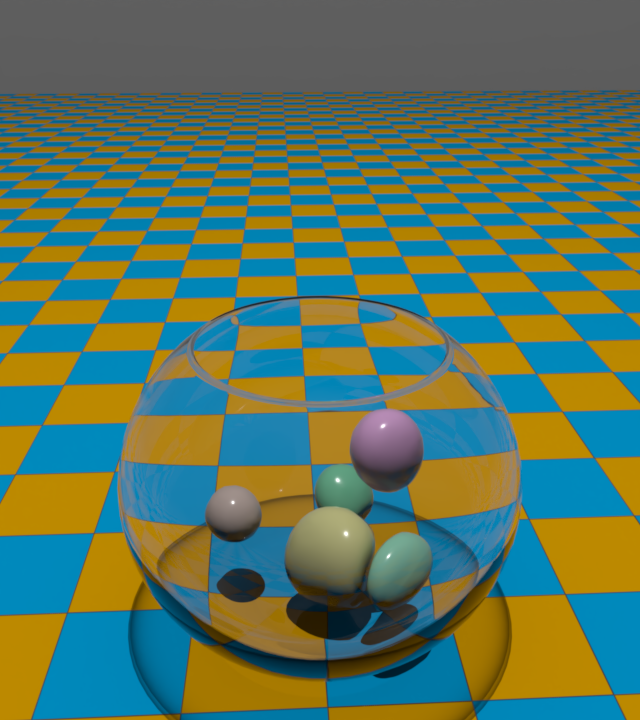}
        \includegraphics[width=0.161\linewidth]{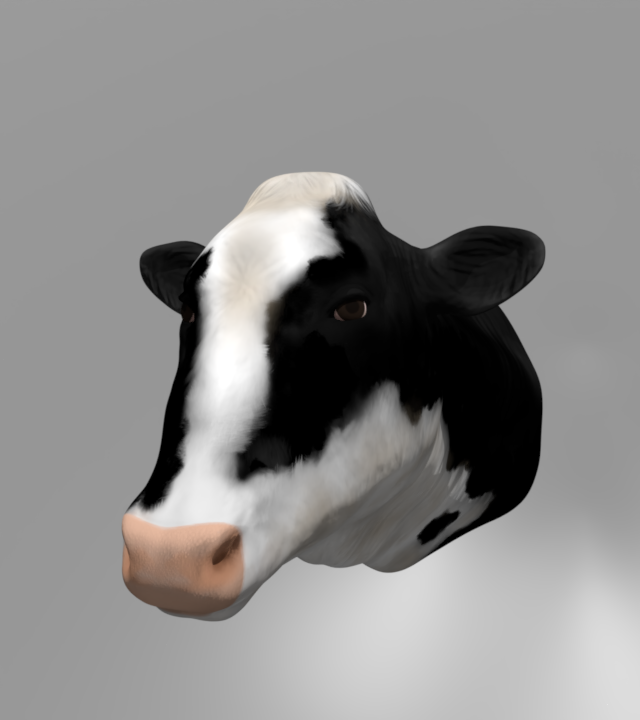}
         \includegraphics[width=0.161\linewidth]{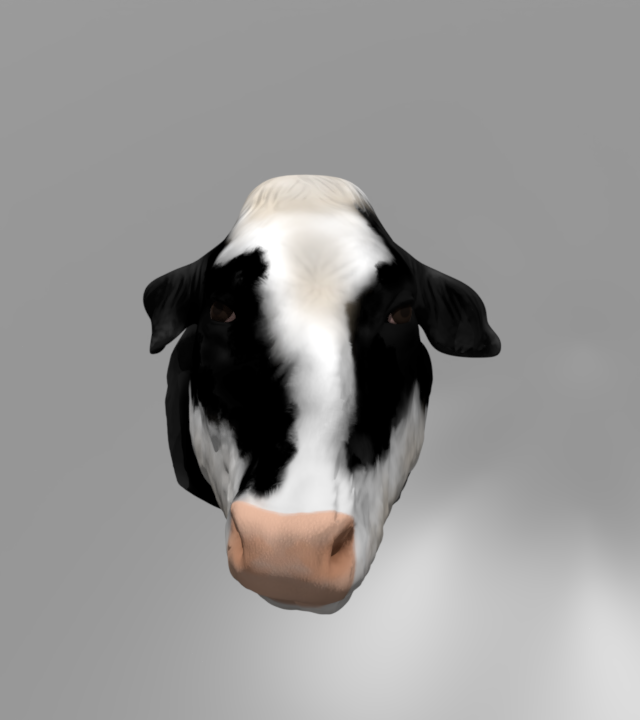}
        \includegraphics[width=0.161\linewidth]{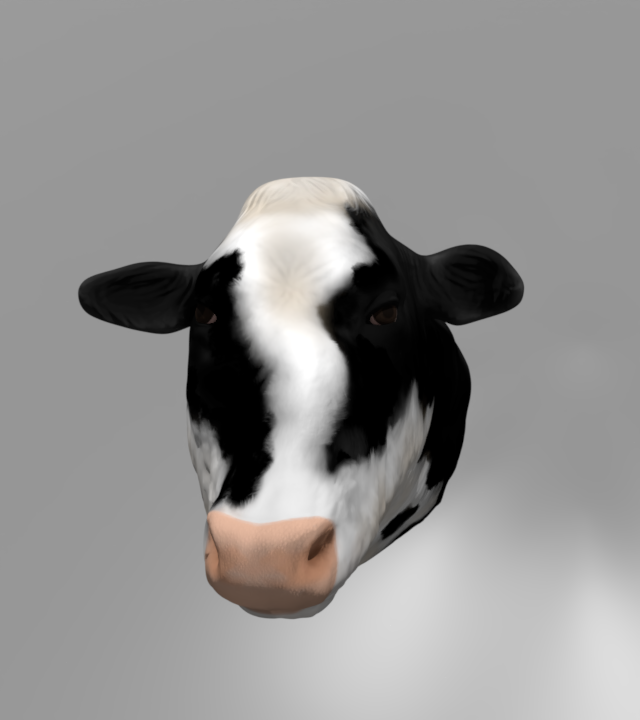}
        \caption{Rheological substances: \textit{pancake} (top-left), \textit{jelly party} (top-right), \textit{bouncing balls} (bottom-left), and \textit{cow} (bottom-right)}
               \hfill \vspace{-1mm}
        \end{subfigure}
         \begin{subfigure}[b]{\linewidth}
        \centering
        \includegraphics[width=0.161\linewidth]{./asset/phystrack/vellum_lucy/2_028.png}
        \includegraphics[width=0.161\linewidth]{./asset/phystrack/vellum_lucy/2_034.png}
        \includegraphics[width=0.161\linewidth]{./asset/phystrack/vellum_lucy/2_091.png}
         \includegraphics[width=0.161\linewidth]{./asset/phystrack/vellum_boxes/2_011.png}
         \includegraphics[width=0.161\linewidth]{./asset/phystrack/vellum_boxes/2_039.png}
         \includegraphics[width=0.161\linewidth]{./asset/phystrack/vellum_boxes/2_072.png}
         \includegraphics[width=0.161\linewidth]{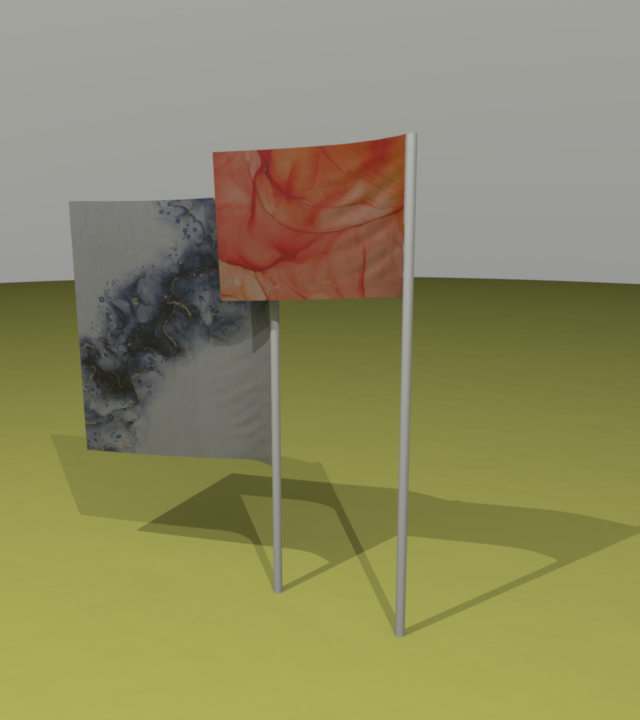}
         \includegraphics[width=0.161\linewidth]{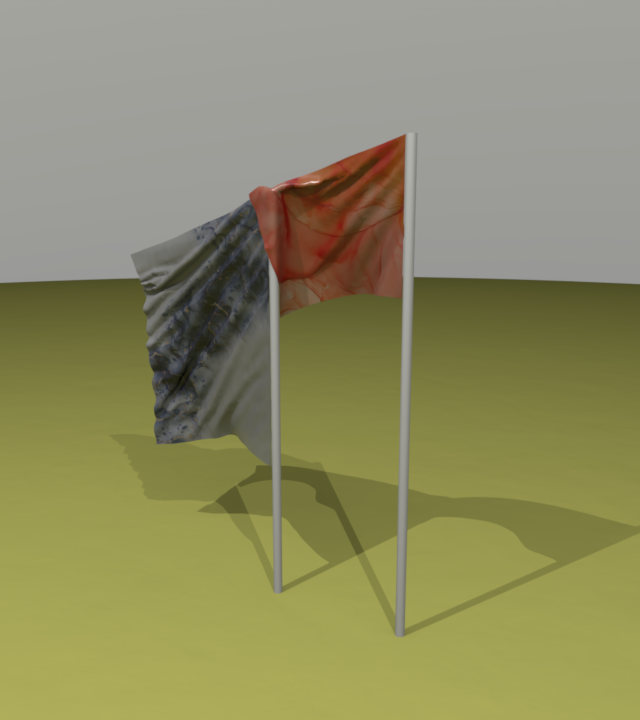}
         \includegraphics[width=0.161\linewidth]{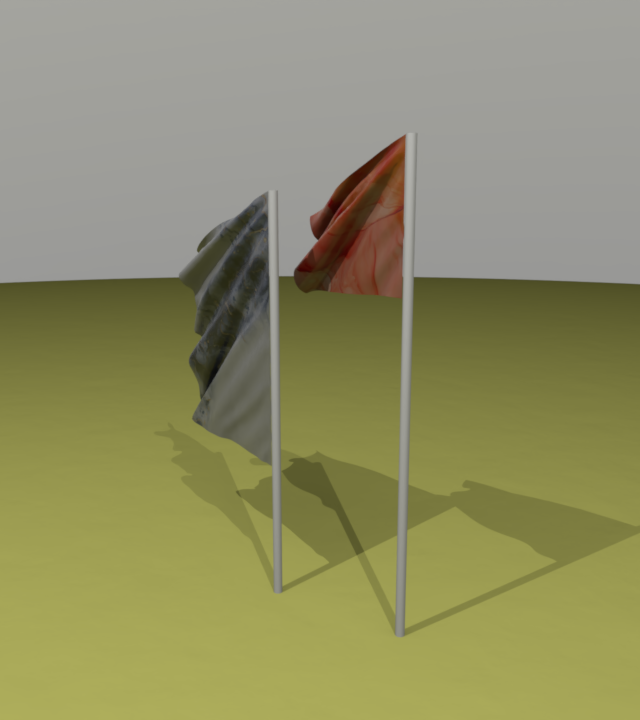}
       \includegraphics[width=0.161\linewidth]{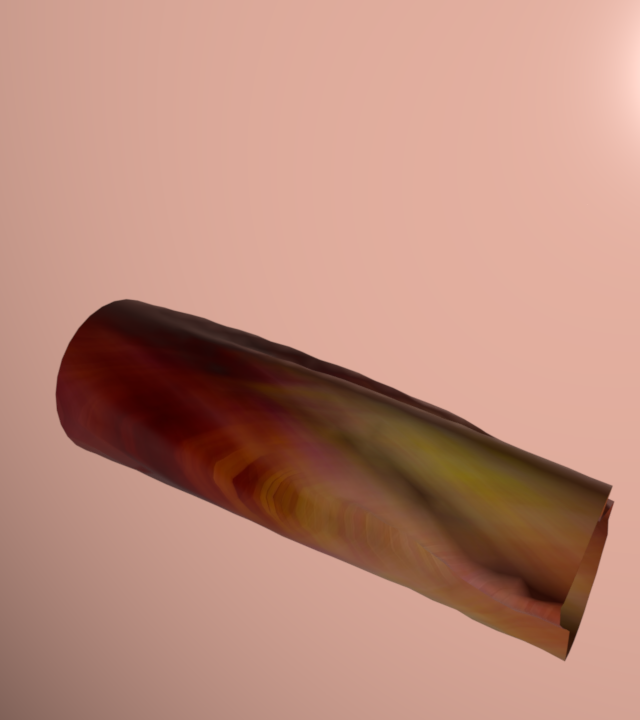}
       \includegraphics[width=0.161\linewidth]{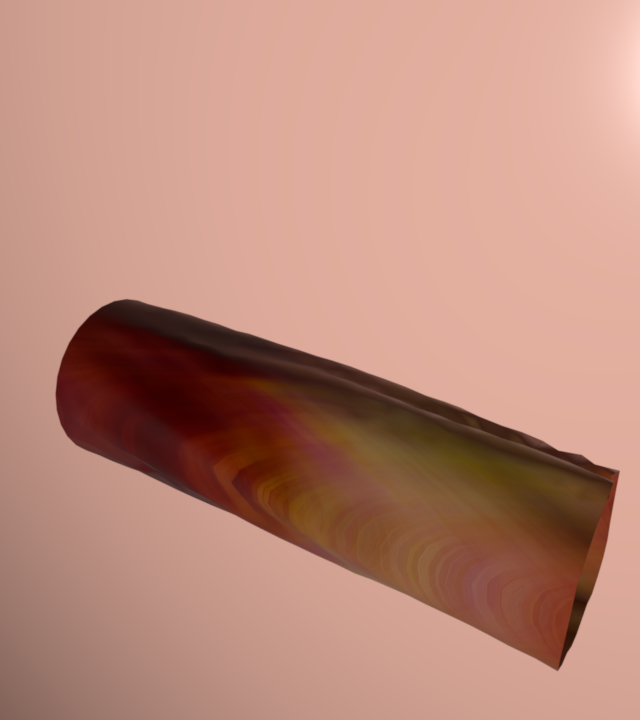}
       \includegraphics[width=0.161\linewidth]{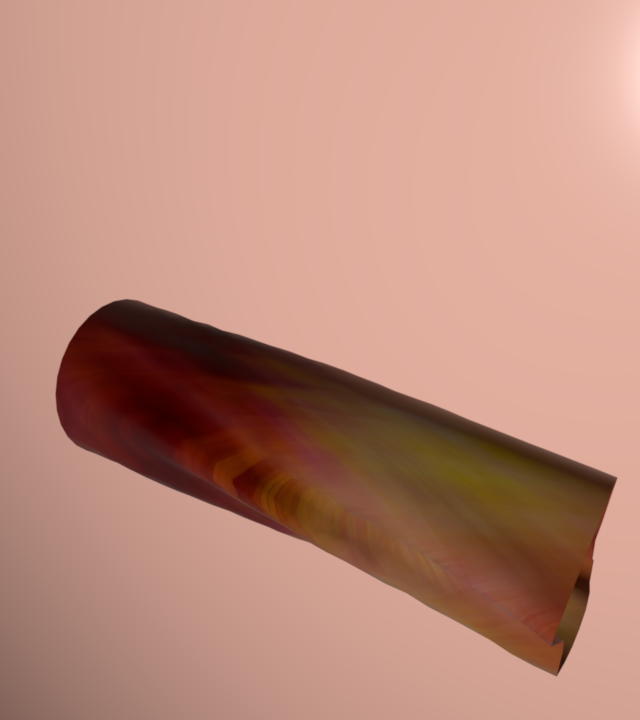}
       \includegraphics[width=0.161\linewidth]{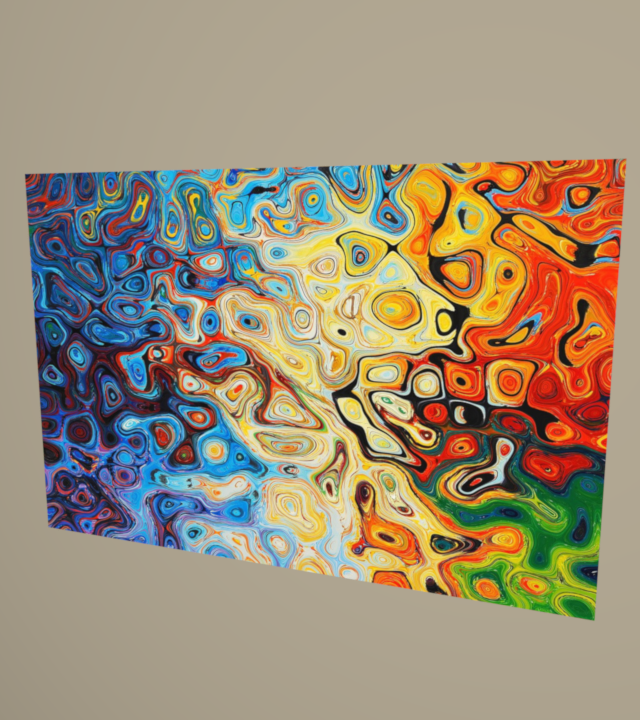}
        \includegraphics[width=0.161\linewidth]{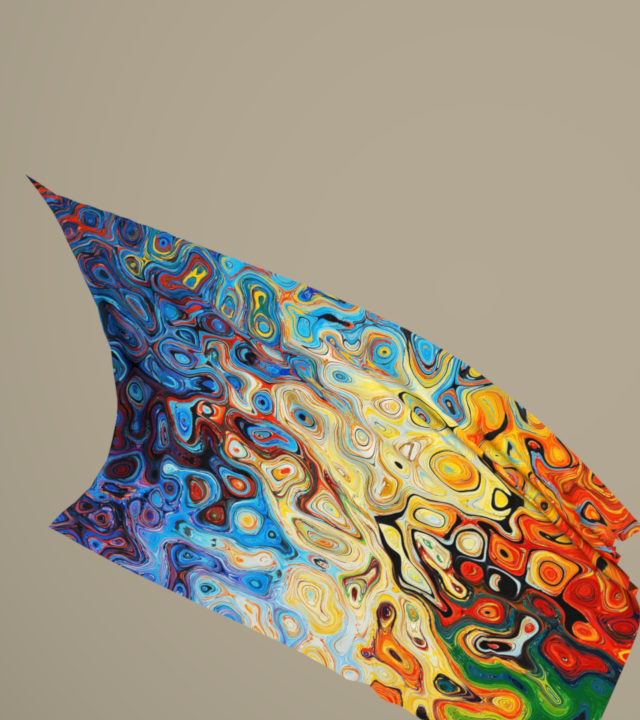}
        \includegraphics[width=0.161\linewidth]{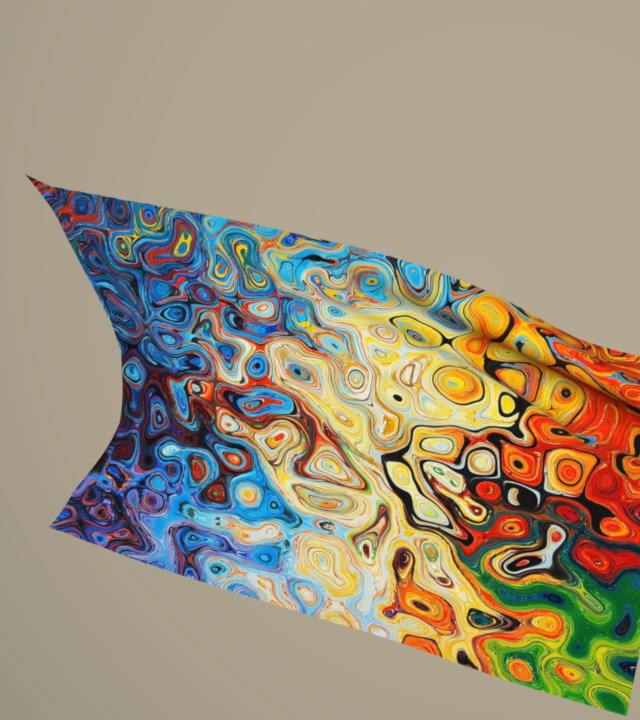}
        \centering
        \caption{Textile:  \textit{lucy} (top-left), \textit{basin} (top-right), \textit{flags} (mid-left), \textit{tube} (mid-right), and \textit{single flag} (bottom)}
        \end{subfigure}     
  \caption{
  Examples from the proposed physics-aware benchmark, {\dataset}. 
  They exhibit complex physical interactions between multiple bodies composed of diverse materials such as liquid, gas, rheological substance, and textile.
This benchmark will foster physics reasoning in dynamic scenes. }
  	\label{fig:data_viz_2}
    \end{center}%
    	      \vspace{-6mm}
\end{figure*}

Figure~\ref{fig:data_viz_1} and Figure~\ref{fig:data_viz_2} visualize more scenes in {\dataset} to supplement the Figure~\cvprb{3} in our main paper. 
As shown in Table~\cvprb{1} and Table~\cvprb{2} of the main paper, textile materials remain underexplored in physics-based datasets.
To facilitate progressive research development in this domain, we include few simpler scenes such as \textit{single flag} and \textit{tube} that employ with only basic wind interactions. 
We believe this approach effectively supports advancements by providing more accessible starting points.
Similarly, our benchmark also includes few simpler gas scenes, \textit{simple smoke} and \textit{falling}, for the same reason as with textile scenarios: to facilitate progressive development. 
For the \textit{cow} scene, internal forces intrinsic to the cow itself introduce unique and more complex dynamics, rather than using only gravity.

\clearpage
\section{Common Setup}
\paragraph{Simulator selection} 
We select SideFX Houdini 20.5 as the foundation of our physics-informed data-generation pipeline because it integrates multiple physics solvers within a unified procedural environment.
By sharing a common computational graph, it ensures consistent multi-body interactions under uniform boundary conditions.
We access simulation data, such as particle positions and flow fields, on a per-frame basis via its Python API.

\paragraph{Rendering}
All frames are rendered at a resolution of $640 \times 720$. 
We use path tracing with 256 samples per pixel and apply NVIDIA OptiX for denoising. 
Scenes are illuminated with 1–3 point lights (intensity: 600–4000); shadows are disabled in textile-focused scenes like \textit{tube-flag} to emphasize geometry. 
All cameras follow spiral trajectories to capture diverse views over time. 
All simulations were conducted on 1 NVIDIA Titan Xp GPUs and Intel Xeon E5-2630 v4 CPUs with 40 cores. 
While most scenes rendered in 2–3 hours, the \textit{lucy} and \textit{hanok} scenes required 12–24 hours due to their complex geometry and increased simulation costs.

\paragraph{Camera Trajectories}
\begin{wrapfigure}{r}{0.54\textwidth}
\vspace{-6mm}
    \begin{center}
        \centering
         \begin{subfigure}[b]{0.49\linewidth}
        \centering
         \includegraphics[width=\linewidth]{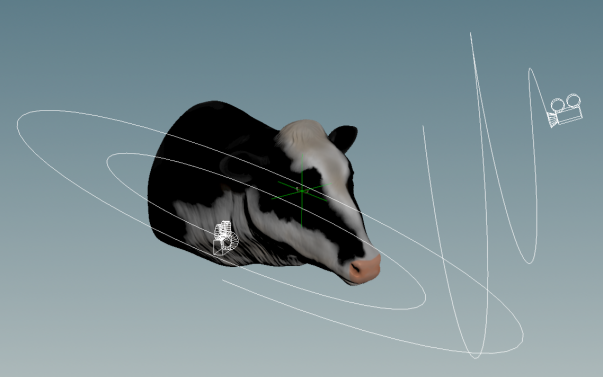}
         \caption{$180^\circ$ scene: 2 cams}
         \label{subfig:half_scene}
        \end{subfigure}
                 \begin{subfigure}[b]{0.49\linewidth}
        \centering
         \includegraphics[width=\linewidth]{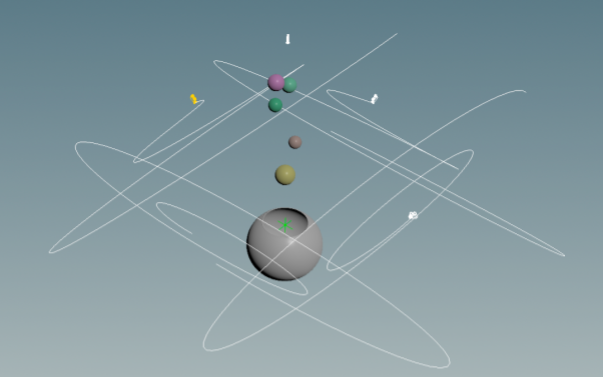}
         \caption{$360^\circ$ scene: 4 cams}
         \label{subfig:full_scene}
        \end{subfigure}
           \caption{Visualization of training camera's traj.
           All cameras follow spiral trajectories to capture diverse views over time.
For the monocular setup, we select one of the training cameras to generate the training data.}  
           \vspace{-3mm}
    \label{fig:cam_traj}
    		\end{center}%
\end{wrapfigure}

Figure~\ref{fig:cam_traj} visualizes our training camera trajectories, where we adopt either 2 or 4 cameras depending on the properties of each scene.
For the \textit{cow}, \textit{single flag}, and \textit{tube} scenes--which resemble $180^\circ$ settings with background walls placed behind the objects--we use only 2 cameras, as shown in Figure~\ref{subfig:half_scene}.
In contrast, for $360^\circ$  scenes, we place 4 cameras at evenly spaced viewpoints around the scene, enabling full coverage from four directions.
All cameras follow spiral trajectories to capture diverse views over time.
For the monocular setup, we select one of the training cameras to generate the training data.

\vspace{6mm}
\section{Material-specific Physics Solver}
\subsection{Liquid}
For liquid scenes, we adopt the Fluid-Implicit Particle (FLIP) solver~\cite{Brackbill1986}  a hybrid particle-grid method. 
FLIP maintains particle velocities throughout the simulation, using the grid solely to compute and apply forces such as pressure and viscosity. 
This approach preserves fine-scale, high-frequency particle velocities, which is crucial for modeling realistic and rapid liquid behavior while adhering to the Navier–Stokes equations. 
While the Material Point Method (MPM)~\citep{Sulsky1994} can also model fluids, its direct velocity aggregation onto the grid limits its ability to capture highly dynamic fluid phenomena like splashing. 
Particle-only solvers~\citep{gingold1977smoothed} are another option, but FLIP is generally better suited for incompressible fluids due thanks to its hybrid grid-based representation.

\vspace{-2mm}
\paragraph{Dynamic interactions}
For scenes where fluid spills onto fixed objects, such as \textit{ice} and \textit{hanok}, we employ a surface operator to simulate multi-body interaction. 
This approach is computationally efficient as interaction primarily occurs near the surface. 
In contrast, for scenes like \textit{ship} and \textit{cereal} where objects fall into liquid, causing both fluid and objects to move and influence each other through force exchange, we use a dynamic operator to accurately model these more complex interactions.

\paragraph{Physics law}
We assume incompressible fluids in our simulations. 
This assumption simplifies the governing equations while maintaining high fidelity for typical liquid behaviors.
Fluid motion is governed by the conservation of momentum and mass. 
We consider gravity as the sole external force.
The momentum equation for our fluid, describing the conservation of momentum, is given by
\begin{align}
\rho \left( \frac{\partial \mathbf{u}}{\partial t} + (\mathbf{u} \cdot \nabla)\mathbf{u} \right) = -\nabla p + \mu \nabla^2 \mathbf{u} + \rho \mathbf{g},
\end{align} 
where $\rho$ is density, $\mathbf{u}$ is velocity, $t$ is time, $p$ is pressure, $\mu$ is dynamic viscosity, and $\mathbf{g}$ is gravitational acceleration.
The incompressibility condition ensures the conservation of mass, stating that the fluid's volume remains constant as follows:
\begin{align} 
\nabla \cdot \mathbf{u}  = 0,
\end{align} 
where $\mathbf{u}$ is velocity.

\paragraph{Physics parameters}
Table~\ref{tab:configuration_liquid} summarizes the density ($\rho$) and viscosity ($\mu$) values, which are the most important factors governing fluid motion. 
For the \textit{ship} scene, we reduced the particle separation from 0.1 to 0.05 to decrease initial particle displacement and improve simulation accuracy. 
Also, the particle separation during dynamic simulation is set to 0.1 for the \textit{cereal} scene and 0.05 for the \textit{ship} scene.

\begin{table}[h]
\centering
\caption{Physics parameters used for liquid materials. }
\label{tab:configuration_liquid}
\scalebox{0.8}{
\setlength\tabcolsep{14pt} 
\begin{tabular}{c|l|cccc}
\toprule
\textbf{Scene name} & \textbf{Object} & \textbf{Density} ($kg/m^3$) & \textbf{Viscosity} \\
\midrule
\multirow{2}{*}{Ship}   & Water  & 800  & 0 \\ 
                        & Ship   & 300  & - \\
\hdashline
\multirow{2}{*}{Cereal} & Water  & 1000 & 0 \\
                        & Cereal & 1000 & - \\
\hdashline
Hanok                   & Snow   & 1000 & 10 \\
Ice                     & Water  & 1000 & 0 \\
\bottomrule
\end{tabular}
}
\end{table}

\subsection{Gas}
For gas (smoke) simulation, we utilize the Pyro solver~\citep{houdini_pyrosolver}. 
Pyro accurately models the temperature field, which is essential for capturing buoyancy effects in gaseous materials. 
It employs grid-based representations of density, velocity, and temperature, ensuring compliance with the Navier–Stokes equations governing fluid mechanics. 
Since ground-truth motion is represented as a velocity field and storing full velocity fields can require up to 2GB per frame, we provide a subsampled set of particle trajectories per scene to facilitate efficient data storage and processing.

\paragraph{Dynamic interactions}
In the \textit{pisa} scene, we reduced the voxel size from the default 0.1 to 0.05 to better capture the tower's intricate details and added a lateral wind of speed 2 to wrap the plume around it. 
Other scenes involving gaseous materials utilized the default simulation settings.

\vspace{-2mm}
\paragraph{Physics law}

Our gas simulations, similar to liquids, are governed by the Navier-Stokes equations. 
However, we specifically account for the effects of buoyancy as an additional external force due to temperature differences.
\begin{align}
\rho \left( \frac{\partial \mathbf{u}}{\partial t} + (\mathbf{u} \cdot \nabla)\mathbf{u} \right) = -\nabla p + \mu \nabla^2 \mathbf{u} + \rho (\mathbf{g} + \mathbf{f}_b),
\end{align} 
where $\mathbf{f}_b$ is the buoyancy force.
The buoyancy force is derived from temperature differences relative to the ambient environment as follows:
\begin{align}
\mathbf{f}_b = k_B \cdot (T - T_{\text{ambient}}) \cdot \mathbf{d}_B,
\end{align}
where $k_B$ is the buoyancy constant, $T$ is the local gas temperature, $T_{\text{ambient}}$ is the ambient temperature, and $\mathbf{d}_B$ is the buoyancy direction.

\vspace{-2mm}
\paragraph{Physics parameters}
In gas simulations, temperature is the most critical physical parameter, as it directly governs buoyancy, expansion, and overall flow dynamics. 
We initialized the temperature of gas materials to 3000K. 
As temperature evolves locally due to gas movement and interactions, its spatiotemporal variation is represented as a temperature field defined on a voxel grid. 
These temperature fields are accessible via the provided simulation source file with a Python API, similar to how flow fields are accessed for generating ground-truth trajectory data.

\subsection{Rheological Substances}
We use MPM~\citep{Sulsky1994} for rheological substances like snow and jelly.
As an extension of FLIP~\citep{Brackbill1986}, MPM is ideal for simulating chunk-based materials. 
It aggregates particle information onto a grid, performs computations, and then reprojects results back to particles, effectively capturing deformation and internal force propagation.

\vspace{-2mm}
\paragraph{Dynamic interactions}
For the \textit{pancake} scene, we reduced the grid size from the default 0.025 to 0.002 to more faithfully capture its thin-sheet dynamics; other scenes use the default grid size. 
To suppress spurious artifacts that can arise from aggregating particle properties onto the grid, we increased the number of samples participating in node calculations by oversampling. 
The oversampling scales are set to 6, 2, 4, and 2 for the \textit{bouncing balls}, \textit{cow}, \textit{jelly party}, and \textit{pancake} scenes, respectively. 
Additionally, for the \textit{bouncing balls} scene, we added a static bowl-shaped collider so that the falling balls rebound off both one another and the bowl’s surface.

\vspace{-2mm}
\paragraph{Physics law}

Simulating viscoelastic materials requires modeling both their elastic (solid-like) and viscous (fluid-like) responses. 
Our approach follows fundamental conservation laws while adopting simplified constitutive models commonly used in MPM-based visual simulations, namely linear elasticity and Kelvin--Voigt viscosity.
The conservation of mass is described by the continuity equation as follows:
\begin{align}
    \frac{D\rho}{Dt} + \rho(\nabla \cdot \mathbf{v})=0,
\end{align}
where $\rho$ is the material density and $\mathbf{v}$ is the velocity field.
Conservation of linear momentum is given by
\begin{align}
    \rho \frac{D \mathbf{v}}{Dt}  = \nabla \cdot  \boldsymbol{\sigma} + \rho  \mathbf{g},
\end{align}
where $\boldsymbol{\sigma}$ is the Cauchy stress tensor and $\mathbf{g}$ is gravitational acceleration.
We adopt a Kelvin--Voigt-type viscoelastic model, where the total stress is decomposed into elastic and viscous components as follows: 
\begin{align}
    \boldsymbol{\sigma} = \boldsymbol{\sigma}_{\text{elastic}} + \boldsymbol{\sigma}_{\text{viscous}}.
\end{align}
The elastic stress is modeled using linear isotropic elasticity as follows:
\begin{align}
    \boldsymbol{\sigma}_{\text{elastic}} =
    \frac{E \nu}{(1 + \nu)(1 - 2\nu)} (\text{tr}(\boldsymbol{\epsilon})) \mathbf{I}
    + \frac{E}{2(1 + \nu)} \boldsymbol{\epsilon},
\end{align}
where $E$ is Young's modulus, $\nu$ is Poisson's ratio, and $\boldsymbol{\epsilon}$ is the strain tensor defined as $\boldsymbol{\epsilon} = \frac{1}{2}(\nabla \mathbf{u} + (\nabla \mathbf{u})^T)$, where $\mathbf{u}$ is the displacement field.
The viscous stress accounts for rate-dependent deformation as follows:
\begin{align}
    \boldsymbol{\sigma}_{\text{viscous}} = 2\mu \left( \mathbf{E} - \frac{1}{3} (\text{tr}(\mathbf{E})) \mathbf{I} \right) + \zeta (\text{tr}(\mathbf{E})) \mathbf{I},
\end{align}
where $\mu$ is the dynamic viscosity and $\zeta$ is the bulk viscosity. The strain-rate tensor $\mathbf{E}$ is defined as $\mathbf{E} = \frac{1}{2}(\nabla \mathbf{v} + (\nabla \mathbf{v})^T)$.

 \vspace{-2mm}
\paragraph{Physics parameters}
Gravity is applied to all scenes; however, for the \textit{cow} scene, we introduce additional internal forces to induce cow motion and increase the complexity of the physical reasoning scenarios. 
Boundaries are set to be open in all directions, allowing objects to move freely without collisions against invisible walls. Table~\ref{tab:configuration_MPM} lists detailed physics parameters, including Young’s modulus ($E$) and Poisson’s ratio ($\nu$), used in the simulations.

\begin{table}[h]
\centering
\caption{Physics parameters used for rheological materials.
}
\label{tab:configuration_MPM}
\scalebox{0.8}{
\setlength\tabcolsep{6pt} 
\begin{tabular}{c|l|cccc}
\toprule
\textbf{Scene name} & \textbf{Object} & \textbf{Type} & \textbf{$E$} & \textbf{$\nu$} & \textbf{Viscosity} \\
\midrule
\multirow{2}{*}{Pancake}        & Honey     & Viscous  & \(2.5 \times 10^5 \) & 0.10 & 0.125 \\
                                & Pancakes  & Chunky   & \(8.0 \times 10^4 \) & 0.23 & -     \\
\hdashline
\multirow{2}{*}{Bouncing balls} & Balls     & Elastic  & \(1.0 \times 10^5\)  & 0.50 & -     \\
                                & Fishbowl  & Static   & -                    & -    & -     \\
\hdashline
Jelly party                     & Jelly     & Elastic  & \(8.0 \times 10^4 \) & 0.45 & -     \\
Cow                             & Cow       & Elastic  & \(5.0 \times 10^4 \) & 0.45 & -     \\
\bottomrule
\end{tabular}
}
\end{table}

\clearpage
\subsection{Textile}
For textile materials, we adopt the Vellum solver~\citep{houdini_vellumsolver}, which is based on the Extended Position Based Dynamics (XPBD) framework~\citep{macklin2016xpbd}. 
XPBD improves upon classical Position Based Dynamics (PBD)~\citep{muller2007position} by integrating a Lagrange multiplier and its update. 
This effectively decouples material stiffness from the solver’s time-step size and iteration count, making Vellum a widely used method for simulating deformable objects, especially cloth.

\vspace{-2mm}
\paragraph{Dynamic interactions}
To simulate interactions between objects and textiles--where both move and exchange forces, as seen particularly in the \textit{basin} scene--we employ the shape match constraint on objects.
This constraint helps maintain the overall shape of rigid objects by driving points toward their rest configuration, allowing the material to preserve its structural integrity while still interacting dynamically with textiles and other objects. 
For the \textit{lucy} scene, we increased the simulation sub-step count fivefold over the default to robustly handle collisions with the statue’s complex geometry. 
To simulate wind-induced fluttering in the \textit{flags}, \textit{single-flag}, and \textit{tube} scenes, we applied external forces using a POP Wind node that blows parallel to the ground plane.

\paragraph{Physics law}

We simulate textile materials using the Vellum solver~\citep{houdini_vellumsolver}, 
which is based on Extended Position-Based Dynamics (XPBD)~\citep{macklin2016xpbd}. 
XPBD extends classical Position-Based Dynamics (PBD) by introducing a compliance parameter, 
allowing constraint stiffness to be decoupled from the time step and iteration count.
Unlike force-based methods, XPBD does not explicitly solve equations of motion. 
Instead, particle positions are updated by iteratively projecting them to satisfy geometric constraints.
The core algorithm for each simulation substep is shown in Algorithm~\ref{algo:xpbd}.

\begin{algorithm}[t]
\caption{Extended Position-Based Dynamics (XPBD) Algorithm (per substep)}
\label{algo:xpbd}
\begin{algorithmic}[1]
\State \textbf{Input:}
\State \quad $\mathbf{p}_i$, $\mathbf{v}_i$: Current position and velocity of particle $i$
\State \quad $\mathbf{w}_i = 1/m_i$: Inverse mass of particle $i$
\State \quad $\mathbf{a}_{\text{ext},i}$: External acceleration (e.g., gravity)
\State \quad $\Delta t$: Simulation time step
\State \quad Iterations: Number of constraint projection iterations

\State \textit{\textcolor{RoyalBlue}{// 1. Predict positions}}
\For{each particle $i$}
\State $\mathbf{p}^*_i \leftarrow \mathbf{p}_i + \Delta t \mathbf{v}_i + \Delta t^2 \mathbf{a}_{\text{ext},i}$
\EndFor

\State \textit{\textcolor{RoyalBlue}{// 2. Constraint projection}}
\For{$k=1$ to Iterations}
    \For{each constraint $C_j$}
        \State compute $\Delta \mathbf{p}_i$ using Eq.~(\ref{eq:xpbd_correction})
        \State $\mathbf{p}^*_i \leftarrow \mathbf{p}^*_i + \Delta \mathbf{p}_i$
    \EndFor
\EndFor

\State \textit{\textcolor{RoyalBlue}{// 3. Velocity update}}
\For{each particle $i$}
\State $\mathbf{v}_i \leftarrow (\mathbf{p}^*_i - \mathbf{p}_i) / \Delta t$
\State $\mathbf{p}_i \leftarrow \mathbf{p}^*_i$
\EndFor
\end{algorithmic}
\end{algorithm}
We primarily employ stretch and bend constraints to model textile behavior. 
For a constraint $C(\mathbf{p})$, where $C(\mathbf{p})$ is a scalar constraint function that measures the violation of a geometric constraint, such as maintaining distances (stretch) or bending angles between particles, with $C(\mathbf{p}) = 0$ indicating that the constraint is satisfied, the position correction is as follows:
\begin{align}
\Delta \mathbf{p}_i = - \Delta \lambda \, \mathbf{w}_i \nabla_{\mathbf{p}_i} C,
\label{eq:xpbd_correction}
\end{align}
where
\begin{align}
\Delta \lambda = \frac{C(\mathbf{p})}{\sum_j \mathbf{w}_j |\nabla_{\mathbf{p}_j} C|^2 + \tilde{\alpha}}, 
\quad \tilde{\alpha} = \frac{\alpha}{\Delta t^2}.
\end{align}
Here, $\alpha$ denotes the compliance parameter (inverse stiffness).

\paragraph{Physics parameters}
For textiles, we primarily adopt the default configuration from the Vellum solver, with detailed modifications summarized in Table~\ref{tab:configuration_textile}. 
In addition to stiffness, we specify a damping ratio for each constraint to reduce oscillations and stabilize the simulated cloth.

\begin{table*}[h]
\centering
\caption{Physics parameters used for textile materials.}
\label{tab:configuration_textile}
\scalebox{0.8}{
\begin{tabular}{c|l|cc|cc}
\toprule
\multirow{2}{*}{\textbf{Scene name}} & \multirow{2}{*}{\textbf{Object}} 
& \multicolumn{2}{c|}{\textbf{Stretch}} & \multicolumn{2}{c}{\textbf{Bend}} \\
& & Stiffness & Damping Ratio & Stiffness & Damping Ratio \\
\midrule
\multirow{2}{*}{Flags} 
& Flag1 (small) & \(1.0 \times 10^{10}\) & \(1.0 \times 10^{-3}\) & \(1.0 \times 10^{-4}\) & \(1.0 \times 10^{-2}\) \\
& Flag2 (big)   & \(1.0 \times 10^{11}\) & \(1.0 \times 10^{-3}\) & \(1.0 \times 10^{-4}\) & \(1.0 \times 10^{-2}\) \\
\hdashline
\multirow{2}{*}{Basin} 
& Boxes         & \(1.0 \times 10^{10}\) & \(1.0 \times 10^{-3}\) & N/A & N/A \\
& Cloth         & \(1.0 \times 10^{13}\) & \(1.0 \times 10^{-3}\) & \(1.0 \times 10^{-4}\) & \(1.0 \times 10^{-2}\) \\
\hdashline
Single flag & Flag  & \(1.0 \times 10^{10}\) & \(1.0 \times 10^{-3}\) & \(1.0 \times 10^{-4}\) & \(1.0 \times 10^{-2}\) \\
Lucy        & Cloth & \(1.0 \times 10^{10}\) & \(1.0 \times 10^{-3}\) & \(1.0 \times 10^{-4}\) & \(1.0 \times 10^{-2}\) \\
Tube        & Tube flag & \(1.0 \times 10^{10}\) & \(1.0 \times 10^{-3}\) & \(1.0 \times 10^{-4}\) & \(1.0 \times 10^{-2}\) \\
\bottomrule
\end{tabular}

}
\end{table*}

\section{Details of Experimental Results}
\label{sec:supple_exp}
\subsection{Implementation details}

We implement recent DyNVS methods, including D-3DGS~\citep{yang2024deformable}, 4DGS~\citep{wu20234d}, STG~\citep{li2024spacetime}, MoSca~\citep{lei2024mosca}, and SOM~\citep{wang2024shape}.
We follow the default training settings provided for each method.
For 4DGS, we adopt the training configuration used for the HyperNeRF~\cite{park2021hypernerf} dataset and reduce the grid learning rate to improve training stability.
For SOM, we apply the recommended hyperparameters for the DyCheck~\cite{gao2022monocular}dataset.
For point cloud initialization, we run COLMAP with dense matching and fusion, followed by uniform downsampling to approximately 40,000 points.
All experiments are conducted on NVIDIA RTX A5000 and A6000 GPUs, with training times ranging from 30 minutes to 2 hours depending on the method.
Note that all experiments are implemented based on their public codes~\footnote{D-3DGS: \url{https://github.com/ingra14m/Deformable-3D-Gaussians}}~\footnote{STG: \url{https://github.com/oppo-us-research/SpacetimeGaussians}}~\footnote{4DGS: \url{https://github.com/hustvl/4DGaussians}}~\footnote{SOM: \url{https://github.com/vye16/shape-of-motion}}~\footnote{MoSca: \url{https://github.com/JiahuiLei/MoSca}}.

\subsection{Average performance: all scenes}
Table~\ref{tab:quantitative_average} presents the average performance of all methods under both monocular and multiview settings, aggregated across all 17 scenes.
For SOM~\citep{wang2024shape}, we observed better performance when scaling the estimated depth map using the COLMAP point cloud obtained from dense matching, compared to sparse matching.

\begin{table*}[h]
\centering
\caption{Average quantitative results for both monocular and multiview settings, averaged across all 17 scenes. 
While multiview setups generally offer better reconstruction performance than monocular ones, even multiview results achieve PSNR scores below 30.
This highlights the substantial difficulty in reconstructing the complex multi-body interactions in our benchmark.
}
\scalebox{0.8}{
\setlength\tabcolsep{13pt} 
\begin{tabular}{l|ccc|ccc}
\toprule
\multirow{2}{*}{\textbf{Method}}			&  \multicolumn{3}{c}{\textbf{Monocular}} &  \multicolumn{3}{c}{\textbf{Multiview}}\\
&				PSNR $\uparrow$  & SSIM $\uparrow$  & LPIPS $\downarrow$ &PSNR $\uparrow$  & SSIM $\uparrow$  & LPIPS $\downarrow$\\
\midrule
D-3DGS~\citep{yang2024deformable} & \cellsecond21.7& \cellbest0.86 & \cellbest0.18 & \cellsecond24.2 & \cellbest0.89& \cellbest0.14 \\
4DGS~\citep{wu20234d} 	&\cellbest22.7& \cellsecond0.85 & \cellsecond0.19 & \cellbest24.4 &  \cellsecond0.87 &  \cellsecond0.17 \\
STG~\citep{li2024spacetime} 	&19.3&0.76&0.30 &21.0&0.79&0.30\\
MoSca~\citep{lei2024mosca}  &  \cellthird19.5 &0.76 &0.34&N/A&N/A&N/A\\
SoM~\citep{wang2024shape} 	&19.3& \cellthird0.80 & \cellthird0.26 &N/A&N/A&N/A\\
\bottomrule
\end{tabular}
}
\label{tab:quantitative_average}
\vspace{-2mm}
\end{table*}

\clearpage
\subsection{Performance breakdown: monocular setting}
We provide per-scene breakdown performance of monocular video reconstruction setting in Table~\ref{tab:breakdown_mono}.
We report the performance of D-3DGS~\citep{yang2024deformable}, 4DGS~\citep{wu20234d}, STG~\citep{li2024spacetime}, MoSca~\citep{lei2024mosca} and SOM~\citep{wang2024shape}, which serve as the most common and recent baselines for the DyNVS task.

\begin{table*}[h]
\centering
\caption{
Per-scene breakdown results for all 17 scenes under the monocular setting.
}
\scalebox{0.8}{
\setlength\tabcolsep{10pt}
\begin{tabular}{l|ccc|ccc|ccc}
\toprule
\multirow{2}{*}{\textbf{Method}} & \multicolumn{3}{c|}{\textbf{Cereal}} & \multicolumn{3}{c|}{\textbf{Ship}} & \multicolumn{3}{c}{\textbf{Hanok}} \\

& PSNR $\uparrow$ & SSIM $\uparrow$ & LPIPS $\downarrow$ & PSNR $\uparrow$ & SSIM $\uparrow$ & LPIPS $\downarrow$ & PSNR $\uparrow$ & SSIM $\uparrow$ & LPIPS $\downarrow$ \\
\midrule
D-3DGS~\citep{yang2024deformable}&\cellthird23.3&\cellsecond0.88&\cellthird0.15&\cellbest25.9&\cellsecond0.90&\cellsecond0.16&\cellbest16.3&\cellbest0.78&\cellbest0.28\\
4DGS~\citep{wu20234d}&\cellbest26.3&\cellbest0.90&\cellbest0.14&\cellbest25.9&\cellbest0.91&\cellbest0.15&\cellsecond15.5&\cellsecond0.74&\cellsecond0.32\\
STG~\citep{li2024spacetime}&15.8&0.54&0.54&22.8&\cellthird0.87&\cellthird0.23&\cellthird14.8&0.60&0.41\\
MoSca~\citep{lei2024mosca}&\cellsecond23.7&\cellthird0.83&\cellbest0.14&22.5&0.84&0.39&\cellthird14.8&0.59&0.42\\
SOM~\citep{wang2024shape}&21.2&0.81&0.22&\cellthird23.8&\cellthird0.87&0.23&14.7&\cellthird0.67&\cellthird0.40\\
\hline
\hline
\multirow{2}{*}{\textbf{Method}} & \multicolumn{3}{c|}{\textbf{Ice}} & \multicolumn{3}{c|}{\textbf{Pisa}} & \multicolumn{3}{c}{\textbf{Box-smoke}} \\

& PSNR $\uparrow$ & SSIM $\uparrow$ & LPIPS $\downarrow$ & PSNR $\uparrow$ & SSIM $\uparrow$ & LPIPS $\downarrow$ & PSNR $\uparrow$ & SSIM $\uparrow$ & LPIPS $\downarrow$ \\
\hline
D-3DGS~\citep{yang2024deformable}&\cellsecond25.3&\cellbest0.92&\cellsecond0.30&\cellbest20.7&\cellbest0.72&\cellbest0.26&20.7&\cellbest0.96&\cellsecond0.13\\
4DGS~\citep{wu20234d}&\cellbest29.2&\cellbest0.92&\cellbest0.29&\cellthird19.3&\cellthird0.67&\cellbest0.26&21.3&\cellbest0.96&\cellbest0.12\\
STG~\citep{li2024spacetime}&\cellthird23.1&\cellthird0.87&\cellthird0.39&\cellsecond20.1&\cellsecond0.68&\cellthird0.33&\cellsecond22.7&\cellthird0.95&\cellthird0.16\\
MoSca~\citep{lei2024mosca}&20.9&0.86&0.61&16.7&0.54&0.57&\cellthird22.1&0.94&0.21\\
SOM~\citep{wang2024shape}&18.7&0.84&0.41&17.1&0.65&0.41&\cellbest23.8&\cellthird0.95&0.18\\
\hline
\hline
\multirow{2}{*}{\textbf{Method}} & \multicolumn{3}{c|}{\textbf{Single smoke}} & \multicolumn{3}{c|}{\textbf{Falling}} & \multicolumn{3}{c}{\textbf{Jelly party}} \\

& PSNR $\uparrow$ & SSIM $\uparrow$ & LPIPS $\downarrow$ & PSNR $\uparrow$ & SSIM $\uparrow$ & LPIPS $\downarrow$ & PSNR $\uparrow$ & SSIM $\uparrow$ & LPIPS $\downarrow$ \\
\hline
D-3DGS~\citep{yang2024deformable}&\cellbest26.5&\cellbest0.97&\cellbest0.08&19.6&\cellbest0.90&\cellbest0.18&\cellbest16.3&\cellbest0.81&\cellbest0.21\\
4DGS~\citep{wu20234d}&\cellsecond26.1&\cellbest0.97&\cellbest0.08&\cellthird20.1&\cellbest0.90&\cellsecond0.19&\cellsecond14.9&\cellsecond0.70&\cellsecond0.28\\
STG~\citep{li2024spacetime}&24.5&\cellbest0.97&\cellthird0.10&\cellsecond20.2&0.77&0.37&11.1&0.51&0.44\\
MoSca~\citep{lei2024mosca}&\cellthird24.8&0.95&0.14&\cellbest21.2&0.72&0.49&12.0&0.51&0.38\\
SOM~\citep{wang2024shape}&23.9&0.95&0.18&15.3&\cellthird0.81&\cellthird0.33&\cellthird14.9&\cellthird0.69&\cellsecond0.28\\
\hline
\hline
\multirow{2}{*}{\textbf{Method}} & \multicolumn{3}{c|}{\textbf{Pancake}} & \multicolumn{3}{c|}{\textbf{Bouncing balls}} & \multicolumn{3}{c}{\textbf{Cow}} \\

& PSNR $\uparrow$ & SSIM $\uparrow$ & LPIPS $\downarrow$ & PSNR $\uparrow$ & SSIM $\uparrow$ & LPIPS $\downarrow$ & PSNR $\uparrow$ & SSIM $\uparrow$ & LPIPS $\downarrow$ \\
\hline
D-3DGS~\citep{yang2024deformable}&\cellbest22.8&\cellbest0.88&\cellsecond0.13&\cellthird19.8&0.77&0.15&\cellsecond21.6&\cellbest0.92&\cellbest0.10\\
4DGS~\citep{wu20234d}&\cellthird17.0&\cellthird0.78&\cellthird0.22&\cellbest22.7&\cellbest0.86&\cellbest0.10&\cellbest23.6&\cellbest0.92&\cellsecond0.11\\
STG~\citep{li2024spacetime}&10.6&0.60&0.39&11.9&0.51&0.60&20.6&\cellthird0.87&\cellthird0.16\\
MoSca~\citep{lei2024mosca}&\cellsecond22.2&\cellsecond0.87&\cellbest0.12&\cellsecond19.9&\cellsecond0.81&\cellsecond0.12&17.1&0.78&0.32\\
SOM~\citep{wang2024shape}&13.9&0.68&0.34&17.4&\cellthird0.79&\cellthird0.14&\cellthird20.7&0.85&0.17\\
\hline
\hline
\multirow{2}{*}{\textbf{Method}} & \multicolumn{3}{c|}{\textbf{Lucy}} & \multicolumn{3}{c|}{\textbf{Basin}} & \multicolumn{3}{c}{\textbf{Flags}} \\

& PSNR $\uparrow$ & SSIM $\uparrow$ & LPIPS $\downarrow$ & PSNR $\uparrow$ & SSIM $\uparrow$ & LPIPS $\downarrow$ & PSNR $\uparrow$ & SSIM $\uparrow$ & LPIPS $\downarrow$ \\
\hline
D-3DGS~\citep{yang2024deformable}&\cellsecond22.8&\cellsecond0.92&\cellsecond0.10&\cellsecond18.2&\cellsecond0.67&\cellbest0.36&23.6&\cellsecond0.94&\cellsecond0.12\\
4DGS~\citep{wu20234d}&\cellbest27.5&\cellbest0.94&\cellbest0.08&\cellthird18.0&\cellbest0.68&\cellsecond0.38&\cellbest31.9&\cellbest0.96&\cellbest0.08\\
STG~\citep{li2024spacetime}&19.6&0.86&0.19&16.8&\cellthird0.66&0.43&24.7&0.90&0.17\\
MoSca~\citep{lei2024mosca}&14.8&0.80&0.49&\cellbest18.4&0.62&\cellthird0.41&\cellthird26.0&0.88&0.17\\
SOM~\citep{wang2024shape}&\cellthird21.0&\cellthird0.90&\cellthird0.17&16.3&0.61&0.43&\cellsecond27.4&\cellthird0.93&\cellsecond0.12\\
\hline
\hline
\multirow{2}{*}{\textbf{Method}} & \multicolumn{3}{c|}{\textbf{Single flag}} & \multicolumn{3}{c|}{\textbf{Tube}} & \multicolumn{3}{c}{\cellcolor{row1}\textbf{Average}} \\

& PSNR $\uparrow$ & SSIM $\uparrow$ & LPIPS $\downarrow$ & PSNR $\uparrow$ & SSIM $\uparrow$ & LPIPS $\downarrow$ & \cellcolor{row1} PSNR $\uparrow$ & \cellcolor{row1} SSIM $\uparrow$ & \cellcolor{row1} LPIPS $\downarrow$ \\
\hline
D-3DGS~\citep{yang2024deformable}&\cellsecond18.3&\cellsecond0.68&\cellsecond0.24&\cellsecond27.7&\cellbest0.96&\cellbest0.05&\cellsecond21.7&\cellbest0.86&\cellbest0.18\\
4DGS~\citep{wu20234d}&\cellthird18.0&\cellthird0.65&0.29&\cellbest28.8&\cellbest0.96&\cellsecond0.08&\cellbest22.7&\cellsecond0.85&\cellsecond0.19\\
STG~\citep{li2024spacetime}&\cellbest26.0&\cellbest0.87&\cellbest0.14&22.1&0.92&\cellthird0.12&19.3&0.76&0.30\\
MoSca~\citep{lei2024mosca}&15.3&0.56&0.50&18.3&0.86&0.25&\cellthird19.5&0.76&0.34\\
SOM~\citep{wang2024shape}&16.3&0.60&\cellsecond0.24&\cellthird22.5&\cellthird0.93&0.15&19.3&\cellthird0.80&\cellthird0.26\\
\bottomrule
\end{tabular}
}
\label{tab:breakdown_mono}
\end{table*}

\clearpage

\subsection{Performance breakdown: multiview setting}
We provide per-scene breakdown performance for the multiview video reconstruction setting in Table~\ref{tab:breakdown_multi}. 
We report results for D-3DGS~\citep{yang2024deformable}, 4DGS~\citep{wu20234d}, and STG~\citep{li2024spacetime}, which serve as the most common and recent baselines for the DyNVS task. 
Note that since MoSca~\citep{lei2024mosca} and SOM~\citep{wang2024shape} are specialized for monocular video setups, we omit its performance in the multiview evaluation.

\begin{table*}[h]
\centering
\caption{
Per-scene breakdown results for all 17 scenes under the multiview setting.
}
\scalebox{0.8}{
\setlength\tabcolsep{10pt} 
\begin{tabular}{l|ccc|ccc|ccc}
\toprule
\multirow{2}{*}{\textbf{Method}} & \multicolumn{3}{c|}{\textbf{Cereal}} & \multicolumn{3}{c|}{\textbf{Ship}} & \multicolumn{3}{c}{\textbf{Hanok}} \\
& PSNR $\uparrow$ & SSIM $\uparrow$ & LPIPS $\downarrow$ & PSNR $\uparrow$ & SSIM $\uparrow$ & LPIPS $\downarrow$ & PSNR $\uparrow$ & SSIM $\uparrow$ & LPIPS $\downarrow$\\
\hline
D-3DGS~\citep{yang2024deformable}&\cellbest28.1&\cellbest0.93&\cellbest0.11&\cellbest28.8&\cellbest0.93&\cellbest0.11&15.6&\cellbest0.76&\cellbest0.31\\
4DGS~\citep{wu20234d}&\cellsecond27.6&\cellsecond0.92&\cellsecond0.13&25.6&\cellsecond0.91&\cellsecond0.15&\cellbest15.8&\cellsecond0.75&\cellsecond0.33\\
STG~\citep{li2024spacetime}&15.8&0.58&0.59&\cellsecond25.9&0.90&0.20&\cellsecond15.7&0.65&0.43\\
\hline
\hline
\multirow{2}{*}{\textbf{Method}} & \multicolumn{3}{c|}{\textbf{Ice}} & \multicolumn{3}{c|}{\textbf{Pisa}} & \multicolumn{3}{c}{\textbf{Box-smoke}} \\
& PSNR $\uparrow$ & SSIM $\uparrow$ & LPIPS $\downarrow$ & PSNR $\uparrow$ & SSIM $\uparrow$ & LPIPS $\downarrow$ & PSNR $\uparrow$ & SSIM $\uparrow$ & LPIPS $\downarrow$\\
\hline
D-3DGS~\citep{yang2024deformable}&16.4&0.86&0.43&22.3&\cellsecond0.75&\cellsecond0.23&\cellsecond20.5&\cellsecond0.96&\cellsecond0.11\\
4DGS~\citep{wu20234d}&\cellbest31.4&\cellbest0.93&\cellbest0.28&\cellsecond22.8&0.74&0.24&\cellbest22.6&\cellbest0.97&\cellbest0.07\\
STG~\citep{li2024spacetime}&\cellsecond25.9&\cellsecond0.89&\cellsecond0.38&\cellbest28.7&\cellbest0.90&\cellbest0.18&19.0&0.95&0.17\\
\hline
\hline
\multirow{2}{*}{\textbf{Method}} & \multicolumn{3}{c|}{\textbf{Single-smoke}} & \multicolumn{3}{c|}{\textbf{Falling}} & \multicolumn{3}{c}{\textbf{Jelly party}} \\
& PSNR $\uparrow$ & SSIM $\uparrow$ & LPIPS $\downarrow$ & PSNR $\uparrow$ & SSIM $\uparrow$ & LPIPS $\downarrow$ & PSNR $\uparrow$ & SSIM $\uparrow$ & LPIPS $\downarrow$\\
\hline
D-3DGS~\citep{yang2024deformable}&26.4&\cellbest0.98&\cellbest0.07&\cellbest25.7&\cellbest0.95&\cellbest0.11&\cellbest18.2&\cellbest0.87&\cellbest0.16\\
4DGS~\citep{wu20234d}&\cellbest28.2&\cellbest0.98&\cellbest0.07&23.4&\cellsecond0.87&\cellsecond0.30&\cellsecond16.1&\cellsecond0.80&\cellsecond0.24\\
STG~\citep{li2024spacetime}&\cellsecond27.0&\cellbest0.98&\cellbest0.07&\cellsecond25.1&0.83&0.34&12.5&0.60&0.39\\
\hline
\hline
\multirow{2}{*}{\textbf{Method}} & \multicolumn{3}{c|}{\textbf{Pancake}} & \multicolumn{3}{c|}{\textbf{Bouncing balls}} & \multicolumn{3}{c}{\textbf{Cow}} \\
& PSNR $\uparrow$ & SSIM $\uparrow$ & LPIPS $\downarrow$ & PSNR $\uparrow$ & SSIM $\uparrow$ & LPIPS $\downarrow$ & PSNR $\uparrow$ & SSIM $\uparrow$ & LPIPS $\downarrow$\\
\hline
D-3DGS~\citep{yang2024deformable}&\cellbest25.3&\cellbest0.90&\cellbest0.08&\cellbest25.7&\cellbest0.91&\cellbest0.07&19.8&0.87&\cellbest0.10\\
4DGS~\citep{wu20234d}&\cellsecond18.7&\cellsecond0.84&\cellsecond0.18&\cellsecond25.0&\cellsecond0.84&\cellsecond0.09&\cellsecond24.1&\cellsecond0.92&\cellbest0.10\\
STG~\citep{li2024spacetime}&11.7&0.65&0.36&12.9&0.59&0.54&\cellbest31.8&\cellbest0.95&0.15\\
\hline
\hline
\multirow{2}{*}{\textbf{Method}} & \multicolumn{3}{c|}{\textbf{Lucy}} & \multicolumn{3}{c|}{\textbf{Basin}} & \multicolumn{3}{c}{\textbf{Flags}} \\
& PSNR $\uparrow$ & SSIM $\uparrow$ & LPIPS $\downarrow$ & PSNR $\uparrow$ & SSIM $\uparrow$ & LPIPS $\downarrow$ & PSNR $\uparrow$ & SSIM $\uparrow$ & LPIPS $\downarrow$\\
\hline
D-3DGS~\citep{yang2024deformable}&\cellbest30.2&\cellbest0.96&\cellbest0.06&\cellbest23.6&\cellbest0.82&\cellbest0.24&\cellsecond30.5&\cellbest0.96&\cellsecond0.09\\
4DGS~\citep{wu20234d}&\cellsecond28.6&\cellsecond0.95&\cellsecond0.07&\cellsecond20.5&\cellsecond0.75&\cellsecond0.33&\cellbest32.8&\cellbest0.96&\cellbest0.08\\
STG~\citep{li2024spacetime}&21.2&0.88&0.19&17.5&0.70&0.45&26.0&0.91&0.17\\
\hline
\hline
\multirow{2}{*}{\textbf{Method}} & \multicolumn{3}{c|}{\textbf{Single flag}} & \multicolumn{3}{c|}{\textbf{Tube}} & \multicolumn{3}{c}{\cellcolor{row1}\textbf{Average}} \\
& PSNR $\uparrow$ & SSIM $\uparrow$ & LPIPS $\downarrow$ & PSNR $\uparrow$ & SSIM $\uparrow$ & LPIPS $\downarrow$ & \cellcolor{row1} PSNR $\uparrow$ & \cellcolor{row1} SSIM $\uparrow$ & \cellcolor{row1} LPIPS $\downarrow$\\
\hline
D-3DGS~\citep{yang2024deformable}&\cellbest21.6&\cellbest0.81&\cellbest0.14&\cellbest32.6&\cellbest0.98&\cellbest0.05&\cellsecond24.2& \cellbest0.89& \cellbest0.14\\
4DGS~\citep{wu20234d}&\cellsecond19.7&\cellsecond0.72&\cellsecond0.21&\cellsecond31.3&\cellsecond0.97&\cellsecond0.07& \cellbest24.4& \cellsecond0.87&\cellsecond0.17\\
STG~\citep{li2024spacetime}&16.9&0.62&0.36&23.8&0.93&0.11& 21.0&0.79& 0.30\\
\bottomrule
\end{tabular}
}
\label{tab:breakdown_multi}
\end{table*}

\clearpage

\subsection{Additional Qualitative Results}

\begin{figure*}[h]
    \begin{center}
        \centering
                        \begin{subfigure}[b]{0.161\linewidth}
        \centering
          \includegraphics[width=\linewidth]{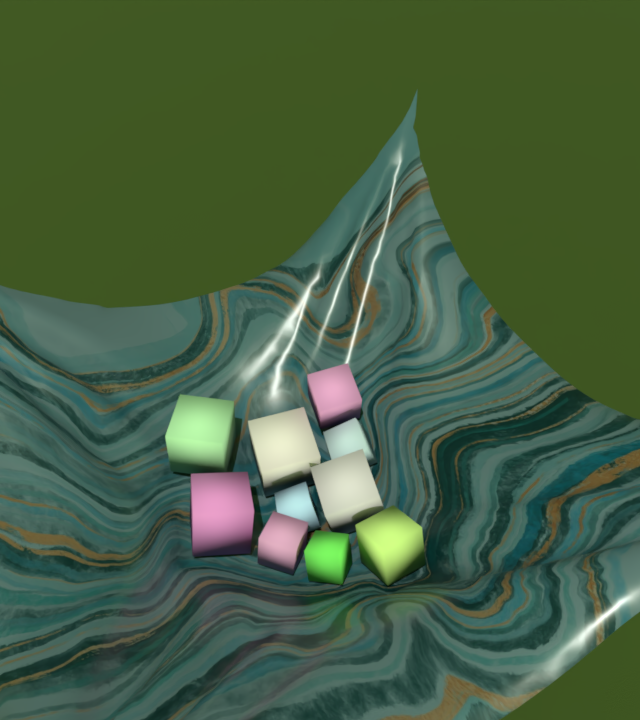}
         \includegraphics[width=\linewidth]{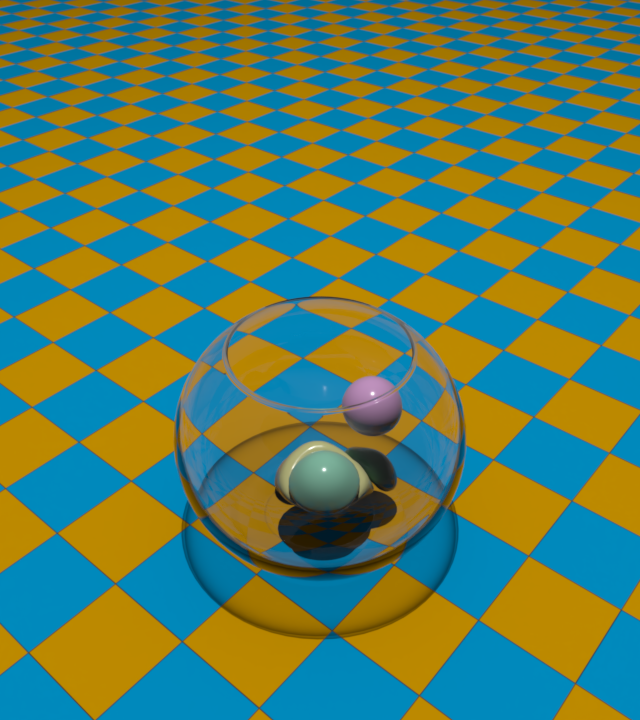}
            \includegraphics[width=\linewidth]{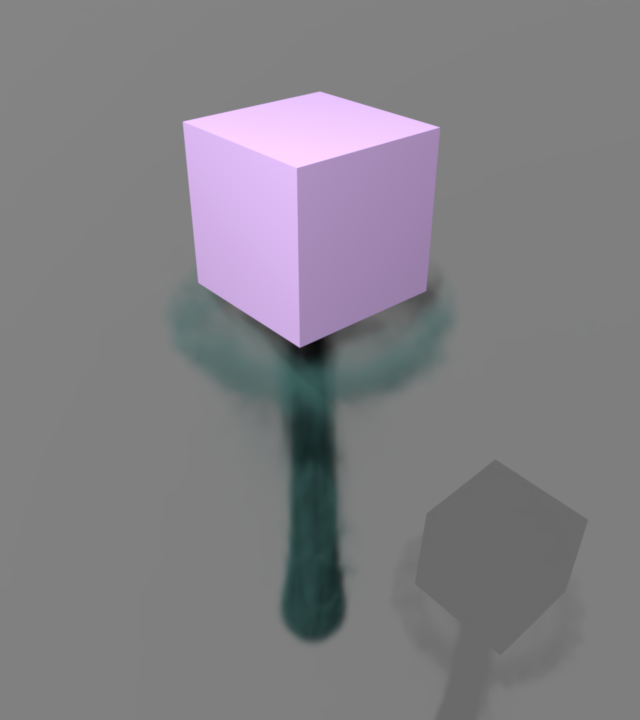}
            \includegraphics[width=\linewidth]{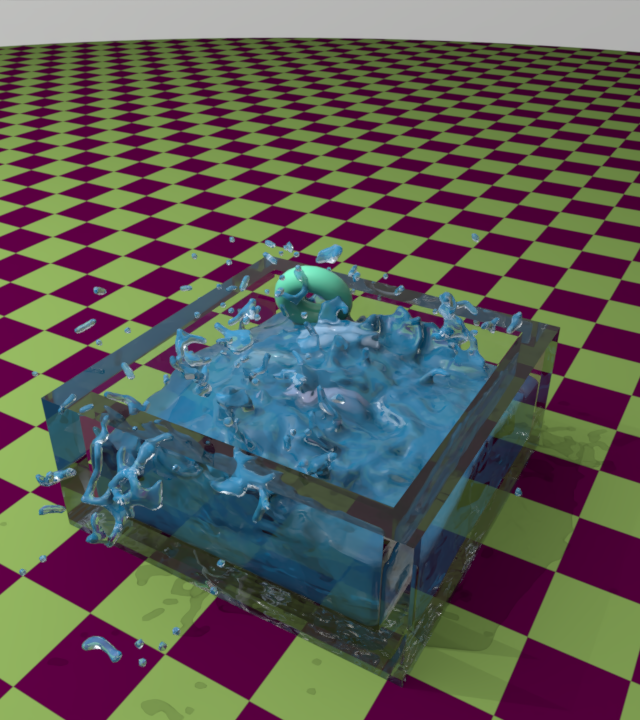}
         \includegraphics[width=\linewidth]{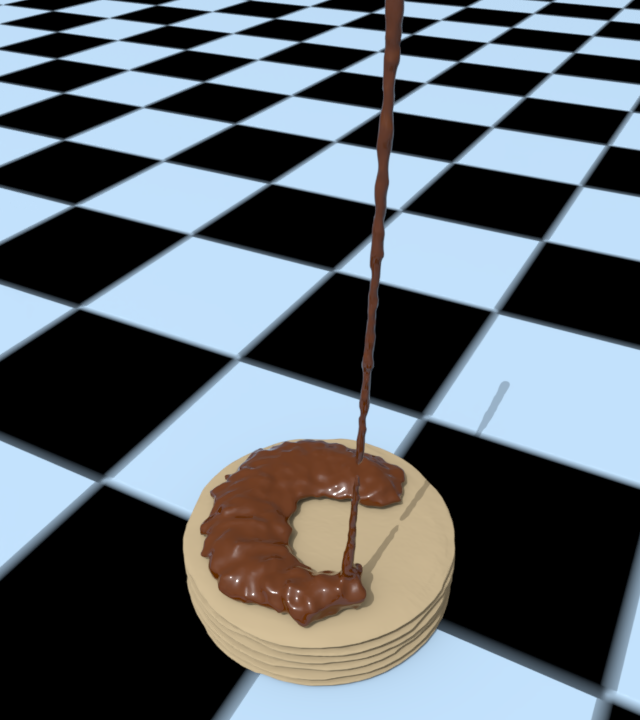}         
        \caption{GT image}
        \end{subfigure}
                        \begin{subfigure}[b]{0.161\linewidth}
        \centering
          \includegraphics[width=\linewidth]{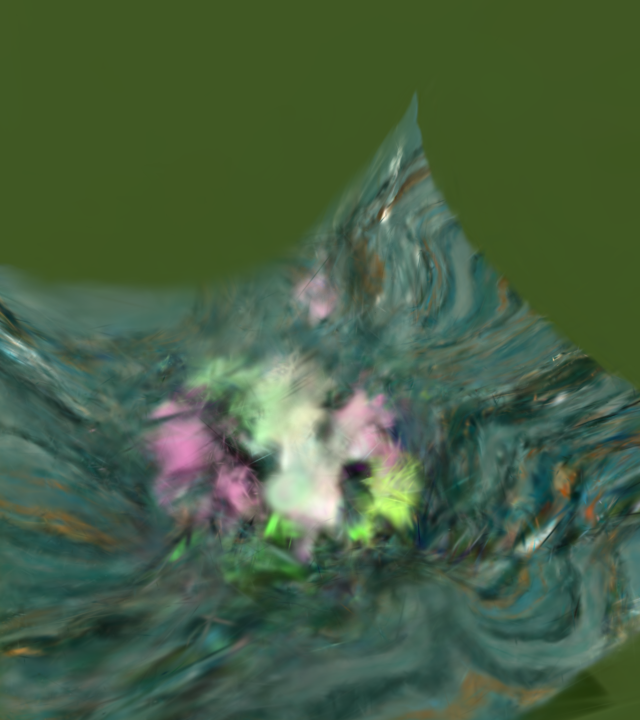}
          \includegraphics[width=\linewidth]{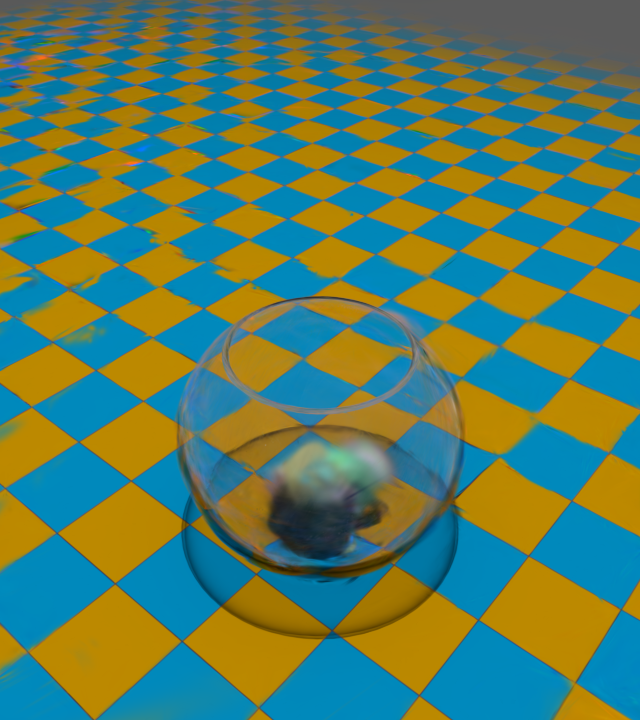}
            \includegraphics[width=\linewidth]{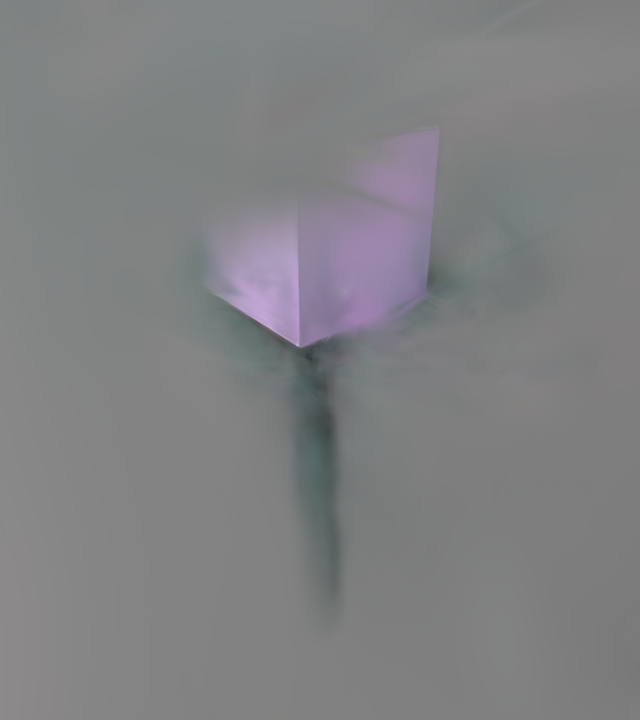}
            \includegraphics[width=\linewidth]{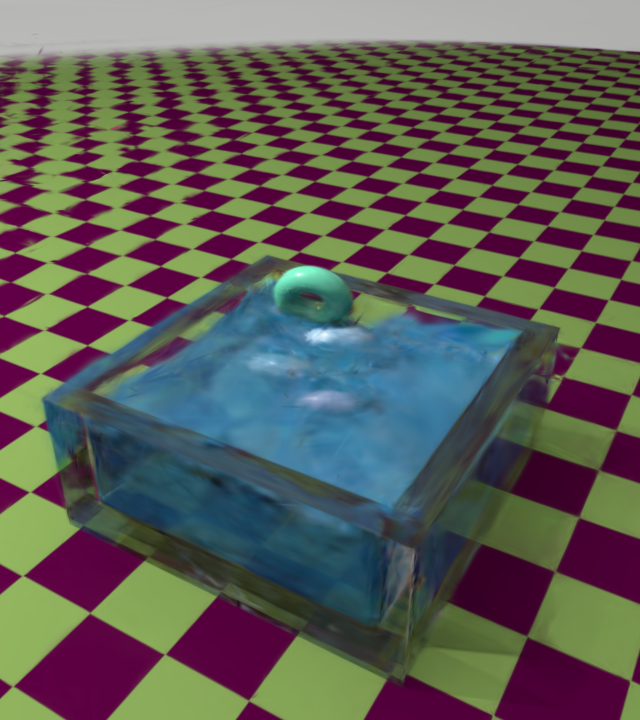}
         \includegraphics[width=\linewidth]{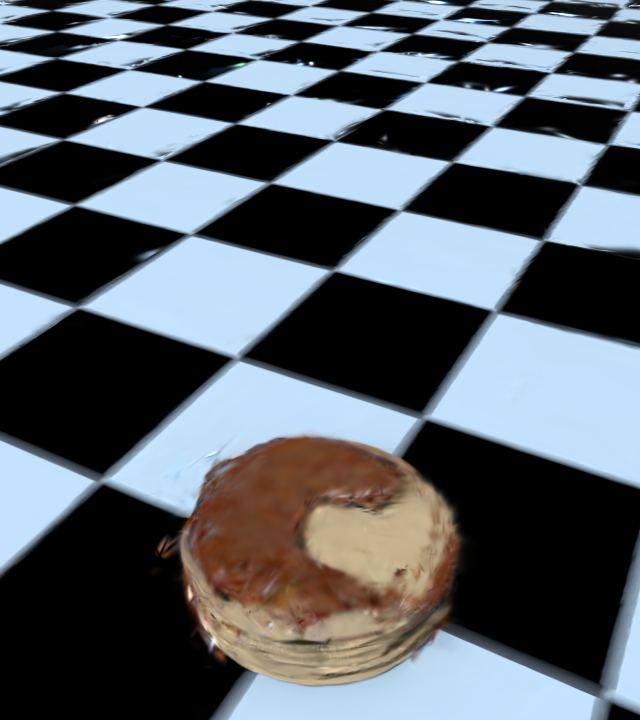}     
         \caption{D-3DGS~\cite{yang2024deformable}}
        \end{subfigure}
                        \begin{subfigure}[b]{0.161\linewidth}
        \centering
          \includegraphics[width=\linewidth]{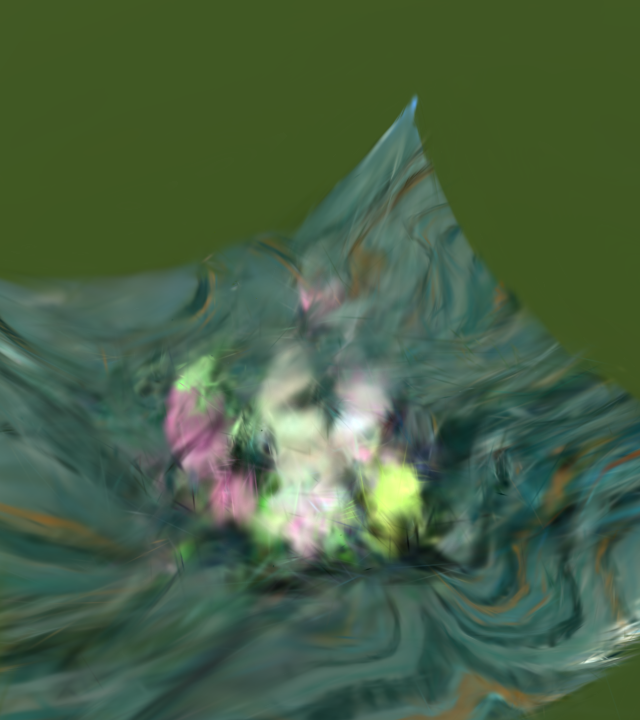}
         \includegraphics[width=\linewidth]{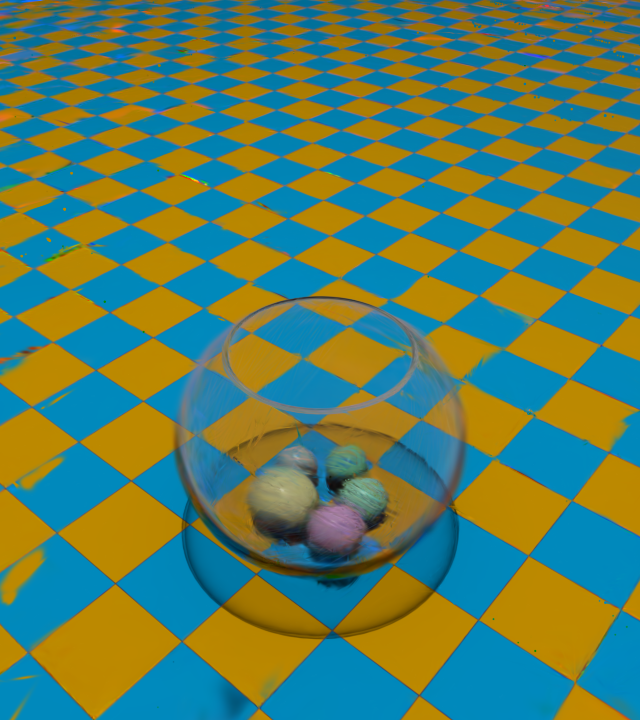}
            \includegraphics[width=\linewidth]{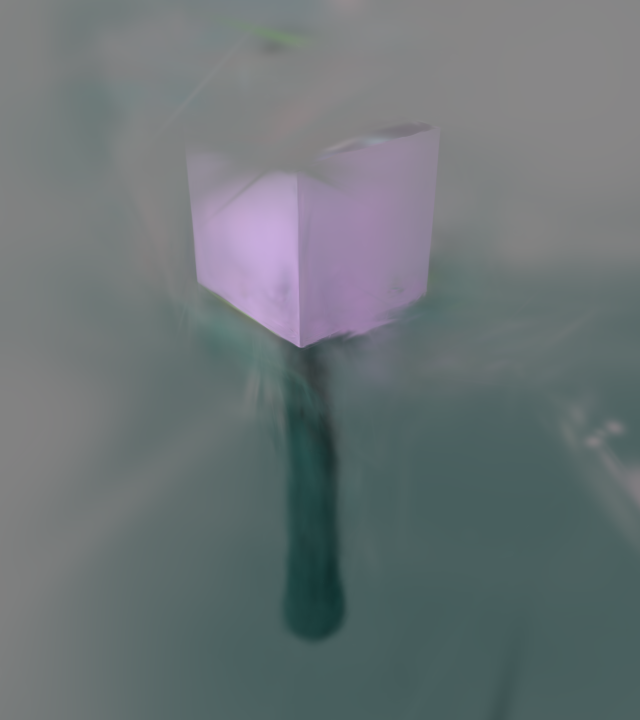}
            \includegraphics[width=\linewidth]{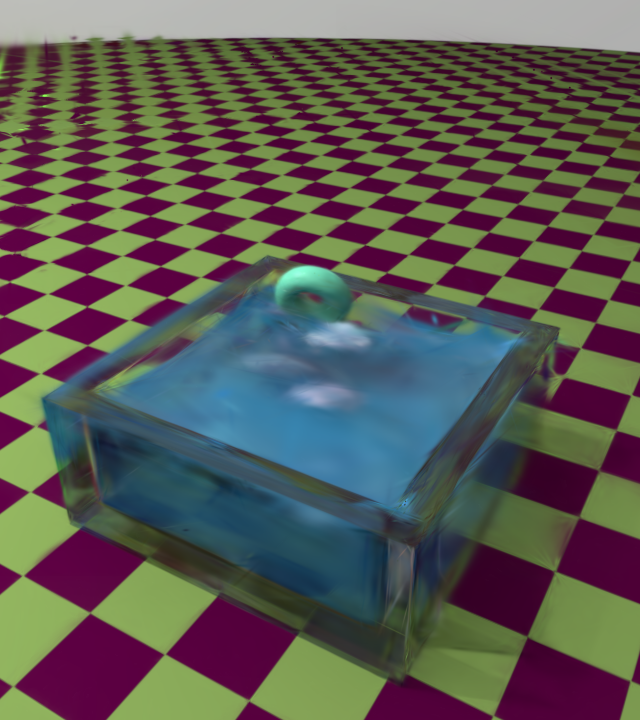}
             \includegraphics[width=\linewidth]{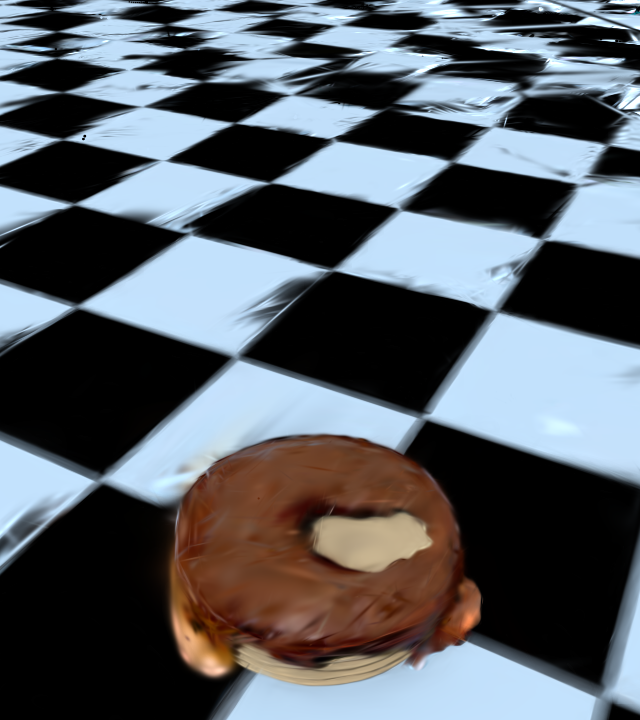}  
          \caption{4DGS~\cite{wu20234d}}
        \end{subfigure}
                        \begin{subfigure}[b]{0.161\linewidth}
        \centering
          \includegraphics[width=\linewidth]{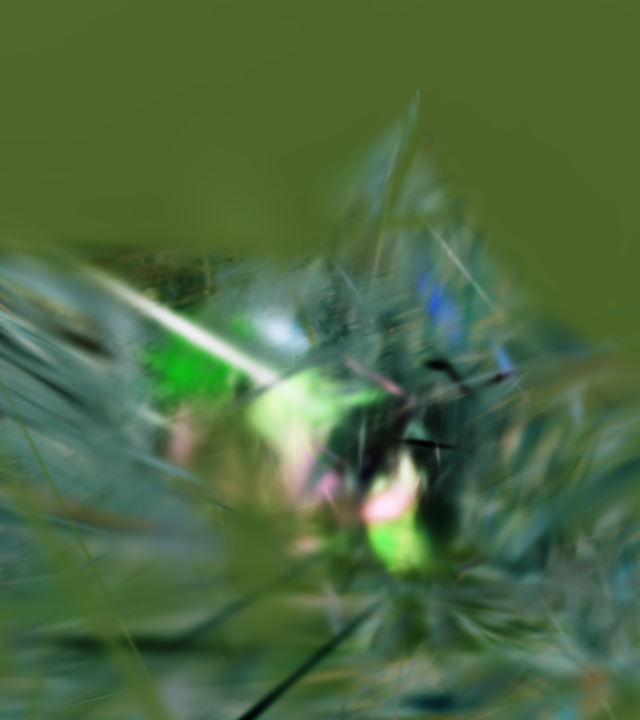}
         \includegraphics[width=\linewidth]{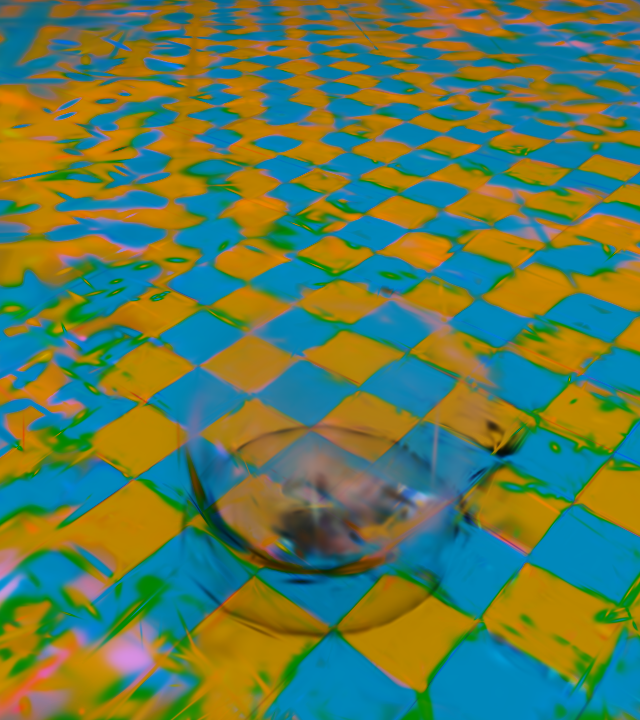}
                                    \includegraphics[width=\linewidth]{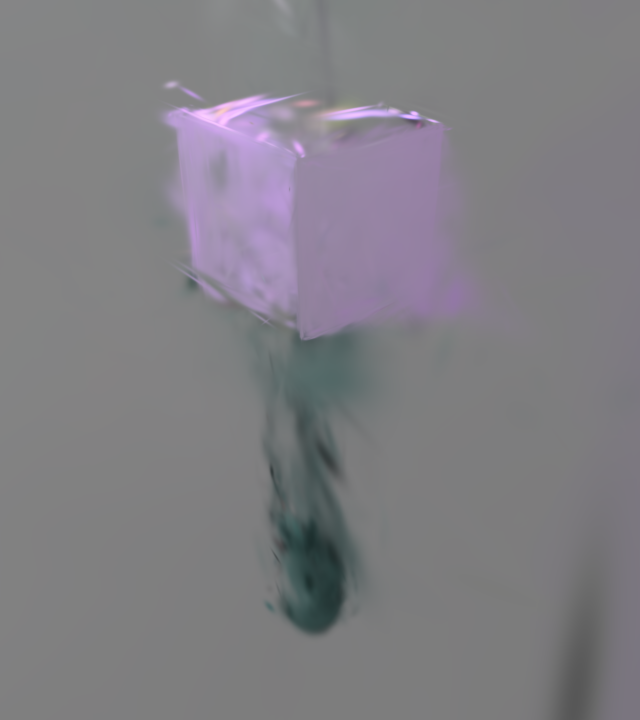}
            \includegraphics[width=\linewidth]{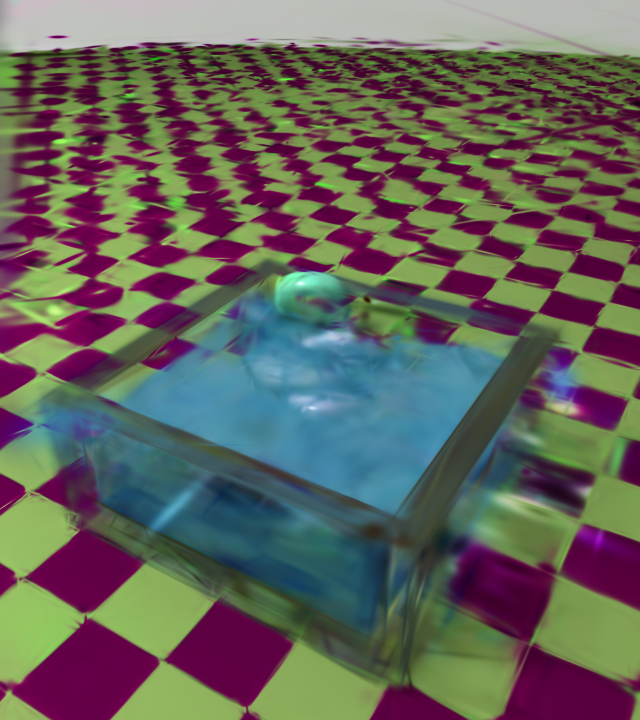}
         \includegraphics[width=\linewidth]{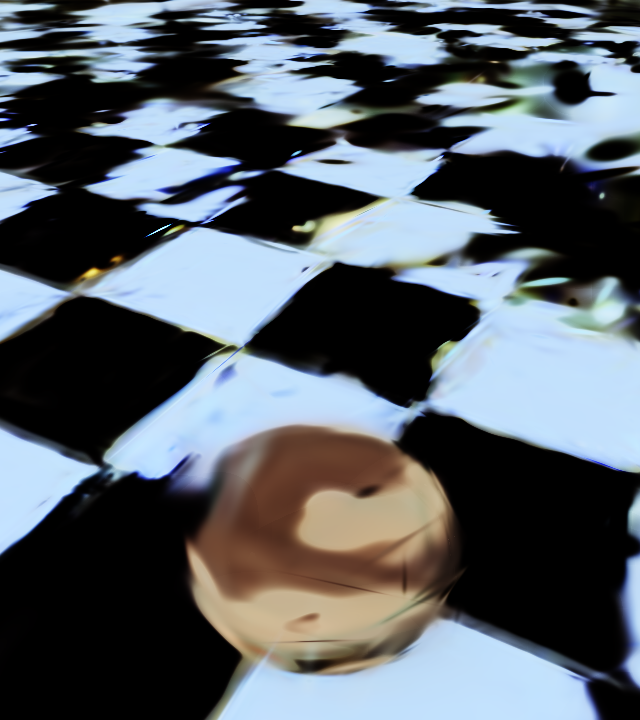}   
        \caption{STG~\cite{li2024spacetime}}
        \end{subfigure}
                                \begin{subfigure}[b]{0.161\linewidth}
        \centering
          \includegraphics[width=\linewidth]{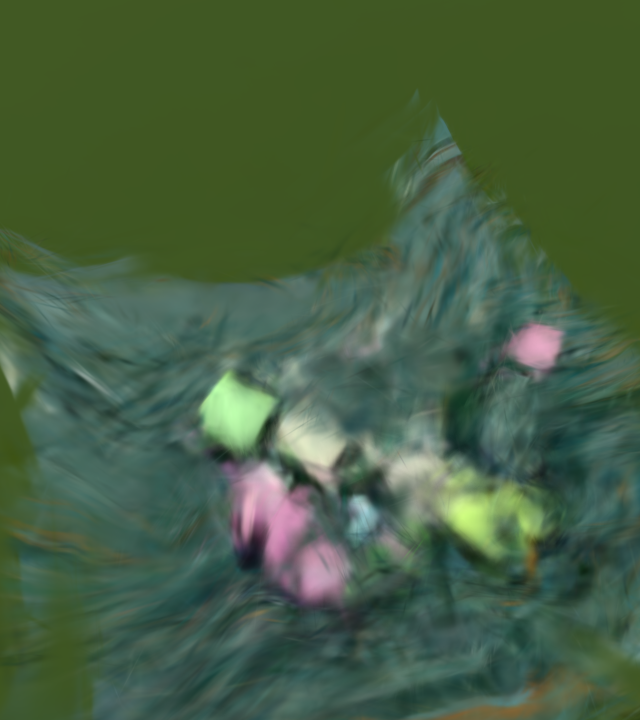}
         \includegraphics[width=\linewidth]{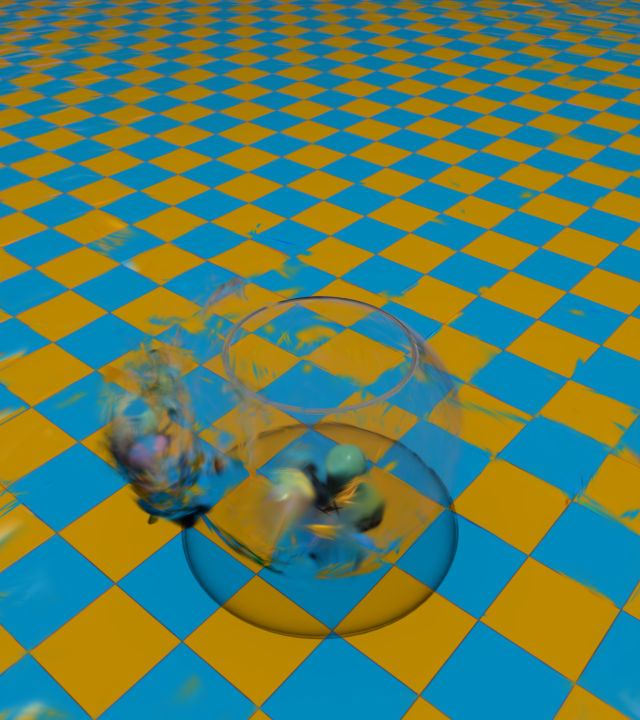}
            \includegraphics[width=\linewidth]{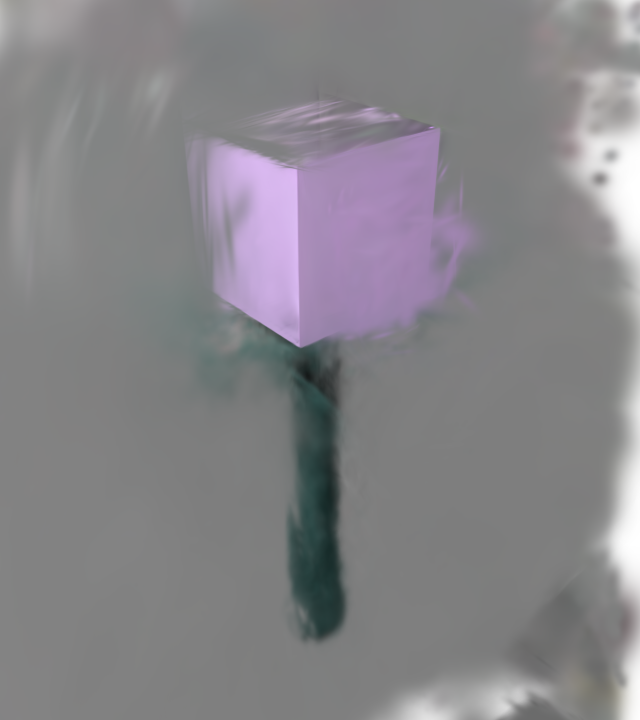}
            \includegraphics[width=\linewidth]{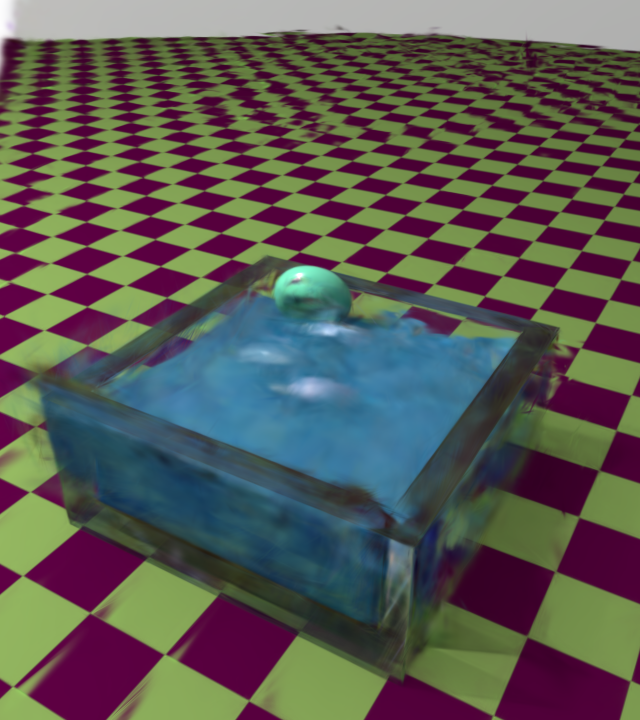}
         \includegraphics[width=\linewidth]{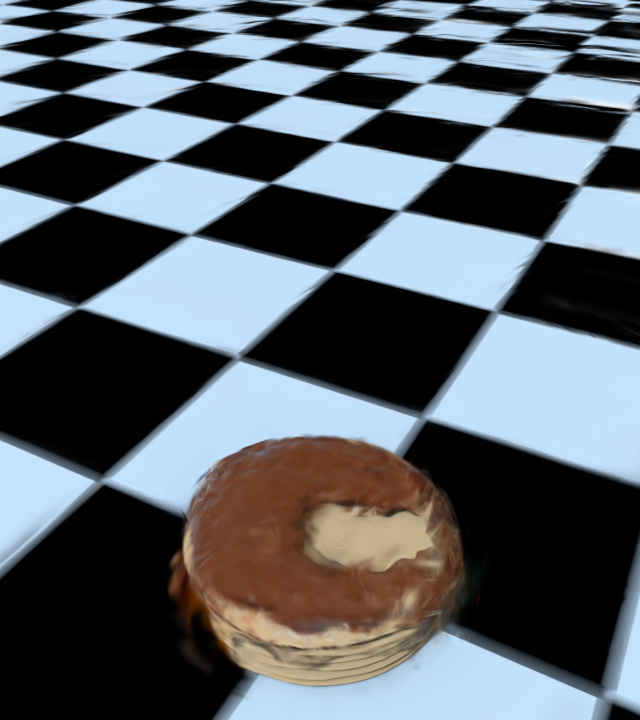}  
      \caption{MoSca~\cite{lei2024mosca}}
        \end{subfigure}
    \begin{subfigure}[b]{0.161\linewidth}
        \centering
          \includegraphics[width=\linewidth]{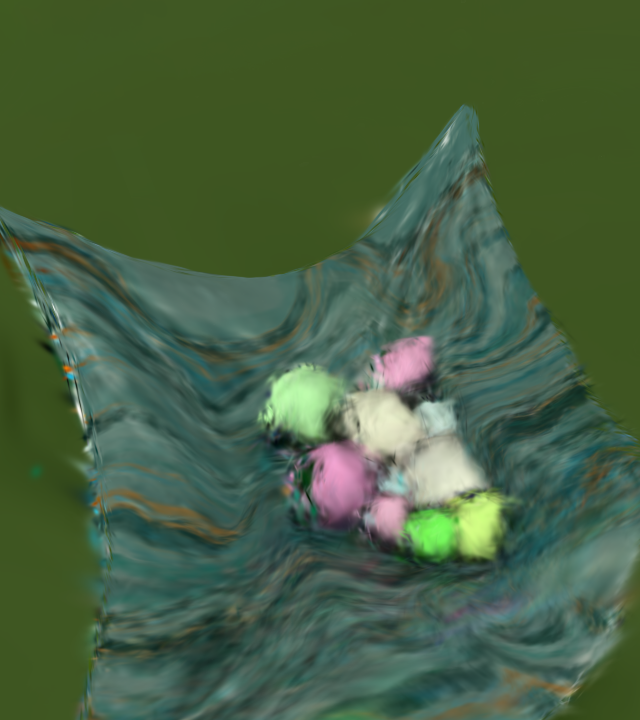}
         \includegraphics[width=\linewidth]{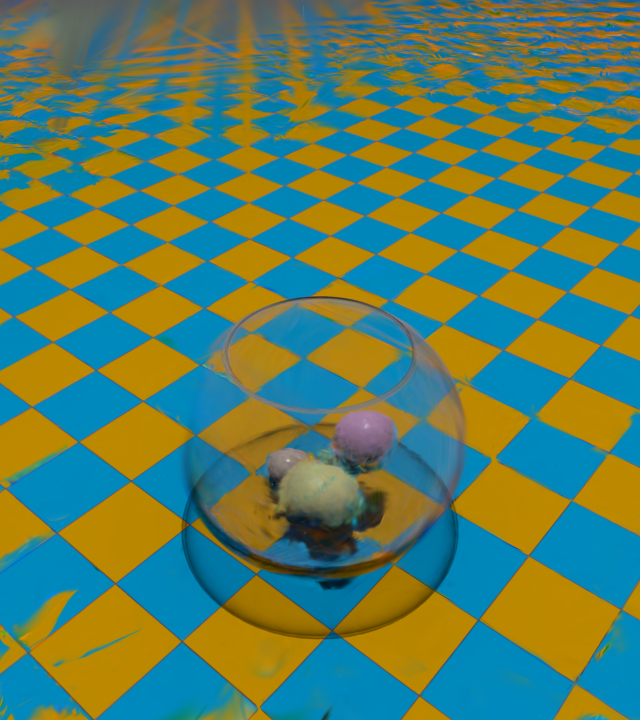}
            \includegraphics[width=\linewidth]{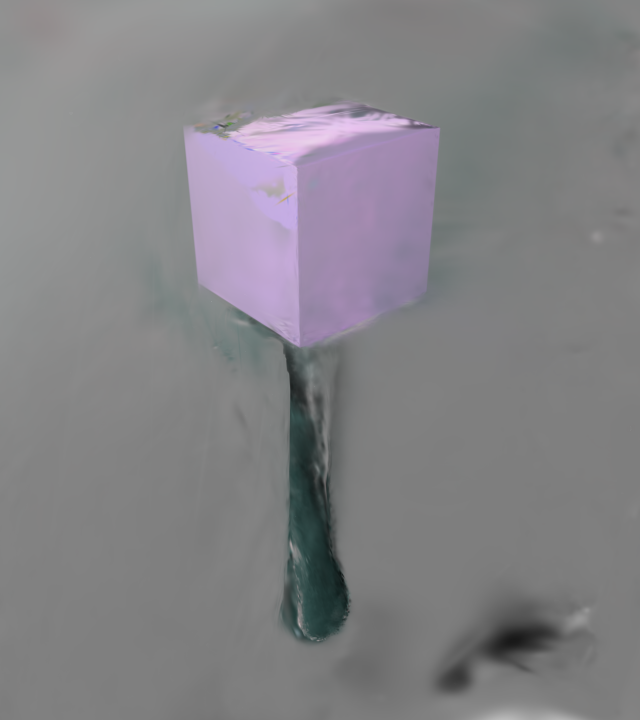}
            \includegraphics[width=\linewidth]{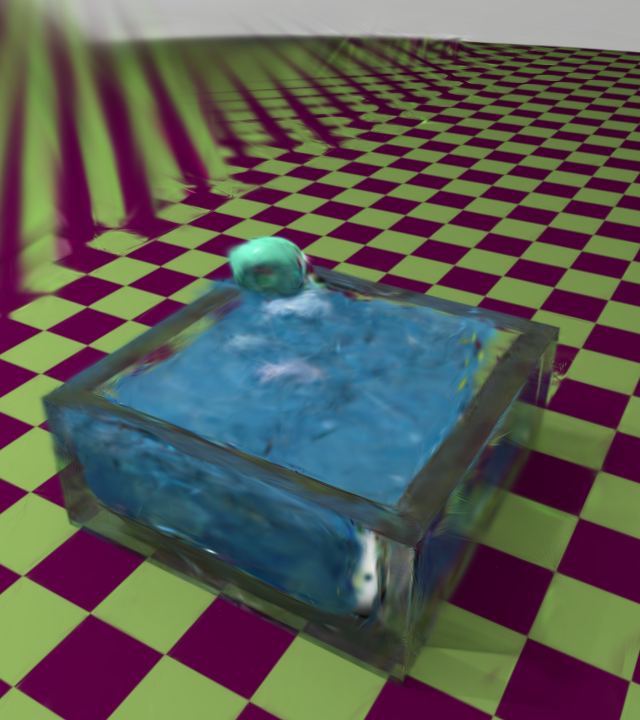}
         \includegraphics[width=\linewidth]{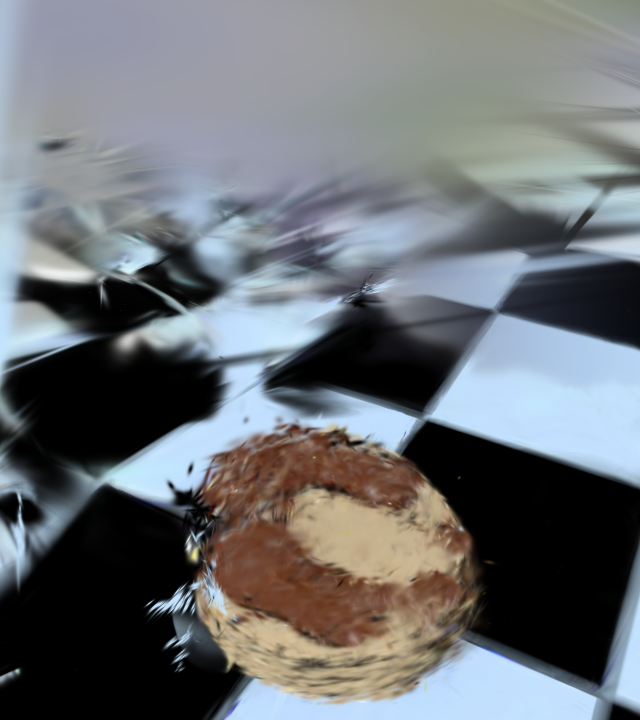}  
      \caption{SOM~\cite{wang2024shape}}
        \end{subfigure}
  \caption{
  Qualitative results of recent DyNVS methods on the \textit{basin}, \textit{bouncing balls}, \textit{box-smoke}, \textit{pancake}, and \textit{ice} scenes with monocular training setup.
These results show that all methods frequently exhibit needle-like artifacts and fail to reconstruct dynamic elements accurately.}
  	\label{fig:qualitative_result_supple}
	\vspace{-4mm}
    \end{center}%
\end{figure*}

Figure~\ref{fig:qualitative_result_supple} and Figure~\ref{fig:qualitative_result_main} present qualitative results from monocular settings, revealing that all methods struggle to accurately capture multi-body interactions. 
This leads to common issues like needle-like artifacts and under-reconstruction of dynamic regions. 
For 4DGS~\citep{wu20234d}, its default grid learning rate often causes NaN loss values; while reducing it stabilizes training, the output resembles static scenes due to poor dynamic capture. 

\begin{figure*}[h]
    \begin{center}
        \centering
                        \begin{subfigure}[b]{0.161\linewidth}
        \centering
	\includegraphics[width=\linewidth]{./asset/supple_result_viz/MPM_falling_jelly_1_019_00036_gt.png}
          \includegraphics[width=\linewidth]{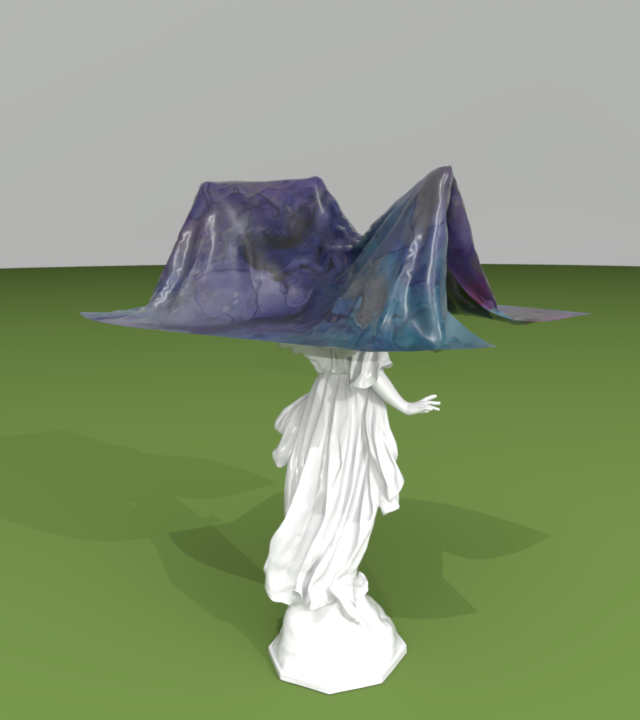}
            \includegraphics[width=\linewidth]{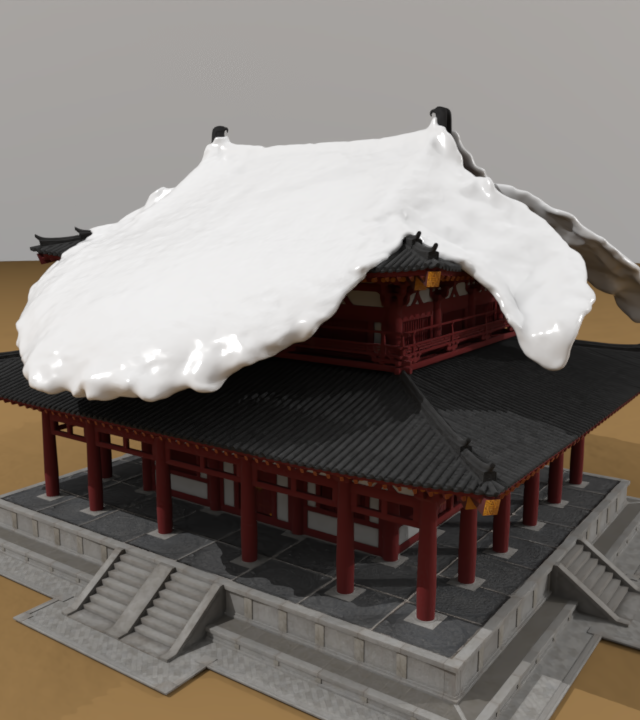}
                      \includegraphics[width=\linewidth]{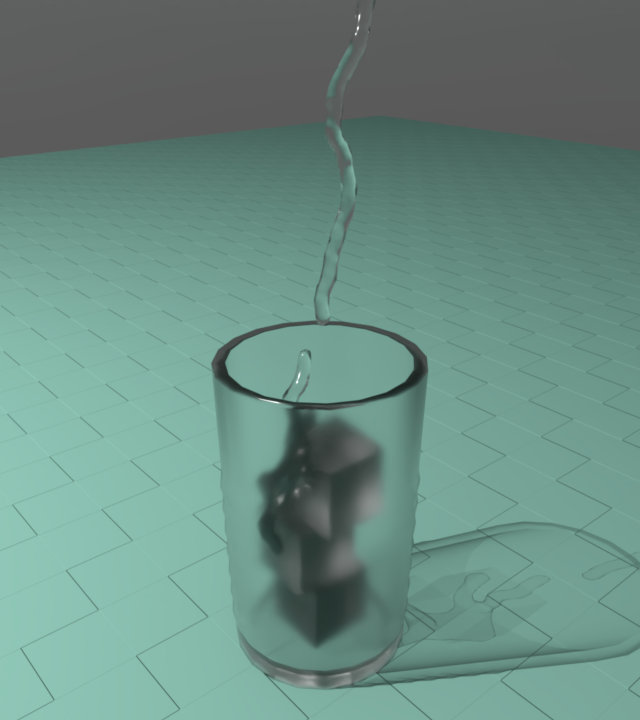}
        \caption{GT image}
        \end{subfigure}
                        \begin{subfigure}[b]{0.161\linewidth}
        \centering
	\includegraphics[width=\linewidth]{./asset/supple_result_viz/MPM_falling_jelly_1_019_00036_deform.png}
          \includegraphics[width=\linewidth]{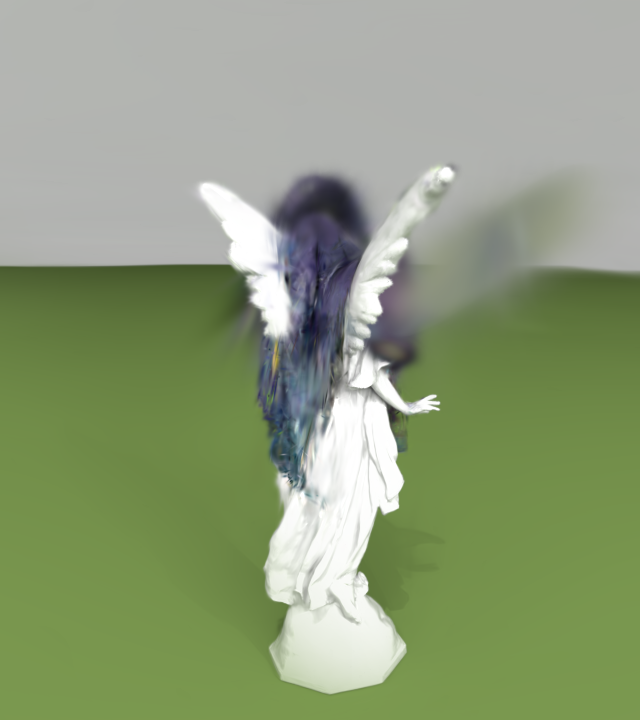}
                        \includegraphics[width=\linewidth]{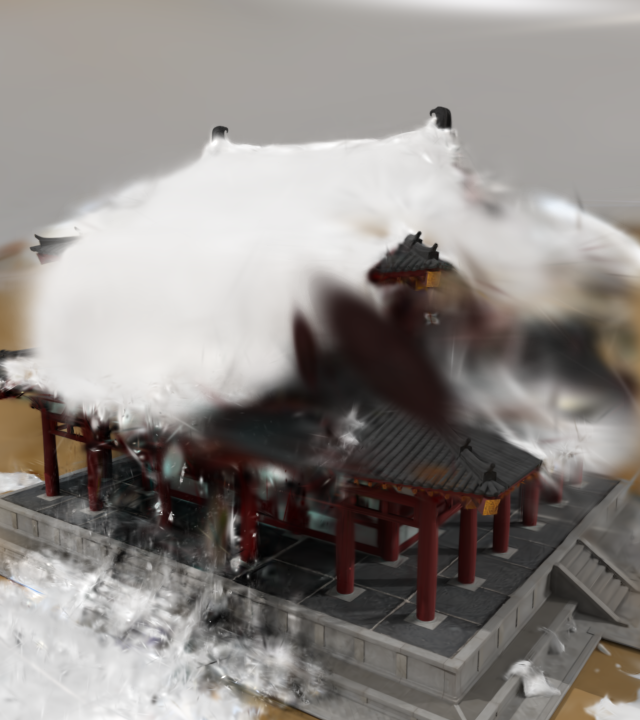}
                          \includegraphics[width=\linewidth]{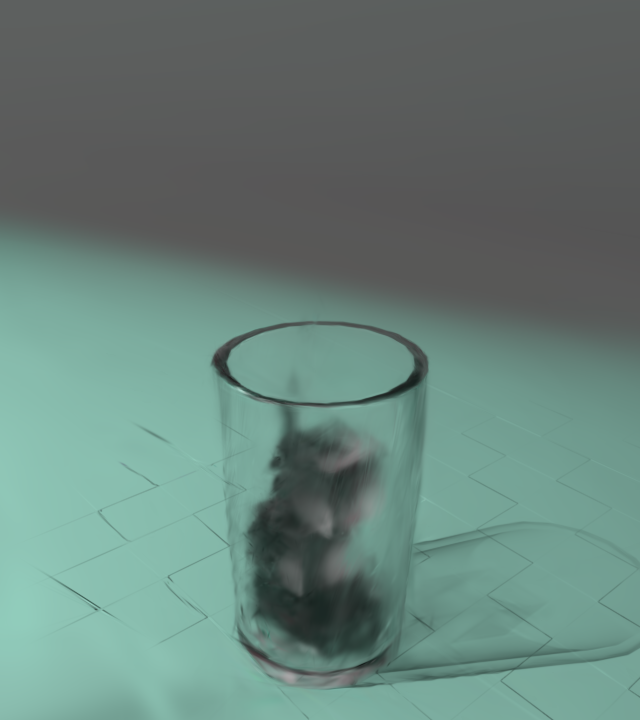}
         \caption{D-3DGS~\cite{yang2024deformable}}
        \end{subfigure}
                        \begin{subfigure}[b]{0.161\linewidth}
        \centering
	\includegraphics[width=\linewidth]{./asset/supple_result_viz/MPM_falling_jelly_1_019_00036_4dgs.png}
          \includegraphics[width=\linewidth]{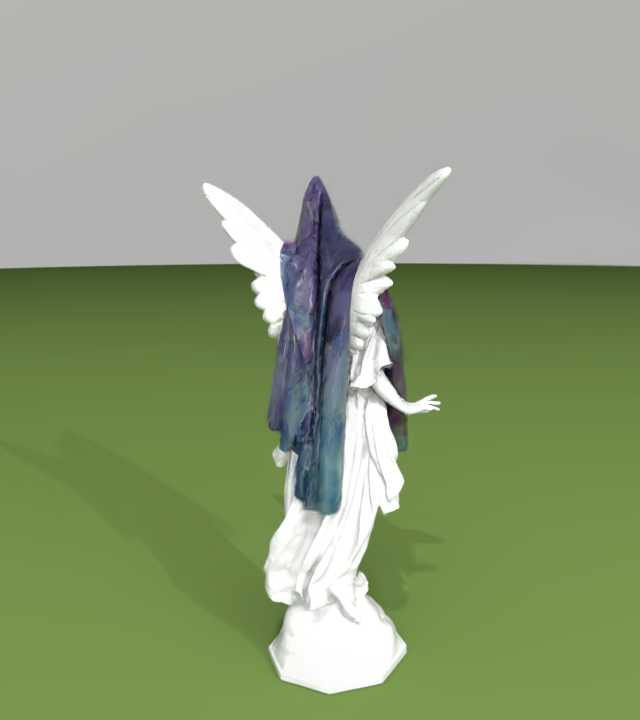}
                        \includegraphics[width=\linewidth]{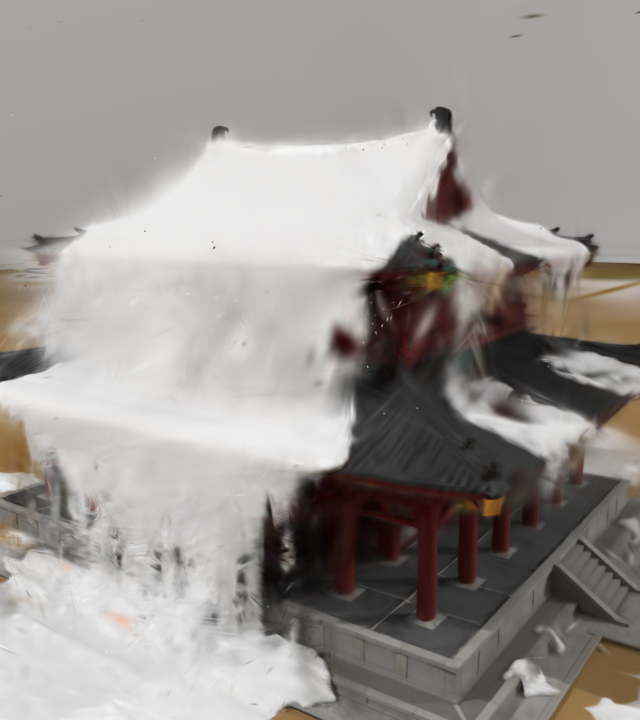}
                         \includegraphics[width=\linewidth]{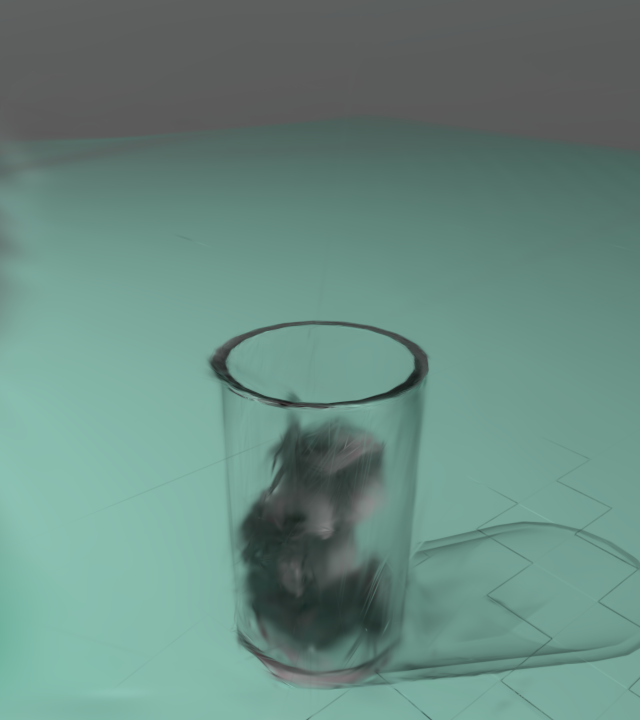}
          \caption{4DGS~\cite{wu20234d}}
        \end{subfigure}
                        \begin{subfigure}[b]{0.161\linewidth}
        \centering
	\includegraphics[width=\linewidth]{./asset/supple_result_viz/MPM_falling_jelly_1_019_00036_spacetime.png}
          \includegraphics[width=\linewidth]{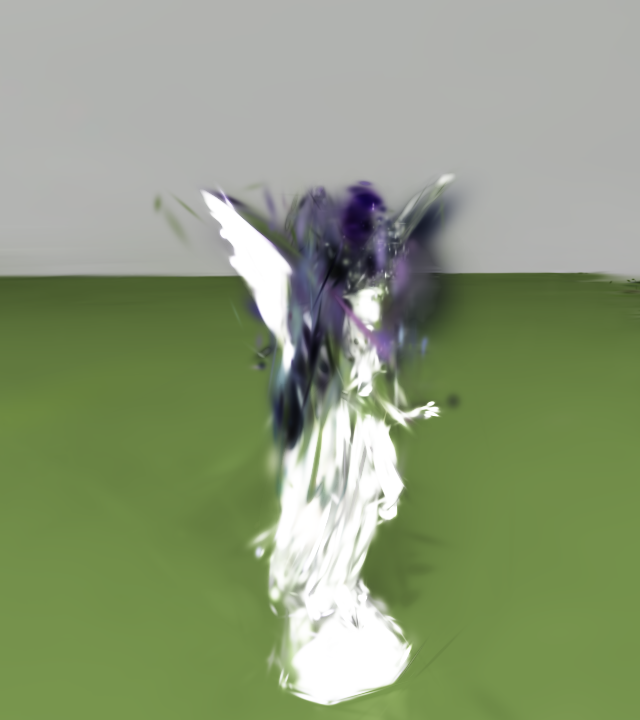}
                        \includegraphics[width=\linewidth]{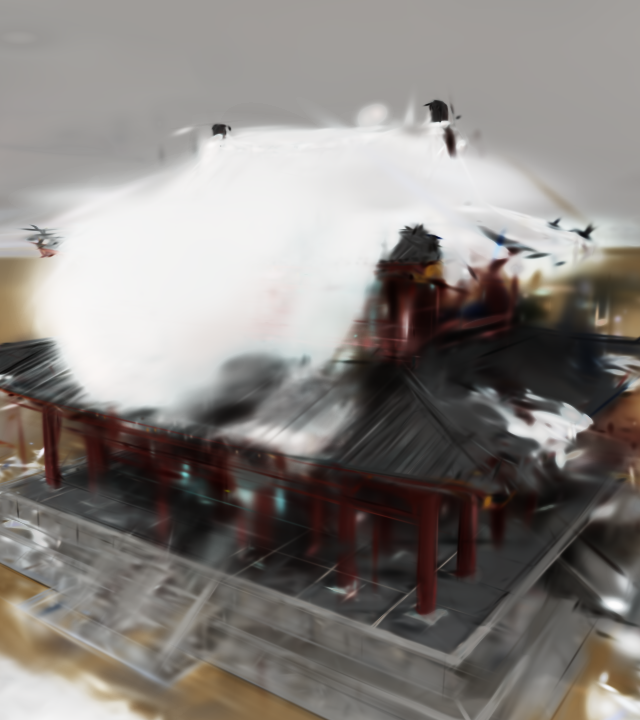}
                         \includegraphics[width=\linewidth]{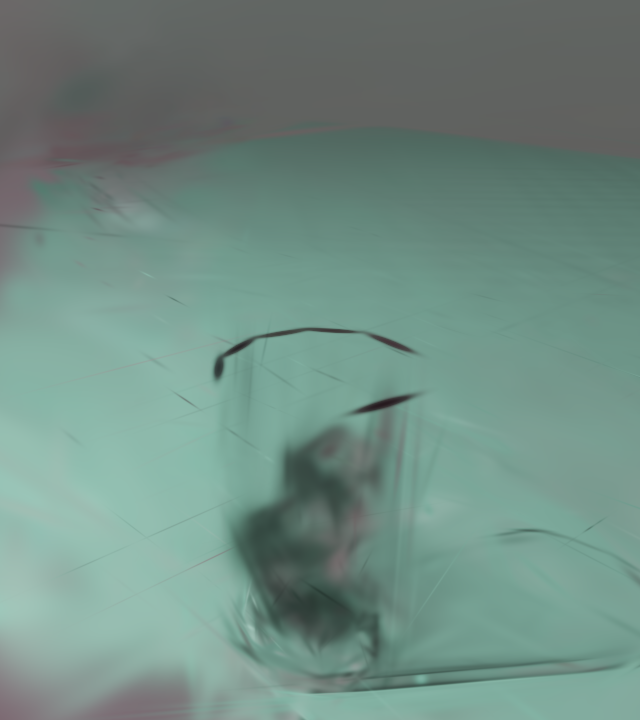}
        \caption{STG~\cite{li2024spacetime}}
        \end{subfigure}
            \begin{subfigure}[b]{0.161\linewidth}
        \centering
	\includegraphics[width=\linewidth]{./asset/supple_result_viz/MPM_falling_jelly_1_019_00036_mosca.png}
          \includegraphics[width=\linewidth]{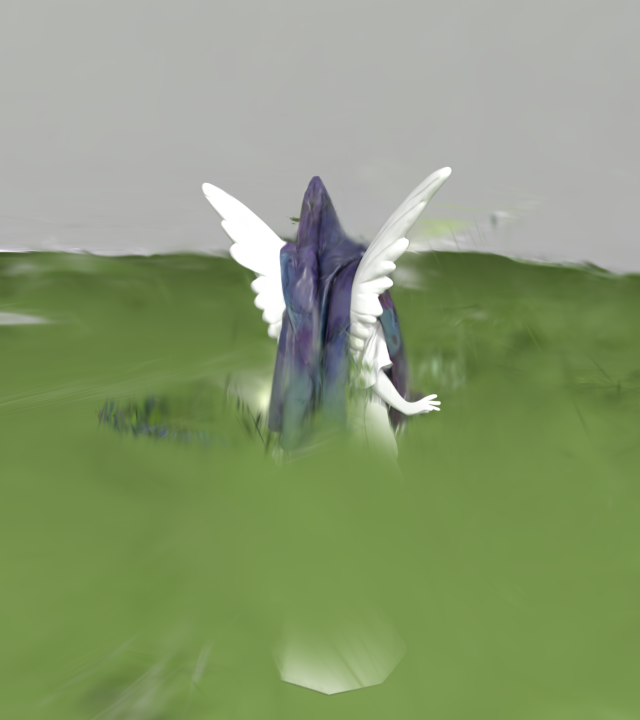}
                        \includegraphics[width=\linewidth]{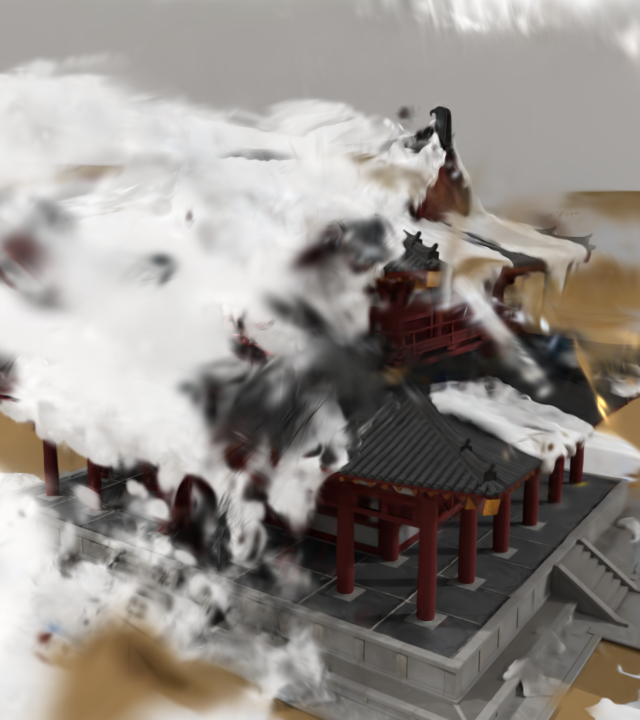}
                        \includegraphics[width=\linewidth]{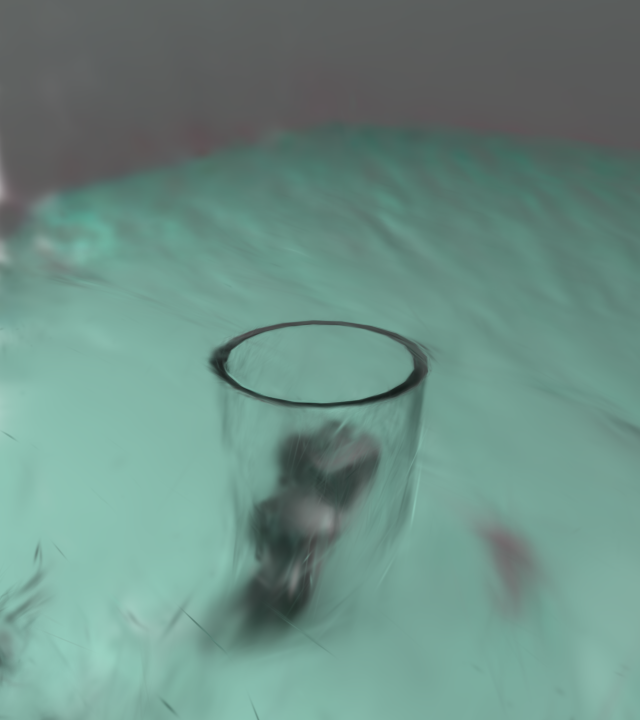}  
         \caption{MoSca~(\citeyear{lei2024mosca})}
        \end{subfigure}
                        \begin{subfigure}[b]{0.161\linewidth}
        \centering
	\includegraphics[width=\linewidth]{./asset/supple_result_viz/MPM_falling_jelly_1_019_00036_som.png}
          \includegraphics[width=\linewidth]{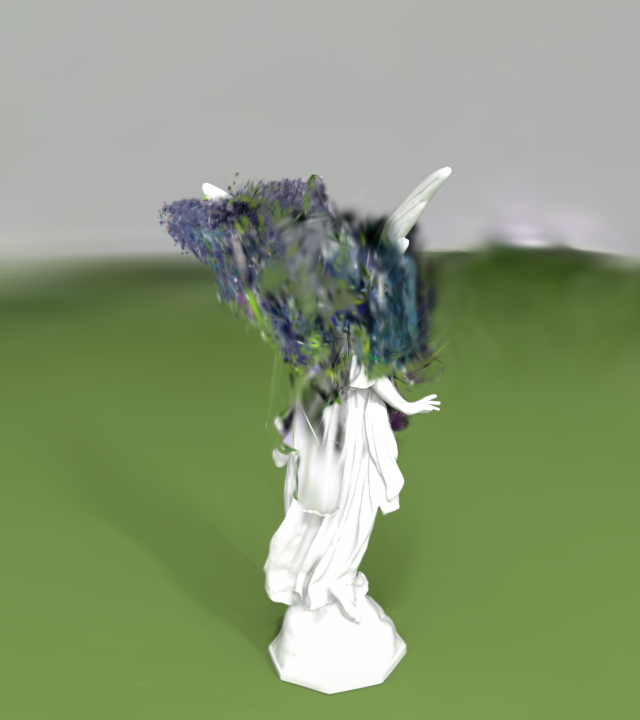}
                        \includegraphics[width=\linewidth]{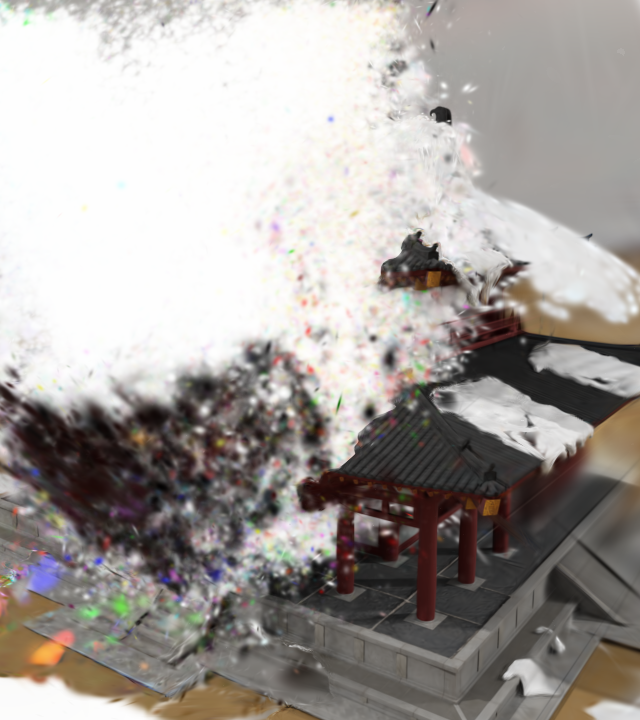}
                                \includegraphics[width=\linewidth]{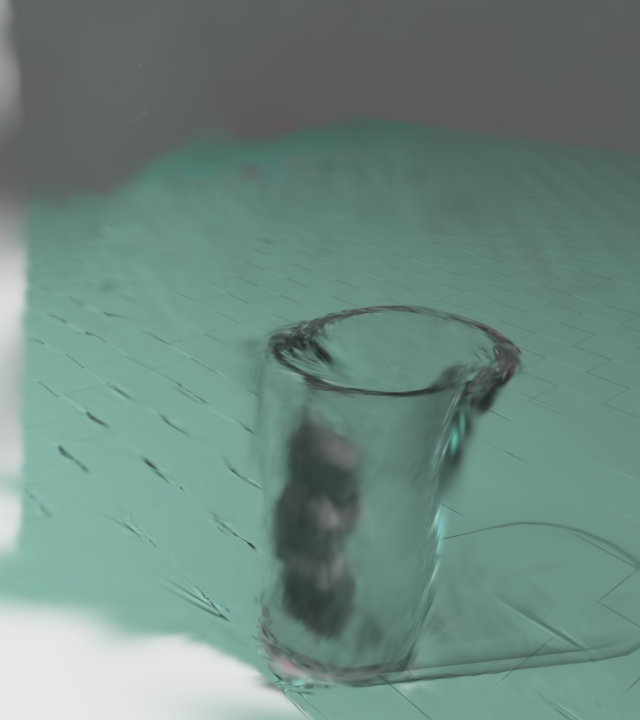}
         \caption{SOM~\cite{wang2024shape}}
        \end{subfigure}
  \caption{
Qualitative results of recent DyNVS methods on the \textit{lucy}, and \textit{hanok} scenes with monocular setup.
All methods struggle to accurately capture multi-body interactions by frequently exhibiting needle-like artifacts and failing to reconstruct dynamic elements accurately.}
  	\label{fig:qualitative_result_main}
	\vspace{-3mm}
    \end{center}%
\end{figure*}

\section{Limitations \& Broad Impact}

\paragraph{Limitations}
While our \textit{\dataset} is built upon a physics simulator to provide physically consistent supervision, it inherits the inherent approximations of the underlying physical models. 
In practice, physics simulators rely on simplified constitutive assumptions and numerical discretizations to make the problem tractable and stable. 
As a result, certain aspects of real-world dynamics, such as complex material responses or fine-scale interactions, may not be fully captured. 
We view this not as a limitation of our dataset alone, but as an inherent characteristic of simulation-based approaches, and an important direction for future work.

\vspace{-2mm}
\paragraph{Broad impact}
From a positive perspective, our work advances research on physically plausible 4D reconstruction. 
This, in turn, significantly improves monocular video reconstruction, an essential technology for future AR/VR applications. 
However, such progress in 4D reconstruction, especially given its potential for scene editing of existing videos, may raise intellectual property concerns regarding the original video content.

\clearpage
\section{License}

Our benchmark is released under the Creative Commons Attribution-NonCommercial (CC BY-NC) license.
The mesh used for the \textit{lucy} scene is sourced from the Stanford 3D Repository, which permits usage for research purposes. Other mesh objects are licensed under CC-BY 4.0, as detailed in Table~\ref{tab:license}. 
For the texture maps in the textile categories, we utilized images from \href{https://pixabay.com}{Pixabay}, which are freely available with contributor consent and comparable to a CC-BY 4.0 license, also summarized in Table~\ref{tab:license}. 
Therefore, all data included in our benchmark comply with usage rights and do not pose intellectual property issues.

\begin{table*}[h]
\centering
\caption{
License of the sources  used for generating data.
The licenses for the mesh objects follow Creative Commons terms, except for the Lucy mesh sourced from the Stanford 3D Repository~\cite{Stanford3DRepo}. 
The images used in the {\dataset} scenes, along with their download sources, are licensed under the \href{https://pixabay.com/service/terms}{Content License} granted by \href{https://pixabay.com}{Pixabay}.
}
\label{tab:mesh-cc}
\scalebox{0.8}{
\setlength\tabcolsep{35pt} 
\begin{tabular}{llccccc}
\toprule
 \textbf{Scene}	& \textbf{Resource name} &\textbf{License}  & \textbf{Access} \\
\midrule
Ice								& Glass cup		&CC BY 4.0		&\href{https://sketchfab.com/3d-models/glass-cup-40acfc3e537d4808b1332854f9de0355}{Sketchfab}\\
Hanok							& Korean building			&CC BY 4.0	& \href{https://sketchfab.com/3d-models/korea-traditional-house-agricultural-d5834a3560d249fc8b059f78673d108c}{Sketchfab}\\
Ship								&Ship			&CC BY 4.0		&\href{https://sketchfab.com/3d-models/ship-j-dab8c8407a114ce9b1f5b1bd8028b467}{Sketchfab}\\
Pisa							&Torre Pisa		&CC BY 4.0		& \href{https://sketchfab.com/3d-models/torre-pisa-1ff02ed0588e4e4f9dcf621a8af07336}{Sketchfab} \\ 
Bouncing balls						& Fish bowl 		&CC BY 4.0		&\href{https://sketchfab.com/3d-models/simple-fish-bowl-dfd5807cadd1480785dfe45349d8182a}{Sketchfab}\\
Cow								& Cow		&CC BY 4.0		& \href{https://sketchfab.com/3d-models/pbr-cow-head-free-3d-model-271e5642793b4239b7ba40b0ad54d082}{Sketchfab}\\
Lucy								&Lucy			& research only	 & \href{https://graphics.stanford.edu/data/3Dscanrep}{Stanford 3D Scan Repo}\\
\hline
\hline
 \textbf{Scene}	& \textbf{Contributor} &\textbf{License}  & \textbf{Access}\\
 \hline
\multirow{2}{*}{Flags} 		& PaftDrunk &\href{https://pixabay.com/service/terms}{Content License}		&\href{https://pixabay.com/ko/illustrations/\%EA\%B0\%9C\%EC\%9A\%94-\%EC\%9E\%89\%ED\%81\%AC-\%EB\%B0\%B0\%EA\%B2\%BD-\%ED\%8C\%8C\%EB\%8F\%84-6305508/}{Pixabay}\\
						&Zoeysmom  &\href{https://pixabay.com/service/terms}{Content License}			&\href{https://pixabay.com/ko/illustrations/\%EC\%9E\%89\%ED\%81\%AC-\%EB\%AC\%BC-\%EC\%83\%89\%EA\%B9\%94-\%EC\%82\%BD\%ED\%99\%94-\%EB\%B0\%B0\%EA\%B2\%BD-7294678/}{Pixabay}\\
Single flag 				&-- &\href{https://pixabay.com/service/terms}{Content License}	&\href{https://pixabay.com/ko/illustrations/\%EC\%A1\%B0\%EC\%A7\%81-\%EA\%B0\%9C\%EC\%9A\%94-\%EA\%B5\%AC\%EC\%A1\%B0-\%ED\%99\%94\%EB\%A0\%A4\%ED\%95\%9C-1909992/}{Pixabay}\\
Lucy 					& WalterClark &\href{https://pixabay.com/service/terms}{Content License} &\href{https://pixabay.com/ko/illustrations/\%EC\%9E\%89\%ED\%81\%AC-\%EB\%B0\%B0\%EA\%B2\%BD-\%EC\%86\%8C\%EC\%9A\%A9\%EB\%8F\%8C\%EC\%9D\%B4-\%ED\%8C\%8C\%EB\%9E\%80\%EC\%83\%89-2229457/}{Pixabay} \\
Tube						& -- &\href{https://pixabay.com/service/terms}{Content License} &\href{https://pixabay.com/ko/illustrations/\%EB\%B0\%B0\%EA\%B2\%BD-\%ED\%85\%8D\%EC\%8A\%A4\%EC\%B2\%98-\%EB\%AC\%B4\%EB\%8A\%AC-\%EC\%84\%A4\%EA\%B3\%84-1872844/}{Pixabay} \\ 
Basin					&Yourialka  &\href{https://pixabay.com/service/terms}{Content License} &\href{https://pixabay.com/ko/illustrations/\%EB\%8C\%80\%EB\%A6\%AC\%EC\%84\%9D-\%EA\%B2\%B0\%EC\%84\%9D-\%EA\%B0\%9C\%EC\%9A\%94-\%EB\%AC\%B4\%EB\%8A\%AC-7728245/}{Pixabay}
\\

\bottomrule
\end{tabular}
}
\label{tab:license}
\vspace{4mm}
\end{table*}

\end{document}